\documentclass[onecolumn]{emulateapj}

\usepackage{color}
\usepackage[pdf]{pstricks}

\RequirePackage{color}

\shortauthors{Komiyama et al.}
\shorttitle{Subaru/HSC deep imaging of the M31 halo}

\begin{document}

\title{Stellar Stream and Halo Structure in the Andromeda Galaxy
From a Subaru/Hyper Suprime-Cam Survey\altaffilmark{1}}

\author{Yutaka~Komiyama\altaffilmark{2,3}, Masashi~Chiba\altaffilmark{4}, 
Mikito~Tanaka\altaffilmark{5}, Masayuki~Tanaka\altaffilmark{2}, 
Takanobu~Kirihara\altaffilmark{6,7}, 
Yohei~Miki\altaffilmark{8,9,10}, 
Masao~Mori\altaffilmark{6,8}, 
Robert~H.~Lupton\altaffilmark{11}, 
Puragra~Guhathakurta\altaffilmark{12}, 
Jason~S.~Kalirai\altaffilmark{13,14}, 
Karoline~Gilbert\altaffilmark{13,14}, 
Evan~Kirby\altaffilmark{15}, 
Myun~Gyoon~Lee \altaffilmark{16}, 
In~Sung~Jang\altaffilmark{16}, 
Sanjib~Sharma\altaffilmark{17}, 
Kohei~Hayashi\altaffilmark{18,19,20} 
}

\altaffiltext{1}{Based on data collected at Subaru Telescope,
        which is operated by the National Astronomical Observatory of Japan}
\altaffiltext{2}{National Astronomical Observatory of Japan, 2-21-1 Osawa, 
	Mitaka, Tokyo 181-8588, Japan \\E-mail: {\it komiyama@subaru.naoj.org}}
\altaffiltext{3}{Graduate University for Advanced Studies (SOKENDAI), 
	2-21-1 Osawa, Mitaka, Tokyo 181-8588, Japan}
\altaffiltext{4}{Astronomical Institute, Tohoku University, Aoba-ku,
	Sendai 980-8578, Japan}
\altaffiltext{5}{Frontier Research Institute for Interdisciplinary Sciences, 
	Tohoku University, Aoba-ku, Sendai 980-8578, Japan}
\altaffiltext{6}{Faculty of Pure and Applied Physics, University of Tsukuba, 
	Tennodai 1-1-1, Tsukuba, Ibaraki, 305-8577, Japan}
\altaffiltext{7}{Current Address: 
	Institute of Management and Information Technologies, 
	Chiba University, 1-33, Yayoi-cho, Inage-ku, Chiba 263-8522, Japan}
\altaffiltext{8}{Center for Computational Sciences, University of Tsukuba, 
	Tennodai 1-1-1, Tsukuba, Ibaraki, 305-8577, Japan}
\altaffiltext{9}{CREST, JST, Tennodai 1-1-1, Tsukuba, Ibaraki, 305-8577, Japan}
\altaffiltext{10}{Current Address: 
	Information Technology Center, The University of Tokyo, 
	5-1-5 Kashiwanoha, Kashiwa, Chiba 277-8589, Japan}
\altaffiltext{11}{Department of Astrophysical Sciences, 
	Princeton University, 4 Ivy Lane, Princeton, NJ 08544}
\altaffiltext{12}{UCO/Lick Observatory and Department of 
	Astronomy and Astrophysics, University of California, 
	1156 High Street, Santa Cruz, CA 95064, USA}
\altaffiltext{13}{Space Telescope Science Institute, Baltimore, MD 21218, USA}
\altaffiltext{14}{Center for Astrophysical Sciences, Johns Hopkins University, 
	Baltimore, MD, 21218}
\altaffiltext{15}{California Institute of Technology, 
	1200 E. California Boulevard, MC 249-17, Pasadena, CA 91125, USA}
\altaffiltext{16}{Department of Physics and Astronomy, 
	Seoul National University, Seoul 151-742, Korea}
\altaffiltext{17}{Sydney Institute for Astronomy, School of Physics, 
	University of Sydney, NSW 2006, Australia}
\altaffiltext{18}{Kavli Institute for Astronomy and Astrohysics, 
	Peking University, Beijing 100871, China}
\altaffiltext{19}{Kavli Institute for the Physics and Mathematics of 
	the Universe (Kavli IPMU, WPI), The University of Tokyo, 
	Chiba 277-8583, Japan}
\altaffiltext{20}{Current Address: National Astronomical Observatory of Japan, 
	2-21-1 Osawa, Mitaka, Tokyo 181-8588, Japan}

\begin{abstract}
We present wide and deep photometry of the northwest part of
the halo of the Andromeda galaxy (M31)
using Hyper Suprime-Cam on the Subaru Telescope.
The survey covers 9.2 deg$^{2}$ field in the $g$, $i$, and $NB515$ bands 
and shows a clear red giant branch (RGB) of M31's halo stars
and a pronounced red clump (RC) feature. 
The spatial distribution of RC stars shows a prominent stream feature,
the North Western (NW) Stream,
and a diffuse substructure in the south part of our survey field.
We estimate the distances 
based on the RC method and obtain 
$(m-M)$ = 24.63$\pm 0.191$(random)$\pm0.057$(systematic)
and 24.29$\pm 0.211$(random)$\pm0.057$(systematic) mag
for the NW stream and diffuse substructure, respectively,
implying that the NW Stream
is located behind M31,
whereas the diffuse substructure is located in front.
We also estimate line-of-sight distances along the NW Stream
and find that the south part of the stream is $\sim$20 kpc
closer to us relative to the north part.
The distance to the NW Stream inferred from the isochrone fitting 
to the color-magnitude diagram favors the RC-based distance, 
but the TRGB-based distance estimated for $NB515$-selected 
RGB stars does not agree with it. 
The surface number density distribution of RC stars 
across the NW Stream is found to be 
approximately Gaussian with a FWHM of $\sim$25 arcmin (5.7 kpc), 
with a slight skew to the south-west side.
That along the NW Stream shows
a complicated structure including variations in number density  
and a significant gap in the stream.

\end{abstract}


\section{Introduction}

Faint stellar halos in disk galaxies like the Milky Way (MW) 
and Andromeda (M31) serve as fossil records of the formation of 
such galaxies through hierarchical assembly and past accretion events. 
In the MW, stars spread over the vast reaches of its halo region 
are characterized by low metal abundance and high velocity dispersion. 
The extreme nature of halo stars, compared to stars comprising 
the disk component, reflects the early dynamical and 
chemical evolution of the MW, 
when its appearance differed significantly from what we see today.
Extensive analyses of such metal-deficient, old populations 
have revealed various fundamental properties of the Galactic halo, 
e.g., the multiple nature in several aspects including
its spatial structure, velocity distribution and 
chemical abundances \citep[e.g.,][]{Feltzing2013}: 
the Galactic halo consists of at least two overlapping components: 
an inner, flattened halo component having high [$\alpha$/Fe] ratio,
and an outer halo characterized by a more spherical shape, lower metallicity
and lower [$\alpha$/Fe] ratio. 
This complicated global structure of the Galactic halo 
together with recent growing observational evidence for an abundance of 
stellar streams and other substructures \citep[e.g.,][]{Belokurov2006} 
suggests that the stellar halo has formed, at least in its outer part, 
largely from an assembly process of many subsystems, 
such as dwarf galaxies, as opposed to an in situ 
dissipative collapse in its inner part. 
Indeed, this formation picture of stellar halos is 
suggested by recent numerical simulations of galaxy formation 
based on standard $\Lambda$CDM cosmology \citep{Font2011}. 
Thus, detailed studies of the halo properties can provide important clues 
to the understanding of galaxy formation.

M31's halo, the target of this work, is another excellent test-bed 
for understanding galaxy formation processes through studies of 
resolved stars in the halo: it provides an external perspective 
of the nearest large galaxy, 
where all of M31's stars are about the same distance from us 
in contrast to the situation in our own Galaxy, 
which allows us to obtain a relatively complete picture of its stellar halo. 

Past observational studies of M31's halo through large 
photometric and spectroscopic surveys of bright RGB/AGB stars 
--- e.g., the PAndAS survey using CFHT/MegaCam \citep{McConnachie2009}, 
the SPLASH survey using KPNO-Mayall/MOSAIC, Keck/DEIMOS and Subaru/Suprime-Cam
\citep{Guhathakurta2005,Gilbert2006,Tanaka2010} --- have revealed 
several characteristic properties of M31's stellar halo,
some similar to and some different from those of the MW halo.
The similarities to the MW halo are in M31's  
outer halo, which has a power-law surface brightness profile, low stellar 
density and metal-poor stars 
\citep{Guhathakurta2005,Kalirai2006,Gilbert2012,Gilbert2014,Ibata2014}, 
whereas the differences are seen in the inner halo of M31, 
which has metal-rich and intermediate-age populations 
in contrast to largely metal-poor and old halo stars in the MW. 
The presence of substructures in M31's halo, such as the Giant 
Southern Stream, and the non-isotropic distribution of satellites 
\citep{Ibata2013}, are also seen in the MW.

These observational results for M31's halo, which provide us
a new view of an ancient stellar halo compared to the MW halo,
have motivated many subsequent simulations of galaxy 
formation. Indeed, in $\Lambda$-dominated cold dark matter models,
each disk galaxy has been developed through a different assembly
and evolutionary path: the collapse epoch, star formation history
and assembly rate  of subsystems are not the same 
from one halo to another and each stellar halo
is thus expected to have a different morphology
\citep[e.g.,][]{Bullock2005, Cooper2010}. Further observational studies
of M31's halo will thus be of great importance for getting new insights
into the formation process of its present structure, especially the
origin of the differences relative to the MW halo.

In this work, we report our new observations of M31's halo 
using Subaru/Hyper Suprime-Cam (HSC).
HSC is a new prime-focus camera on the Subaru Telescope 
with a 1.5~deg diameter field of view 
\citep{Miyazaki2012, Miyazaki2017, Komiyama2017, Kawanomoto2017, Furusawa2017}. 
This combination of HSC and Subaru thus allows us to survey large areas
of M31's halo with only a small number of pointings and to
explore much deeper domains of the halo, including 
faint red giants and horizontal-branch stars, than 
earlier studies based on 4-m telescopes have revealed.
This work will focus on
the detailed structure and stellar populations of the northwestern (NW)
stream which is full of substructures including dips and gaps, which may
have been induced by interaction with orbiting subhalos 
\citep{Carlberg2011, Carlberg2012}.
Our HSC observations thus provide new insights into the origin
of these halo structures.

The paper is organized as follows. 
In Section~\ref{sec:data}, we describe our observations
of M31's halo with Subaru/HSC and the method for data analysis.
Our target fields are those covering large parts of the NW stream 
(Figure~\ref{fig:pointing}). In Section~\ref{sec:stelpop}, 
the spatial distribution of halo stars in our survey regions
and their distribution in color-magnitude diagram are presented.
Section~\ref{sec:basicprop} is devoted to the detailed analysis of 
stellar populations inside/outside the NW stream, 
their distance distribution, and the three-dimensional 
structure of the NW stream. 
In Section~\ref{sec:discussion}, we discuss
the nature of the NW stream in comparison to the distribution of
globular clusters and a recent numerical simulation of 
the stream. Finally, we present our conclusions
in Section~\ref{sec:conclusion}.

\section{Data and Method}\label{sec:data}

\subsection{Hyper Suprime-Cam Observations}

\begin{figure}[t!]
\centering
\includegraphics[width=80mm]{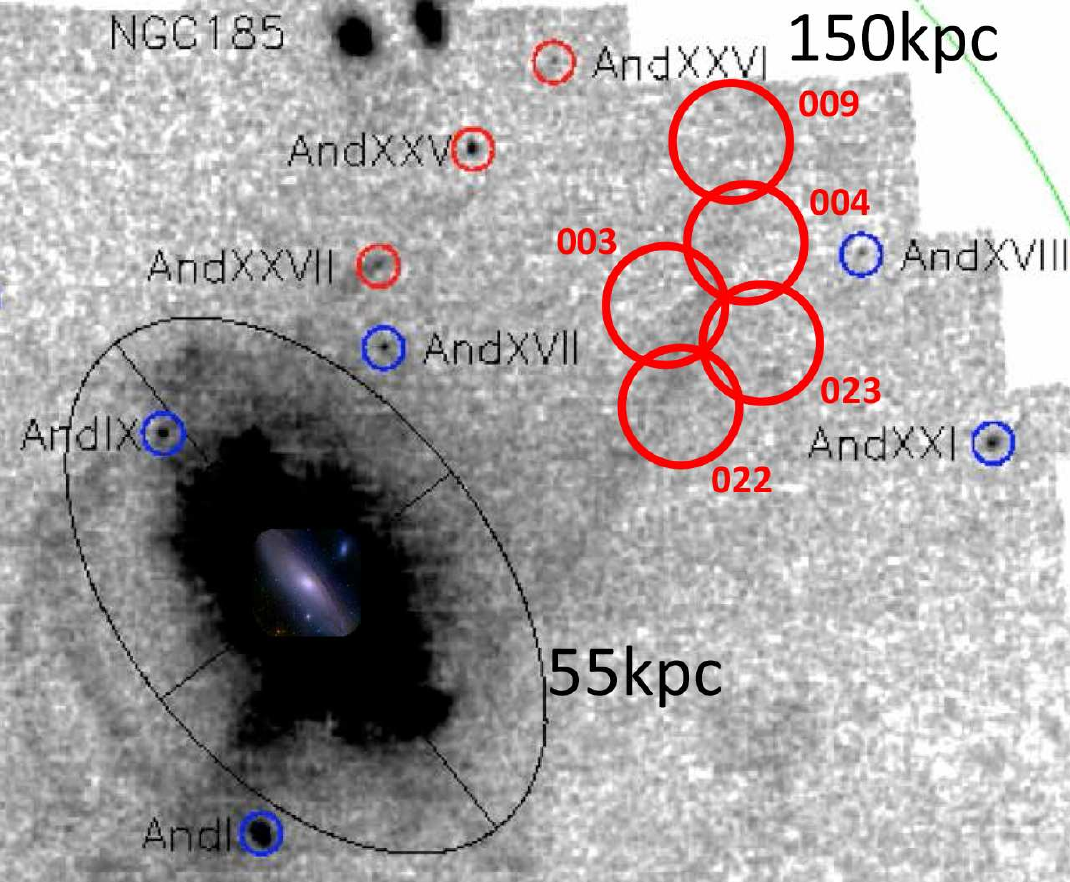}
\caption{
The HSC deep survey fields (numbered large red circles) 
overlaid on Figure 1 of \citet{Richardson2011}
showing a central image of M31 and surrounding star count map 
(grayscale) with some dwarf satellites indicated 
(small blue and red circles).
}
\label{fig:pointing}
\end{figure}

We obtained HSC $g$- and $i$-band images during 9 nights 
in 2014 and 6 nights in 2015. 
The observing conditions were variable and we devoted observing time 
in relatively good condition (i.e., fair transparency and better seeing FWHM 
than 0\farcs8) to a deep survey in the halo of M31 
which we report on in this paper. 
Figure~\ref{fig:pointing} shows the target fields, which consist of   
5 contiguous pointings (Fields 003, 004, 009, 022 and 023),  
covering 9.2 deg$^{2}$ area in the halo of M31. 
In each field, dithering with radius of 2 arcmin was performed 
to cover the gaps between CCDs. 
During the observation, the transparency was good to fair but  
the seeing FWHM varied between 0\farcs45 and 0\farcs8. 
Total exposure times for each field are 
80 min and 80$-$128 min in the $g$ and $i$ bands, respectively. 

\begin{figure}[t!]
\centering
\includegraphics[width=80mm]{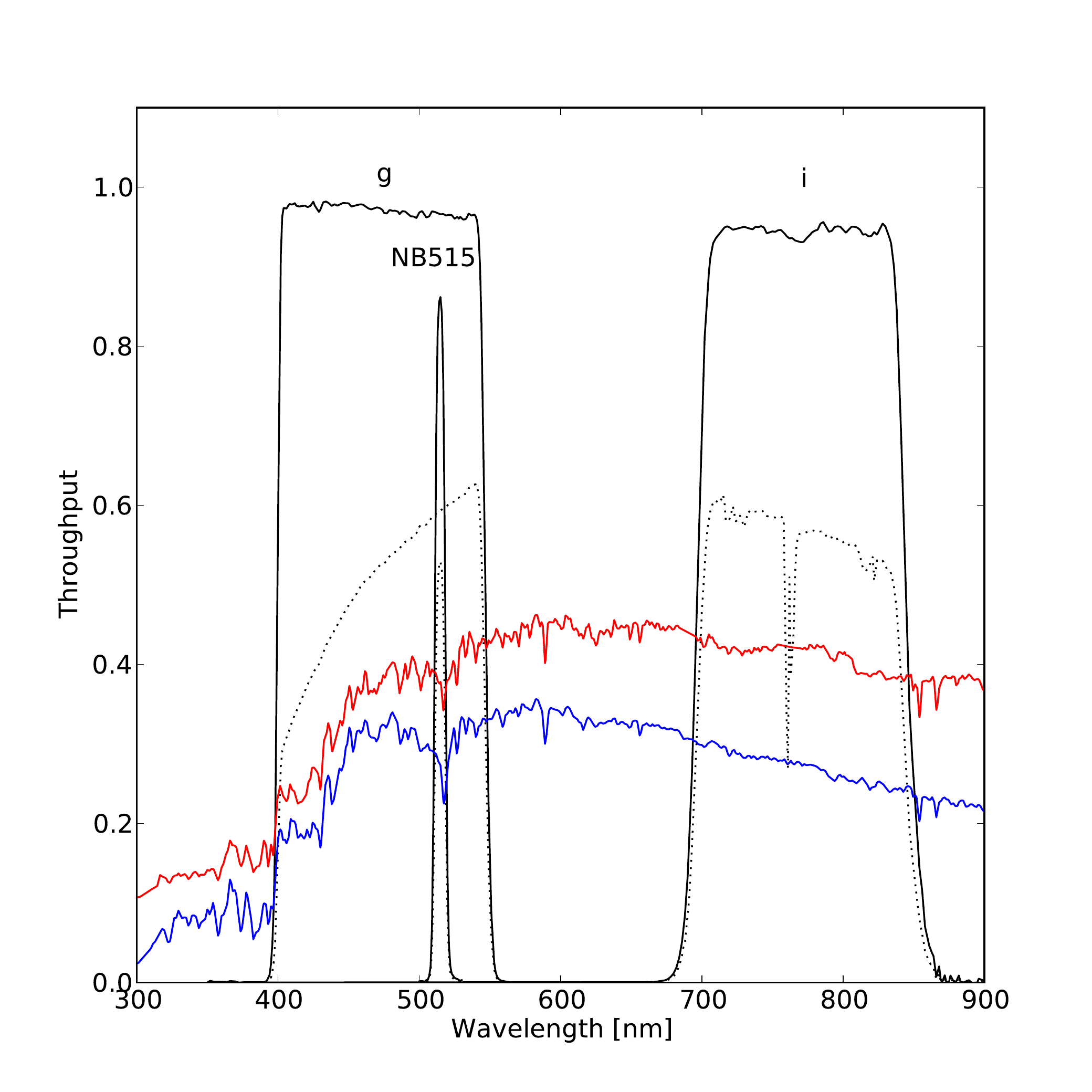}
\caption{
Response curves of filters used in the survey. 
Solid lines show response curves for filters themselves 
and dotted lines show total response. 
The spectra of K1V (blue) and K1III (red) stars from 
Pickles stellar spectral flux library \citep{Pickles1998} are overlaid. 
Note that both spectra are averaged for 5 bins (i.e., 2 nm) 
and the spectrum of K1III star is shifted upwards for clarity. 
}
\label{fig:filter}
\end{figure}

We also obtained short exposure images for the 5 target fields using 
the narrow-band filter $NB515$, which has a bell-shaped transmission curve 
centered on 515 nm with a band width (FWHM) of 7.7 nm and 
samples Mgb features of late-type stars (see Figure~\ref{fig:filter}).  
It is equivalent to the DDO51 filter which  
is useful to distinguish between giant and dwarf stars 
based on the depth of stellar surface gravity sensitive 
spectral absorption features \citep{Morrison2000,Majewski2000}. 
Although the $NB515$ exposure times are as short as 16 min for each field, 
the data are deep enough to separate bright RGB stars of M31 
from foreground main sequence (MS) dwarf stars in the MW. 

The details of the observations are summarized in Table~\ref{tab:obs} 
and the filter response curves are plotted in Figure~\ref{fig:filter}.

\begin{deluxetable}{lllll}
\tablecaption{The details of the observation.}
\tablewidth{0pt}
\tablehead{ \colhead{Field} & \colhead{RA, Dec (J2000)} & \colhead{Filter} & \colhead{Exp. Time} & \colhead{Seeing FWHM} }
\startdata
003 & 00$^{\rm h}$16$^{\rm m}$31$^{\rm s}$.7 & $g$ & 12$\times$400 s & $0\farcs7-0\farcs8$ \\
    & $+$44$^{\circ}$43$'$30$''$             & $i$ & 31$\times$240 s & $0\farcs5-0\farcs8$ \\
    &                                        & $NB515$ & 4$\times$240 s & $0\farcs6-0\farcs75$ \\ 
\tableline
004 & 00$^{\rm h}$10$^{\rm m}$04$^{\rm s}$.7 & $g$ & 12$\times$400 s & $0\farcs6-0\farcs8$ \\
    & $+$45$^{\circ}$27$'$47$''$             & $i$ & 30$\times$240 s & $0\farcs45-0\farcs8$ \\
    &                                        & $NB515$ & 4$\times$240 s & $0\farcs55-0\farcs6$ \\ 
\tableline
009 & 00$^{\rm h}$10$^{\rm m}$24$^{\rm s}$.5 & $g$ & 12$\times$400 s & $0\farcs65-0\farcs8$ \\
    & $+$46$^{\circ}$49$'$07$''$             & $i$ & 20$\times$240 s & $0\farcs6-0\farcs8$ \\
    &                                        & $NB515$ & 4$\times$240 s & $0\farcs55-0\farcs6$ \\ 
\tableline
022 & 00$^{\rm h}$16$^{\rm m}$04$^{\rm s}$.2 & $g$ & 12$\times$400 s & $0\farcs5-0\farcs8$ \\
    & $+$43$^{\circ}$22$'$15$''$             & $i$ & 32$\times$240 s & $0\farcs45-0\farcs8$ \\
    &                                        & $NB515$ & 4$\times$240 s & $0\farcs85-0\farcs95$ \\ 
\tableline
023 & 00$^{\rm h}$09$^{\rm m}$45$^{\rm s}$.8 & $g$ & 12$\times$400 s & $0\farcs5-0\farcs7$ \\
    & $+$44$^{\circ}$06$'$28$''$             & $i$ & 30$\times$240 s & $0\farcs45-0\farcs7$ \\
    &                                        & $NB515$ & 4$\times$240 s & $0\farcs85-0\farcs9$  
\enddata
\label{tab:obs}
\end{deluxetable}

\subsection{Reduction and Photometry}\label{sec:reduction}

The HSC data are processed with hscPipe v4.0.1 \citep{Bosch2017}, 
a branch of the Large Synoptic Survey
Telescope pipeline \citep{Ivezic2008,Juric2015} 
calibrated against Pan-STARRS 1 (PS1) 
photometry and astrometry \citep{Schlafly2012,Tonry2012,Magnier2013}.
The hscPipe is a standard data reduction pipeline 
optimized for the data from wide-field mosaic CCD cameras  
in general and HSC in particular. 
Each CCD frame is calibrated against PS1 
in the course of single frame reduction (i.e., bias subtraction, 
correction for cross-talk, cosmic ray removal, 
flat fielding, sky subtraction). 
Then, all the frames are coadded into a large image 
based on the mosaicking solution which is calculated using 
all the frames in each band. 
Finally, objects are detected and their parameters 
such as magnitudes, positions, sizes, etc. are measured. 

The resultant photometric catalogs in all 3 bands are 
matched with a matching radius of 0\farcs5. 
In this study, we analyze those objects that are classified as point sources
(i.e., $extendedness = 0$ in the hscPipe output catalog) 
in both $g$ and $i$ bands and PSF magnitudes are used 
throughout this study except in Section~\ref{sec:gc}. 
Note that the $NB515$ data were calibrated to $g$-band via hscPipe 
and the $NB515$ magnitude may be offset systematically 
from the correct value by $\le 0.1$ mag but the absolute 
value is not important for our purposes.

Our survey is substantially deeper than previous ground-based surveys of M31. 
We assess how deep our data go by 
measuring the detection completeness and estimating the extinction 
by interstellar dust, as described below, 
before we start further analysis. 

\begin{figure*}[t!]
\centering
\begin{tabular}{ccc}
\includegraphics[width=50mm]{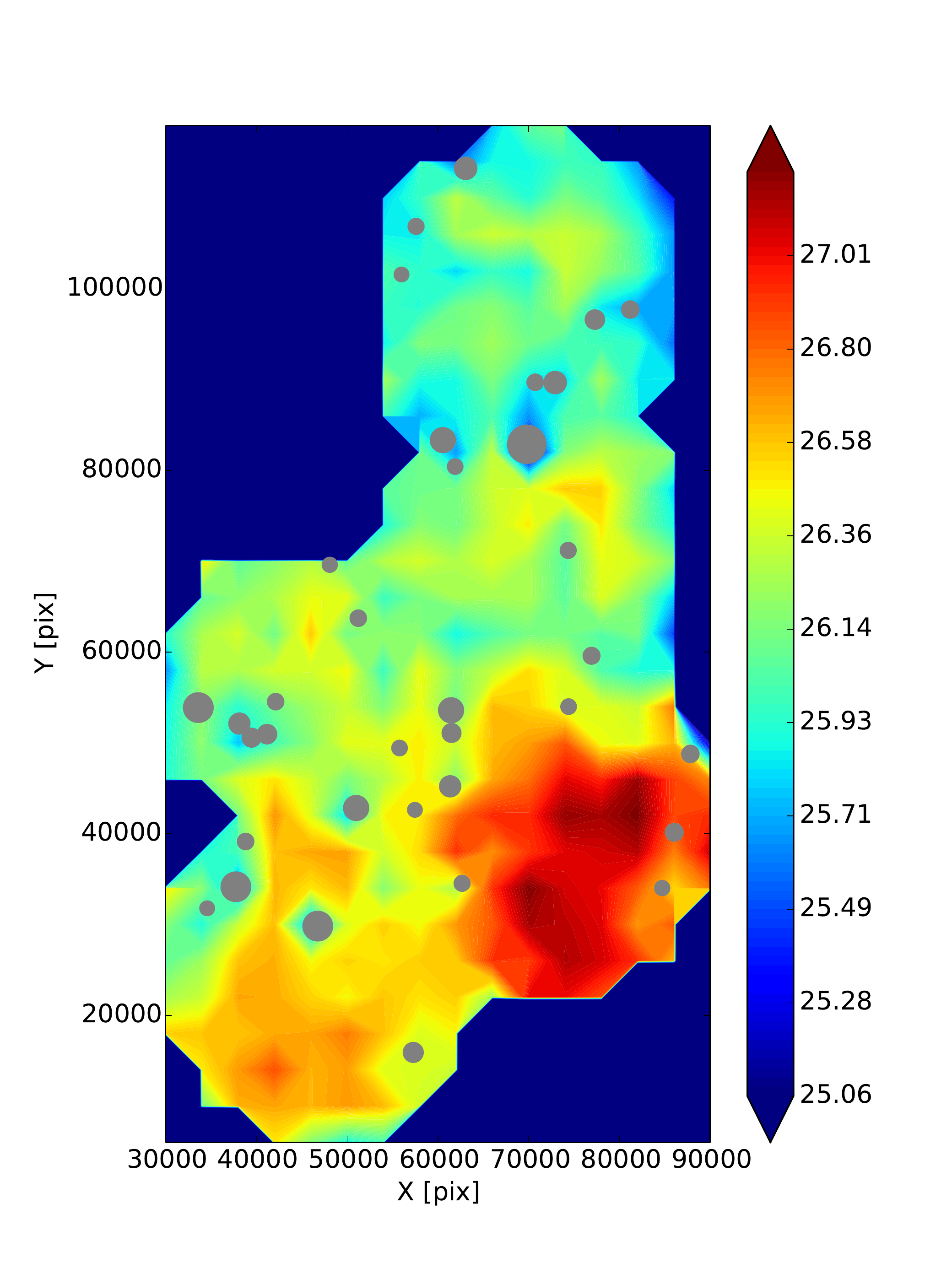} 
&
\includegraphics[width=50mm]{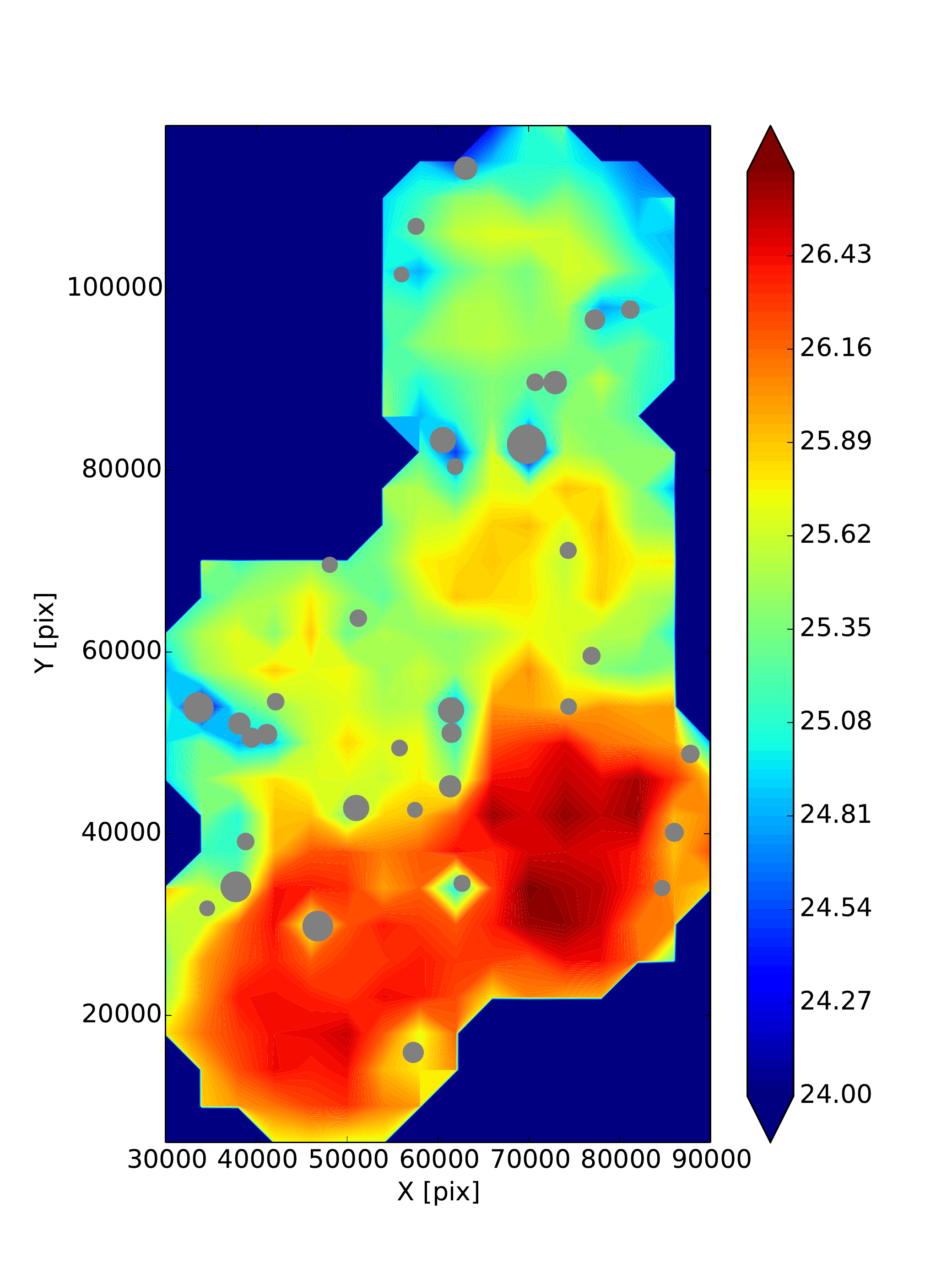} 
&
\includegraphics[width=50mm]{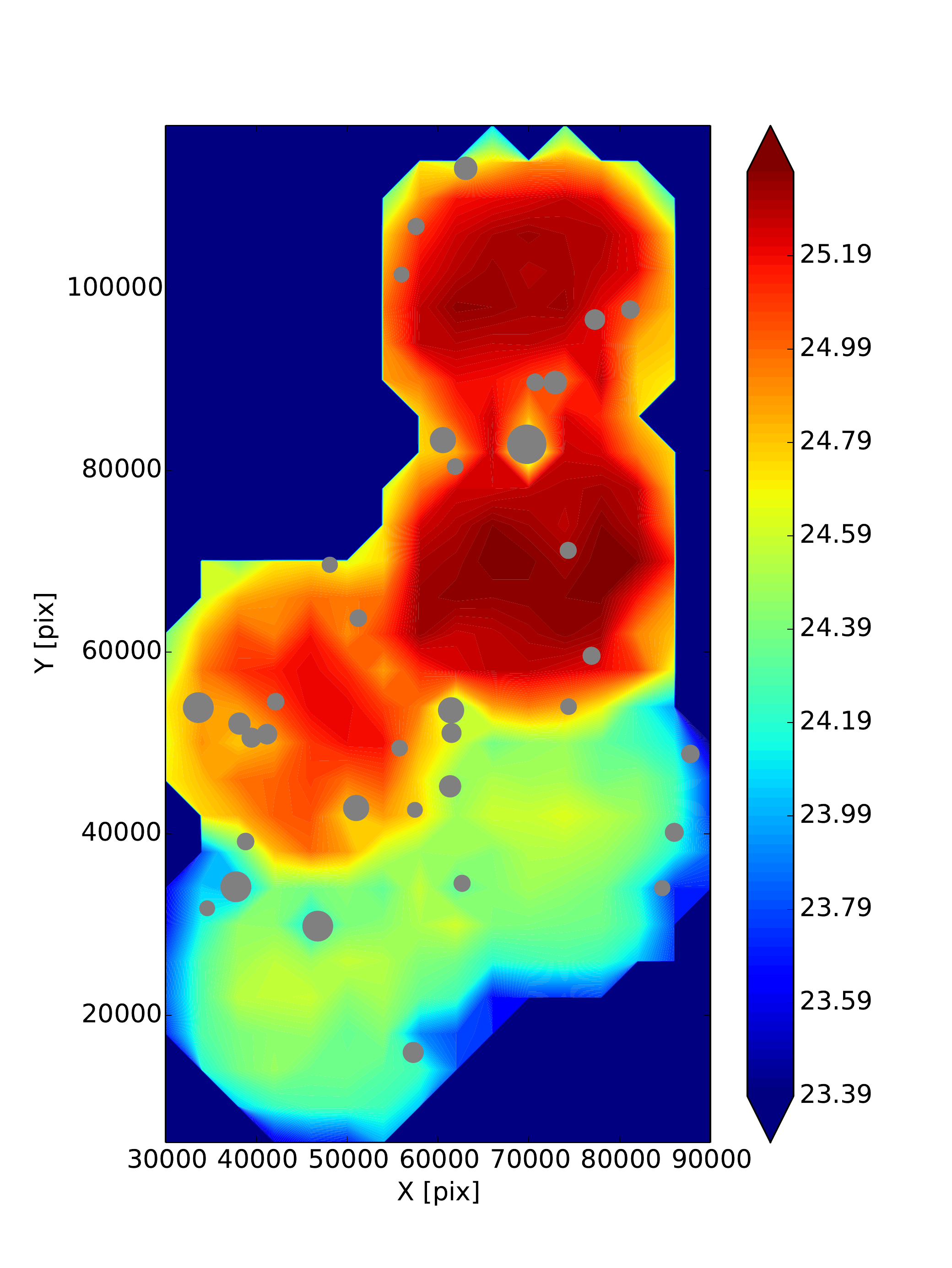} \\
(a) $g$ band & (b) $i$ band & (c) $NB515$ band \\
\end{tabular}
\caption{
The 2 dimensional maps of the 50\% completeness magnitude 
in $g$ (a), $i$ (b) and $NB515$ bands (c). 
The masked areas by bright stars are shown in gray circles. 
}
\label{fig:completeness}
\end{figure*}

The detection completeness is estimated using a PSF model 
obtained by the hscPipe software for every 4000$\times$4000 pixels grid, 
which is referred to as a `patch' in hscPipe. 
A set of 500 artificial stars 
based on the PSF model and a specific apparent magnitude is
embedded in a patch and the hscPipe object detection algorithm is applied. 
This process is repeated in steps of 0.25 mag and 
the completeness, the fraction of artificial stars that are detected, 
is obtained for all the patches. 
Figure~\ref{fig:completeness} shows the apparent magnitude 
that corresponds to 50\% 
completeness in $g$, $i$ and $NB515$ bands over our survey field. 
The mean magnitudes of 50\% completeness in $g$, $i$ and 
$NB515$ band are calculated to be 26.31, 25.69 and 24.71 mag, respectively, 
indicating that our survey is substantially deeper than 
the previous ground-based studies. 
Field 023 is the deepest among the survey fields in $g$ and $i$ bands 
and the north part is relatively shallow compared to 
the south part likely due to the unstable weather 
and variable seeing conditions during the observations. 
Patches that contain bright stars are shallower 
than neighboring patches, which indicates that the completeness 
is degraded by the large halo of a bright star. 
Therefore, areas occupied by the large halos around bright stars
are masked and excluded from the subsequent analysis. 
The $NB515$ band imaging in 
the north and middle part is deeper than the south part. 
This is due to variable seeing conditions between the different pointings  
(see Table~\ref{tab:obs}).

\begin{figure}[t!]
\centering
\includegraphics[width=80mm]{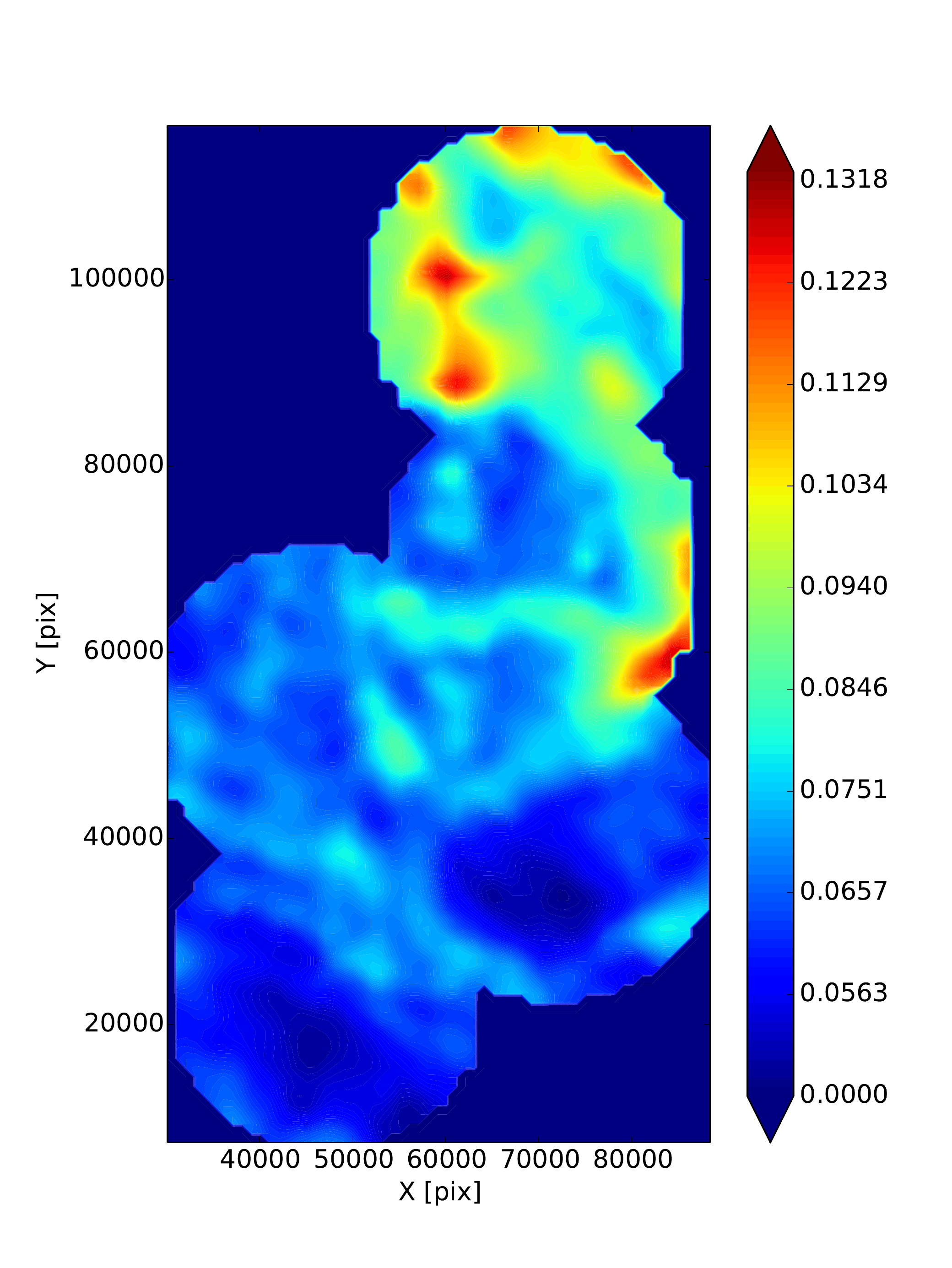}
\caption{
The 2 dimensional E($B-V$) map for our survey field. 
}
\label{fig:extinction}
\end{figure}

The Galactic extinction is corrected using 
the new estimate from \citet{SchlaflyFinkbeiner2011}
which is based on the dust map by \citet{Schlegel1998}. 
Figure~\ref{fig:extinction} shows the 2 dimensional E($B-V$) map 
for our survey field, where 
E($B-V$) values are taken from NASA/IPAC Infrared Science Archive
\footnote{http://irsa.ipac.caltech.edu/applications/DUST/}. 
The variation of E($B-V$) across our survey field is clearly seen 
in the sense that the northest part (field 009)  
suffers from heavy extinction while the majority part of 
other 4 fields are less affected. 
Therefore, the extinction correction should be made 
star by star according to the position on the sky. 
The mean extinctions in $g$, $i$ and $NB515$ bands are calculated to be
0.27, 0.14 and 0.24 mag, respectively. 
The accuracy in the estimate of extinction based on this method 
is discussed in Section~\ref{sec:3Dstr}.

The spatially varying completeness and extinction resulted in
inhomogeneity in the survey depth (i.e., limiting magnitude) 
across the field, in the sense that 
the south part (fields 022 and 023) is the deepest and 
the survey depth gets slightly worse at the middle part (fields 003 and 004)
and significantly degraded at the north part (field 009) 
for $g$ and $i$ bands, while the opposite trend is seen in $NB515$ band. 
We will take the spatial variation into account in the following analysis. 
Nonetheless, our data are considerably deep and bring
wealth of information to understand the nature of the M31 halo.

\subsection{Color-Magnitude Diagram}

\begin{figure}[t!]
\centering
\includegraphics[width=80mm]{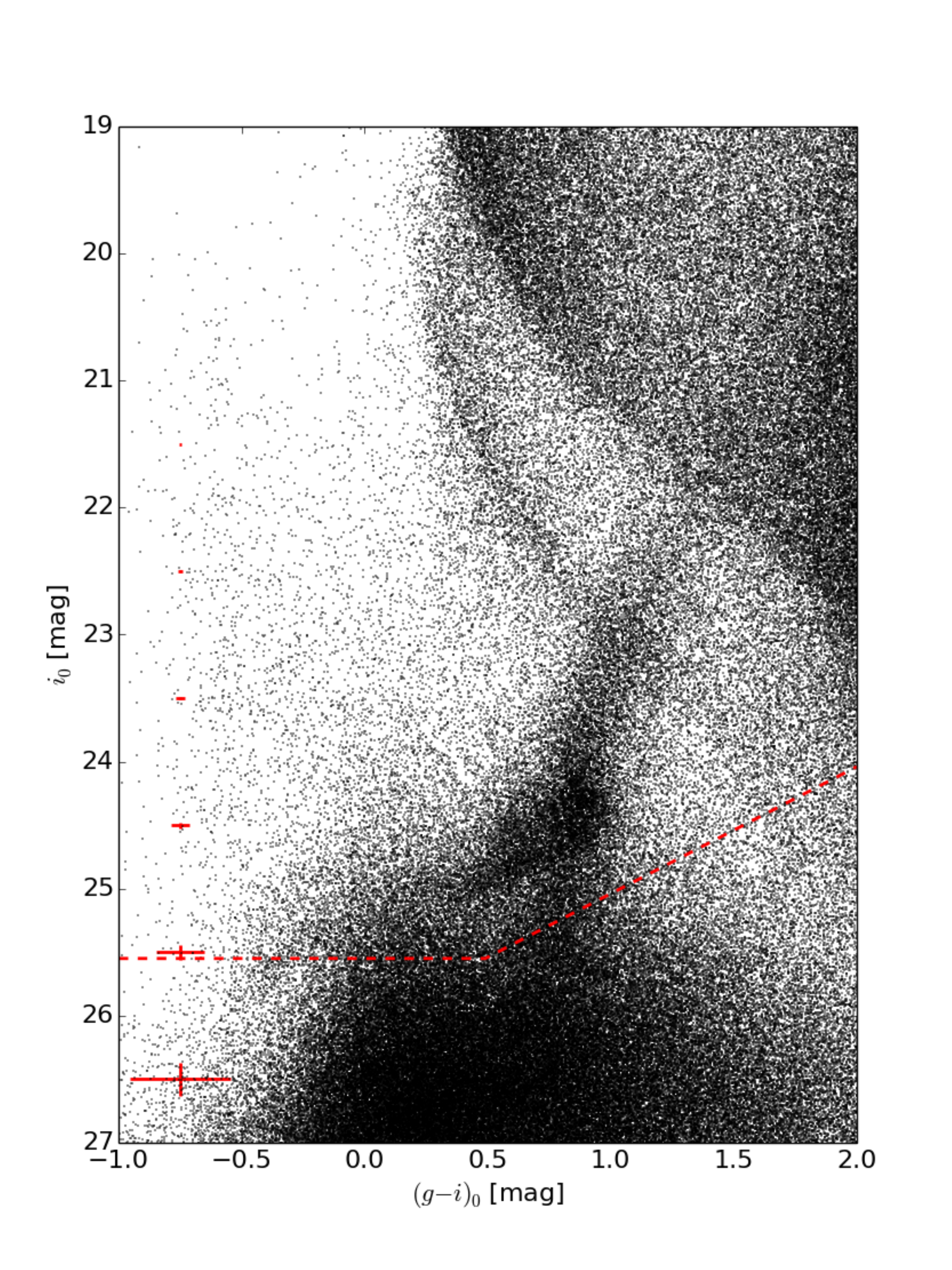}
\caption{
Color-magnitude diagram of catalogued stellar objects in our survey field. 
The photometric error in each band, which is 
a mean value of every 1 mag bin, is calculated and 
the errors in the $i_{0}$ magnitude and color 
[calculated for $(g-i)_{0} = 1.0$] are plotted as red crosses. 
The 50\% completeness limit, which is the mean value 
over the survey field, is plotted as a red dashed line. 
}
\label{fig:cmd}
\end{figure}

Figure~\ref{fig:cmd} is the extinction corrected 
color-magnitude diagram (CMD) 
of 401,834 catalogued stellar objects down to $i$ = 27.0 mag 
in our survey fields. 
The photometric error in each band, which is 
the mean value of every 1 mag bin, is calculated and 
the errors in the $i_{0}$ magnitude and color 
[calculated for $(g-i)_{0} = 1.0$] are plotted as red crosses in the figure. 
The 50\% completeness limit, which is the mean value 
over the survey field, is plotted as a red dashed line. 

The CMD clearly shows the characteristic features: 
dwarf stars of the MW disk lie on the brightest part of the CMD, 
extending from $(g-i)_{0} \simeq 0.3$ and $i_{0} \simeq 19$ 
to the redder and fainter part of the CMD. 
A narrow but significant sequence seen at $\sim$1 mag 
fainter from the distribution of the MW disk stars is corresponding to 
a diffuse stellar stream in the MW halo which is reported by 
\citet{Martin2014}.  
The RGB in the M31 halo is seen as a broad sequence which is perpendicular 
to the above mentioned MW stream and crossing  
at $(g-i)_{0} \simeq 1.1$ and $i_{0} \simeq 23$. 
It is also noted that a significant red clump feature 
is seen at $(g-i)_{0} \simeq 0.8$ and $i_{0} \simeq 24.5$ 
and diffuse but distinct blue horizontal branch (BHB)
is also seen at $(g-i)_{0} \simeq -0.3$ and $i_{0} \simeq 25.5$. 
We call those object found in 
$0.3<(g-i)_{0}<1.0$, $24.0<i_{0}<25.1$ and $g_{0}<25.6$ 
as red clump population (RC) in the following analysis. 
Below $i_{0} = 26$, the photometric error becomes large 
and it is difficult to distinguish characteristic features 
such as sub-giant branch (SGB) and main-sequence (MS)
stars in the M31 halo.

\subsection{NB515 and 2-Color Diagram}

To extract secure M31's RGB stars behind foreground Galactic dwarf stars,
we use a narrow-band filter, $NB515$, in combination with 
$g$ and $i$ band filters.
This filter has been designed by our team for this purpose and
we have already performed a test for distinguishing 
M31's RGB/AGB stars from foreground Galactic dwarfs using the prototype filter 
of $NB515$ in Suprime-Cam and confirmed its validity.
As shown in Figure~\ref{fig:filter}, $NB515$ is sensitive to 
the MgH + Mgb absorption features around 515 nm
which is strongly dependent on surface gravity 
\citep[][and references therein]{Majewski2000}.
By measuring the absorption strength of this feature with $NB515$ filter, 
one can discriminate K giant in the halo of M31 from K dwarfs 
in the disk of Milky Way with same apparent magnitude. 
This is illustrated in Figure~\ref{fig:filter} 
where the spectra of K1V (blue) and K1III (red) stars from 
Pickles stellar spectral flux library \citep{Pickles1998} 
are well discriminated by the absorption feature, i.e., $NB515-g$ color 
in the sense that the $NB515-g$ color at fixed $g-i$ color 
is larger for dwarfs and smaller for giants. 

\begin{figure*}[t!]
\centering
\begin{tabular}{cc}
\includegraphics[width=80mm]{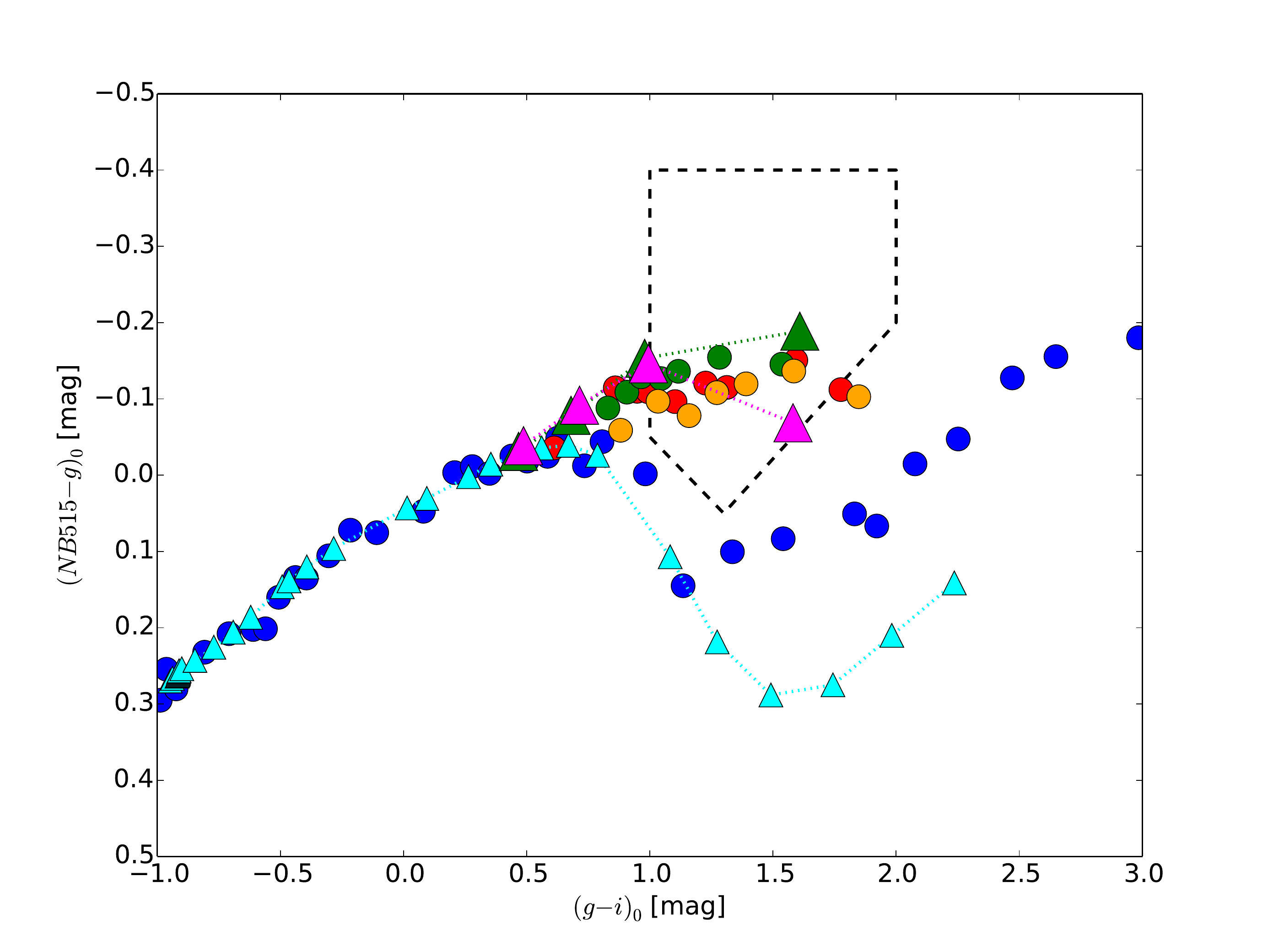} 
&
\includegraphics[width=80mm]{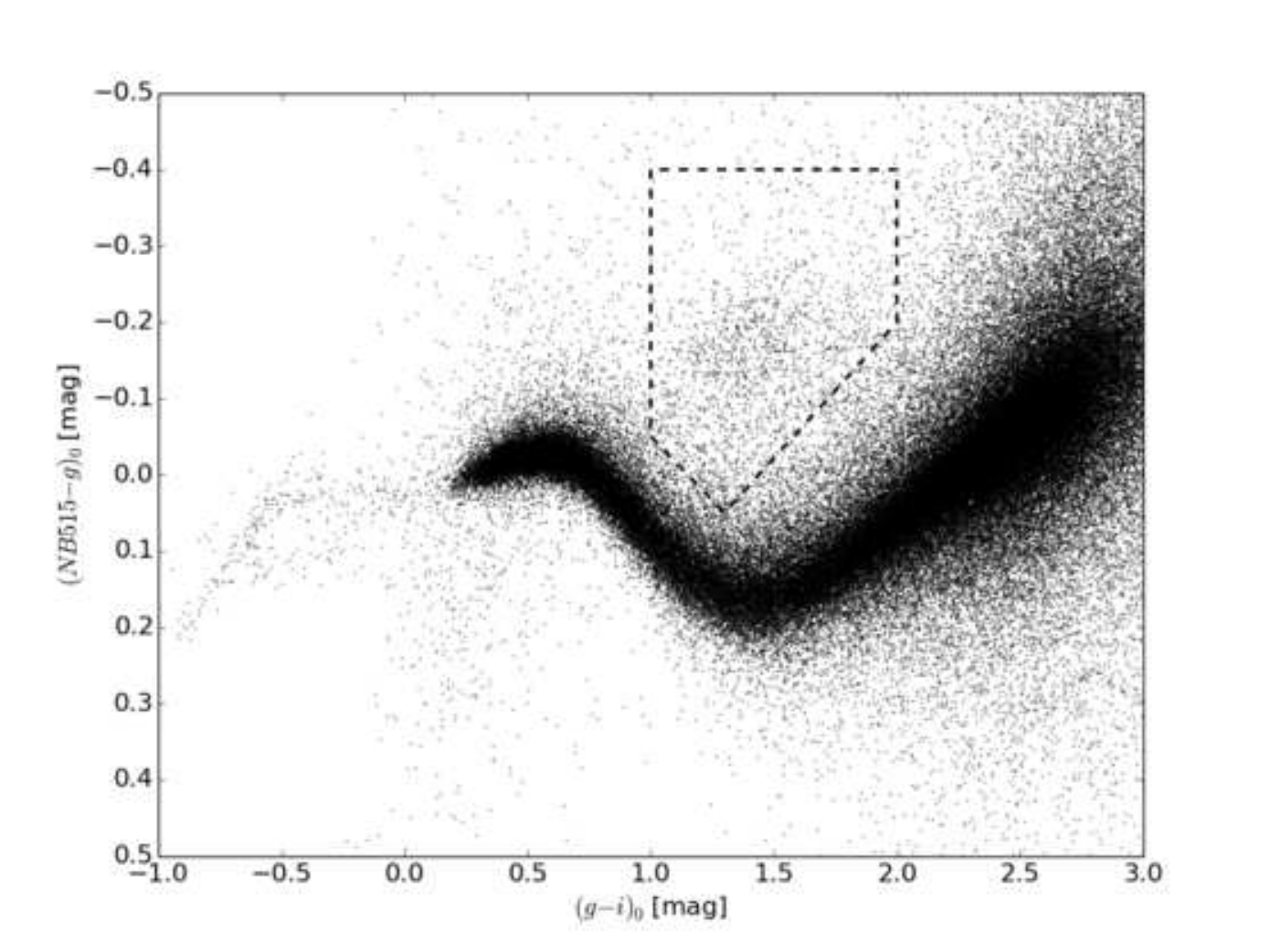} \\
(a) & (b) \\
\end{tabular}
\caption{
(a) $(g-i)$-$(NB515-g)$ 2-color diagram for template and model stars. 
Blue, red, green and orange filled circles show
normal dwarfs, normal G-K type giants, metal-week G-K giants
and metal-rich G-K giants from Pickles atlas \citep{Pickles1998}, respectively,
and cyan, green and magenta filled triangles with dotted line show
dwarfs with solar metallicity, G-K giants with [Fe/H] = -2.0
and those with [Fe/H] = -1.0 from ATLAS9 \citep{Castelli2004}, respectively.
(b) $(g-i)$-$(NB515-g)$ 2-color diagram for stellar objects with $19<i_{0}<22$. 
The stars enclosed in the dashed line are regarded as giant star 
candidates in the M31 halo (NBGs). 
}
\label{fig:2col}
\end{figure*}

We investigate how dwarfs and giants are distributed 
in the 2-color diagram making use of template spectra  
in Figure~\ref{fig:2col} (a). 
We use Pickles stellar spectral flux library \citep{Pickles1998} 
and ATLAS9 stellar atmosphere models \citep{Castelli2004}, 
and $g-i$ and $NB515-g$ colors are calculated by 
convolving the system throughput as shown in Figure~\ref{fig:filter}. 
Blue, red, green and orange filled circles show 
normal dwarfs, normal G-K type giants, metal-week G-K giants 
and metal-rich G-K giants from Pickles atlas, respectively, 
and cyan, green and magenta filled triangles with dotted line show 
dwarfs with solar metallicity, G-K giants with [Fe/H] = -2.0 
and those with [Fe/H] = -1.0 from ATLAS9, respectively. 
The suggested models\footnote{http://www.stsci.edu/hst/observatory/crds/castelli\_kurucz\_atlas.html}
for dwarfs of ATLAS9 are used for dwarfs with solar metallicity, 
while $T_{eff}$ and $\log(g)$ are chosen from 
suggested models for G-K giants and those with different metallicities 
are used to calculate $g-i$ and $NB515-g$ colors. 
The figure indicates that giants can clearly be discriminated 
from dwarfs for $g-i > 1.0$. 
Our criteria to select giants, which are represented by the dashed line, 
fairly well enclose giants except for normal to metal-rich K5 giants. 

Figure~\ref{fig:2col} (b) shows the 2-color diagram
for stellar objects with $19<i_{0}<22$ in our survey field. 
The thick '$\surd$'-shape sequence is composed of MS dwarf stars of the MW. 
On the other hand, the giant stars in the M31 halo 
are populated at around our criteria which are shown as the dashed line. 
The criteria are set reasonably to select giants in the M31 halo 
while eliminating contaminations from MS dwarfs. 
In this study, we regard those objects 
enclosed by the dashed line in Figure~\ref{fig:2col} 
to be giant star candidates in the M31 halo 
and call them as narrow-band selected giants (NBGs).

\section{Stellar Populations in the Survey Fields}\label{sec:stelpop}

\subsection{Spatial Distribution}

\begin{figure}[t!]
\centering
\includegraphics[width=74mm]{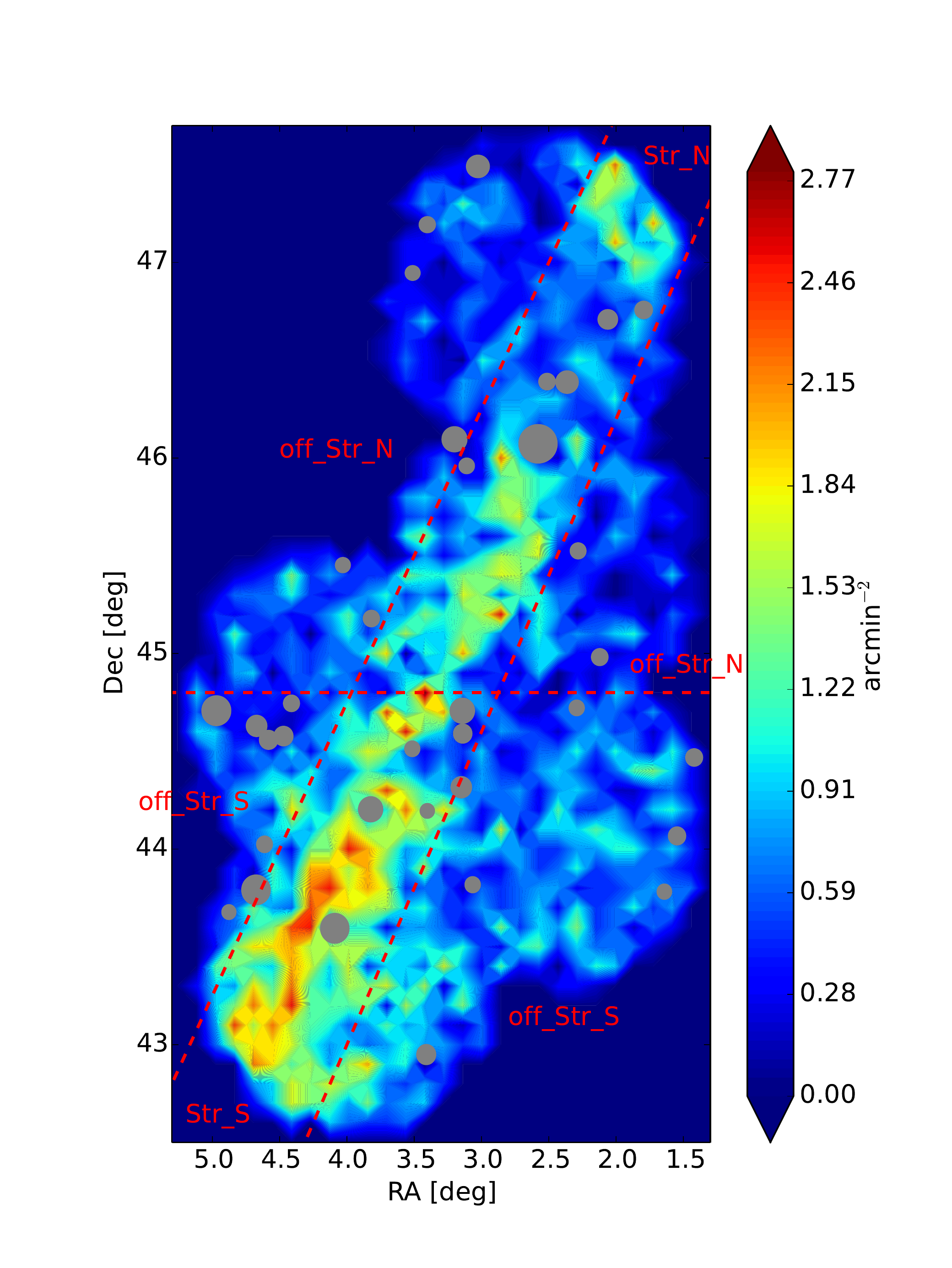} \\
(a) NBG \\
\includegraphics[width=74mm]{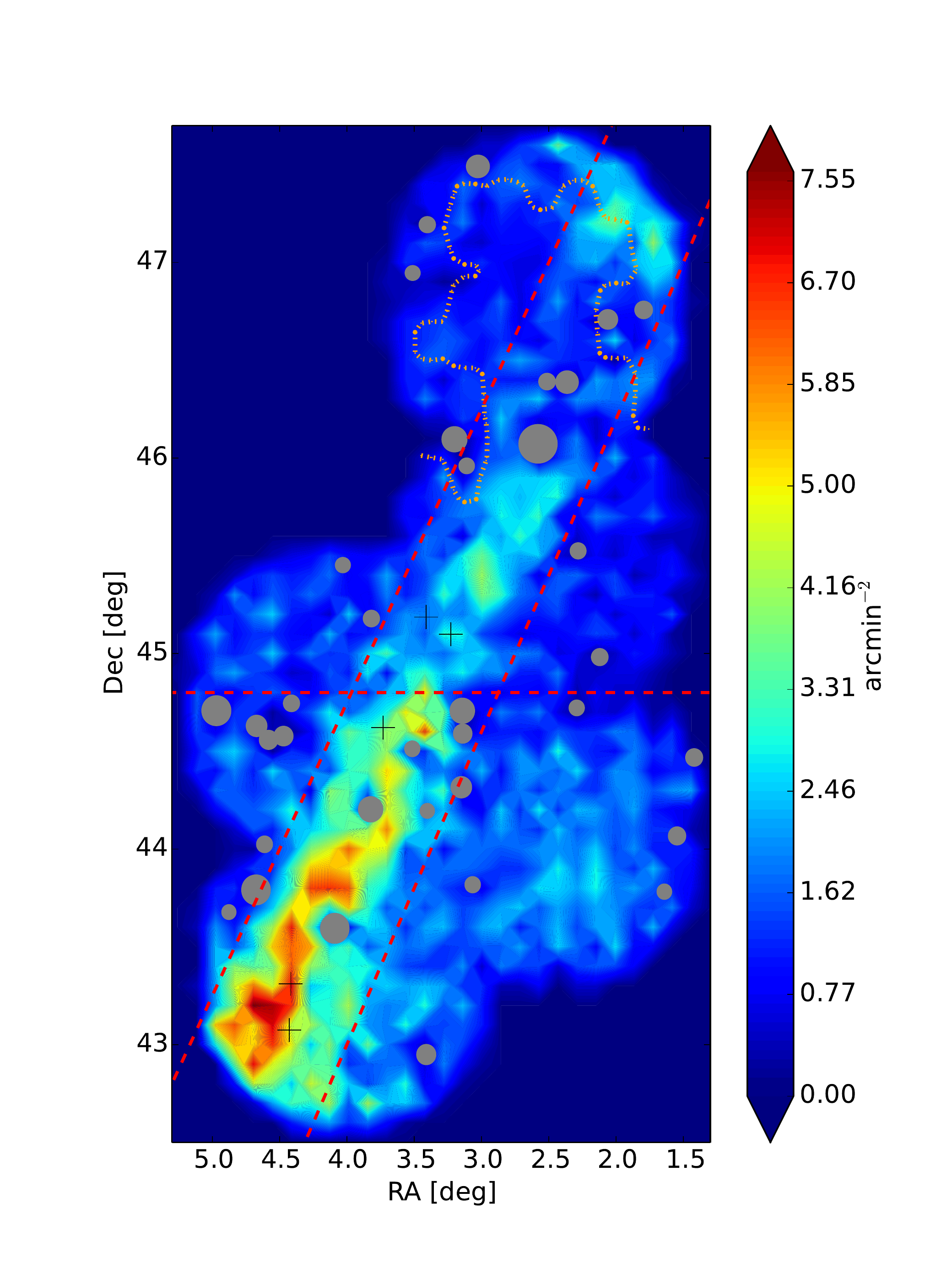} \\
(b) RC \\
\caption{
Surface number density maps of NB-selected RGB stars (a) and RC (b). 
Areas near bright stars where the object detection is not 
well carried out are masked as gray circles. 
The boundaries between Stream/Off-Stream North/South regions 
are shown in red dashed lines and each region is labeled in the panel (a). 
In the panel (b), the region where 50 \% completeness is achieved 
at magnitudes of $i_{0} = 25.1$ mag and $g_{0} = 25.6$ mag, 
which is the fainter boundary of RC selection, 
is shown in orange dotted line, indicating that 
most of the survey field is more than 50 \% complete to detect RC stars. 
Crosses represent 5 globular clusters reported by 
\citet{Huxor2014} in the survey field. 
}
\label{fig:space}
\end{figure}

Figure~\ref{fig:space} shows the distributions of NBGs 
with $19<i_{0}<23$ and RCs. 
A distinct stream feature is clearly seen 
in the figure, which is already shown to exist 
by the PAndAS survey and called as the NW Stream 
\citep{McConnachie2009,Richardson2011}. 
Our data reveal the much more detailed properties of the NW Stream; 
The stars consisting of the stream are not uniformly distributed and 
instead show a clumpy distribution. 

Besides the NW Stream, it is pointed out that 
the surface densities of NBGs and RCs in the off-stream region 
seem to be higher at 
the southern part compared to the northern part. 
This trend may be due to the different limiting magnitude among 
the survey field, in particular for RC population. 
We calculate the 50\% completeness contour 
to detect RC population (i.e., those stars with 
$0.3<(g-i)_{0}<1.0$, $24.0<i_{0}<25.1$ and $g_{0}<25.6$), 
which is plotted as orange dotted line in Figure~\ref{fig:space} (b).
The figure indicates that most survey field is more than 
50\% complete, except for the periphery of field of view of HSC pointing 
at the north part where the extinction is heavy. 
We therefore conclude that this decreasing trend toward north is real. 
The same trend is also seen for NBGs, 
supporting that this trend is real. 

In the following, 
we analyze the data divided into 4 regions, 
Stream North/South and Off-Stream North/South, 
and investigate the stellar populations in the 4 regions separately. 
The boundaries are shown as red dashed lines 
and each region is labeled in Figure~\ref{fig:space}. 
The boundary between north and south ($\delta = 44.8$)
is set so that the numbers of stars are 
almost same between north and south in the off-stream region.

\subsection{Color-Magnitude Diagrams of 4 Regions}\label{sec:4cmd}

\begin{figure*}[t!]
\centering
\begin{tabular}{cc}
\includegraphics[width=80mm]{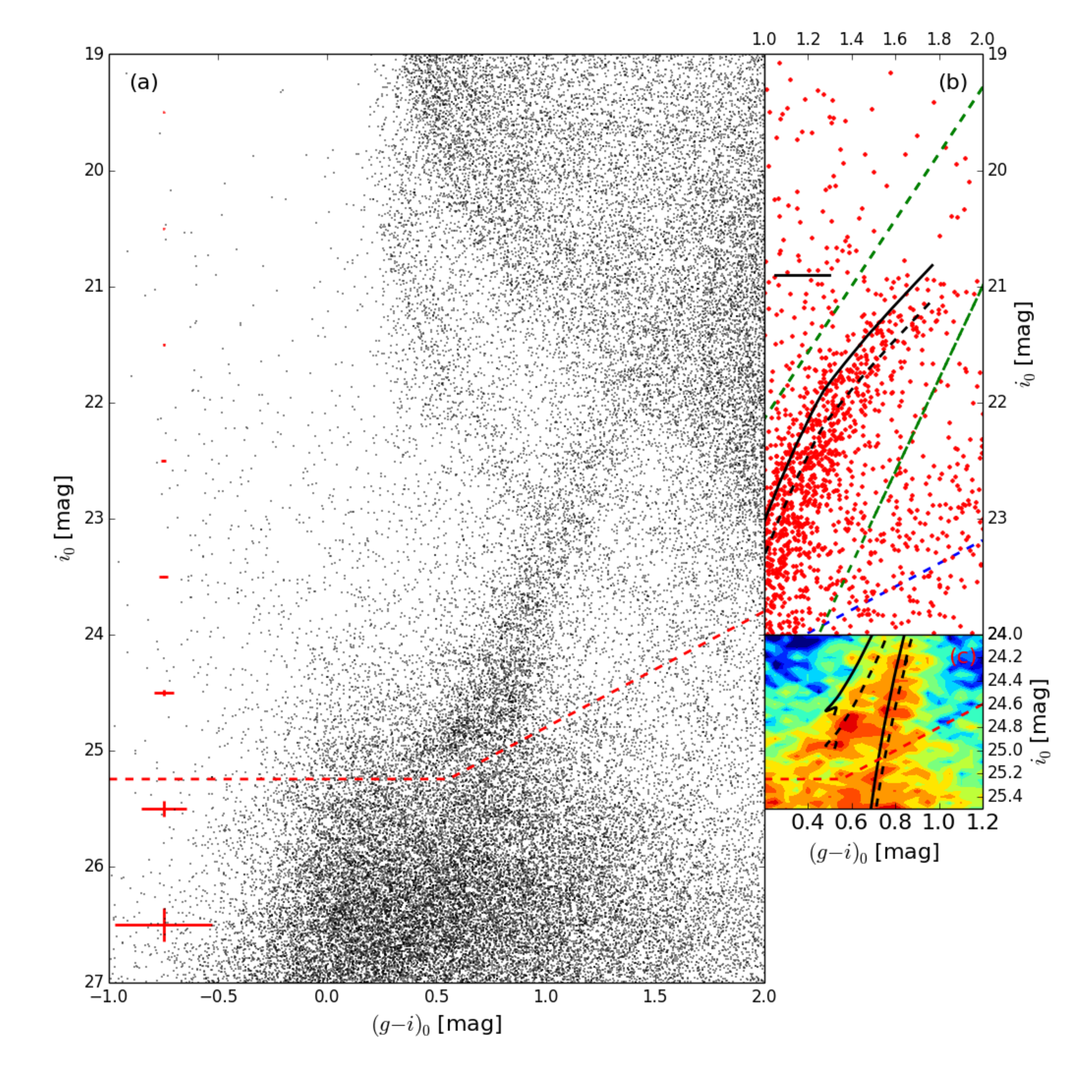} 
& 
\includegraphics[width=80mm]{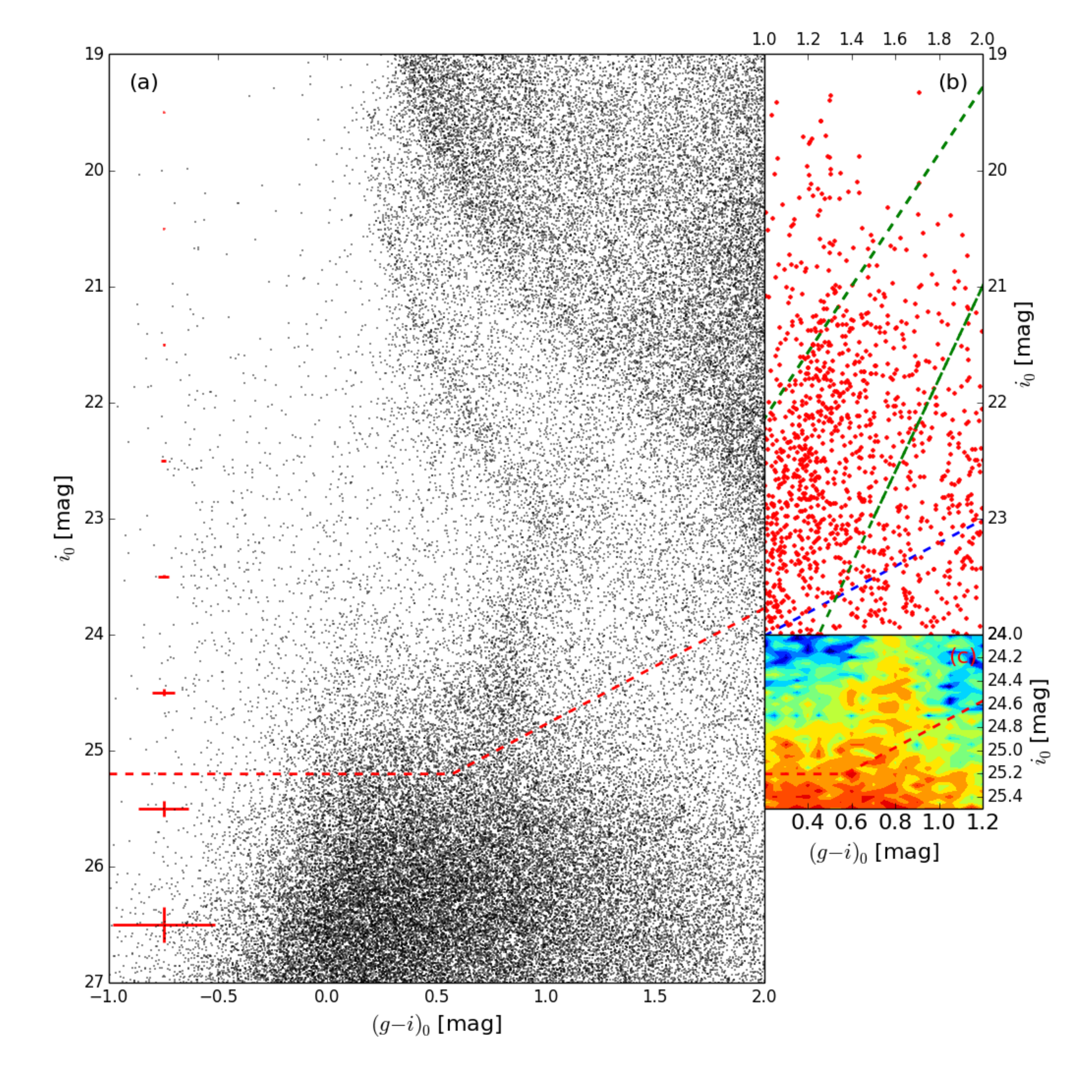} \\
(1) Stream North & (2) Off-Stream North \\
\includegraphics[width=80mm]{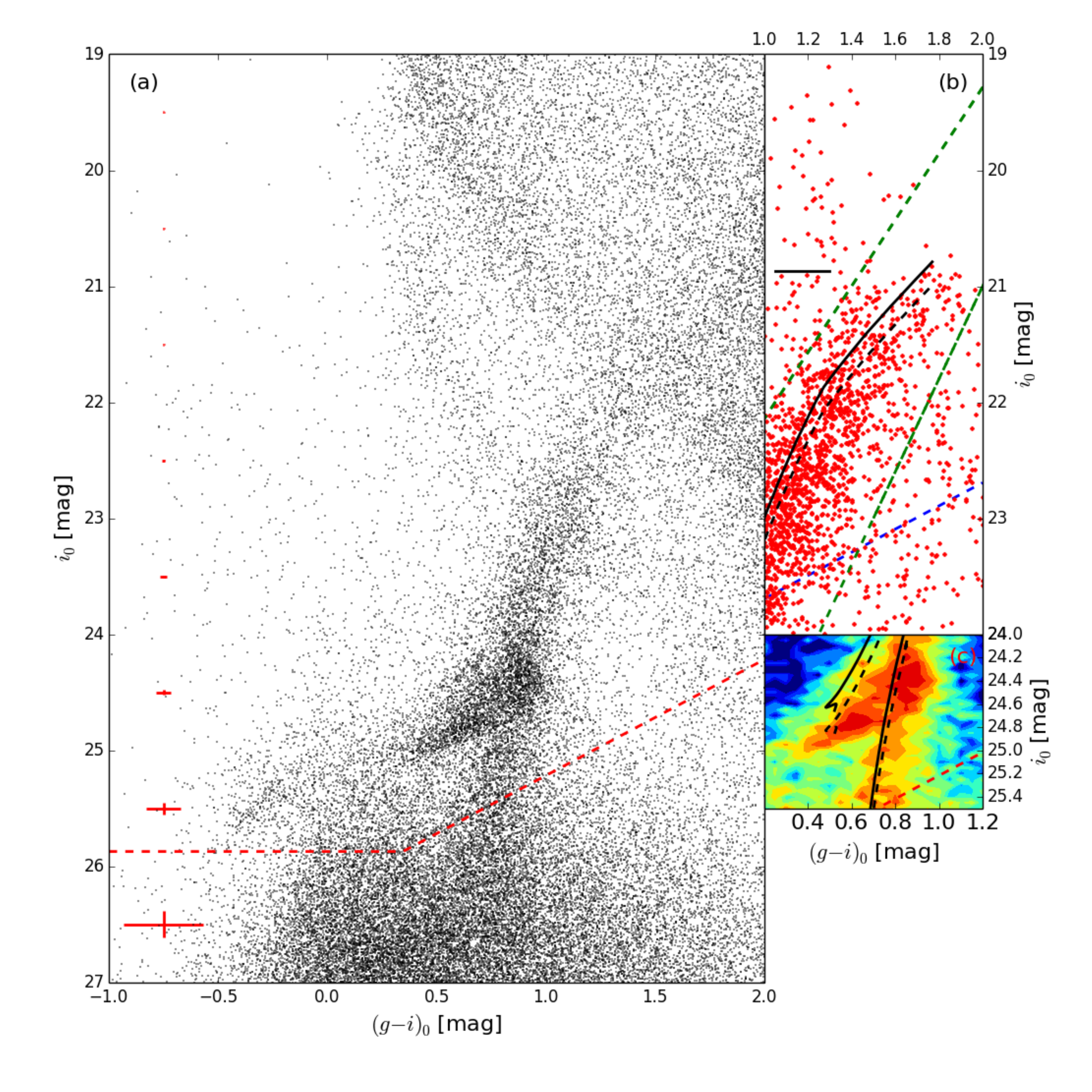}
&
\includegraphics[width=80mm]{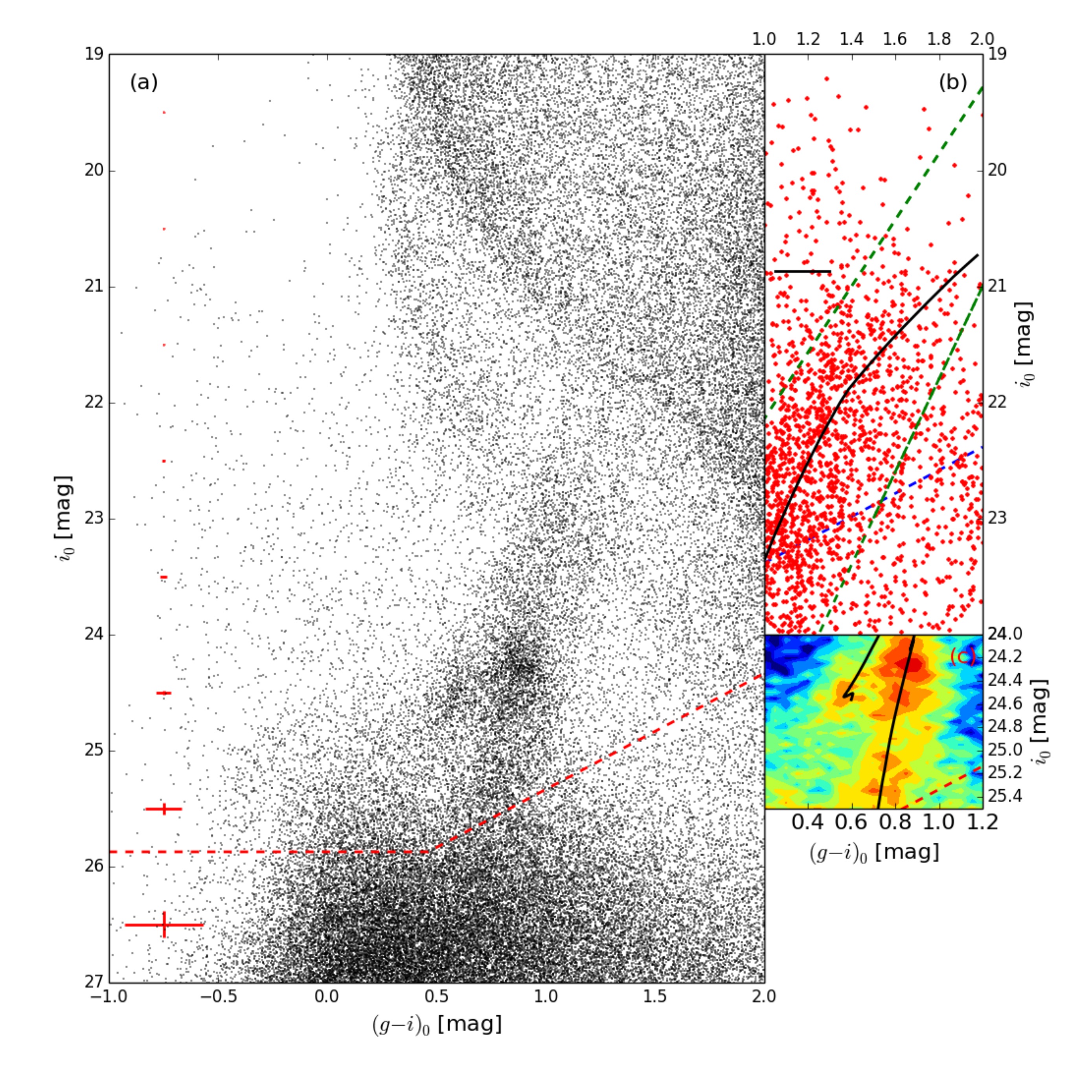} \\
(3) Stream South & (4) Off-Stream South \\
\end{tabular}
\caption{
CMDs of 4 regions in our survey field. 
From top-left to bottom-right, CMDs of 
(1) Stream North, (2) Off-Stream North, (3) Stream South, 
and (4) Off-Stream South, respectively, are plotted. 
For each of the parts (1)--(4), 
panel (a) shows the overall CMD. 
The 50\% completeness limit in $g$ and $i$ bands, 
which is the mean value in each region, and mean error bars 
for every 1 mag intervals are plotted as a red dashed line and 
red crosses, respectively. 
Panel (b) shows the zoomed view of RGB region 
with NBGs plotted in red. 
The corresponding 50\% completeness limit in $NB515$ 
is plotted as a blue dashed line. 
The RGB selection used to determine the TRGB is plotted as 
green dashed lines and the TRGB magnitude is indicated 
in horizontal black bar. 
The best-fit isochrone is plotted in black thick line. 
Panel (c) shows the zoomed density map of RC region, 
together with the best-fit isochrone with the distance modulus 
determined by TRGB method (back thick line) 
and that with increasing distance modulus from TRGB (back dashed line) 
by 0.3 (Stream North) and  0.2 (Stream South). 
}
\label{fig:4cmd}
\end{figure*}

Figure~\ref{fig:4cmd} shows CMDs of 4 regions. 
In panel (a), overall CMD for each region is plotted 
with the 50\% completeness limit in $g$ and $i$ bands, 
which is the mean value in each region, and mean error bars 
for every 1 mag intervals being plotted as a red dashed line and 
red crosses, respectively, in the respective figure. 
Panel (b) shows only NBGs and panel (c) shows 
the zoomed density map of RC regions. 
The corresponding 50\% completeness limit in $NB515$, 
which is translated from the narrow-band selection criteria,  
is plotted as a blue dashed line in panel (b). 
Note that the 50\% completeness limit in $NB515$ is 
only effective for NBGs (red points). 

The CMD of the Stream North clearly shows 
a narrow RGB, populated RC, and a hint of distinct BHB, 
although the survey depth is shallow,  
suggesting that the old and metal-poor population 
is dominated in the Stream North. 
The peak of RC is found at $(g-i)_{0} \simeq 0.6$ and $i_{0} \simeq 24.8$ 
from panel (c). 

On the other hand, the CMD of the Stream South shows 
a different appearance; wider RGB,  
more abundant RC with multiple peaks, 
and a tight sequence of BHB. 
The differences such as the tight BHB sequence may be 
explained by the difference of the limiting magnitude, 
but additional stellar populations which are 
different from that found in the Stream North region  
is strongly suggested to exist in the Stream South region. 
Two distinct peaks of RC are found  
at $(g-i)_{0} \simeq 0.6$ and $i_{0} \simeq 24.8$ and 
at $(g-i)_{0} \simeq 0.8$ and $i_{0} \simeq 24.4$. 

The CMD of the Off-Stream South also shows 
a wider RGB and populated RC, but the morphology of RC is 
different from those of Stream South/North. 
In addition, no tight sequence of BHB is observed in the Off-Stream South. 
The peak of RC is found at $(g-i)_{0} \simeq 0.8$ and $i_{0} \simeq 24.3$,  
which seems to be coincide with  
one of the RC peaks found for the Stream South. 
This finding indicates that the CMD of the Stream South 
can be reproduced by a combination of CMDs of 
the Stream North and the Off-Stream South.

Although the survey depth of the Off-Stream North region is shallow, 
the CMD of the Off-Stream North is different from 
that of the Off-Stream South (e.g., populated RC and RGB), 
suggesting the presence of 
the diffuse but significant substructure in the Off-Stream South region.  
It is also noted that the number of NBGs, which are not affected 
by the survey depth, is smaller compared to that found 
in each of the other 3 regions [see panels (b)]. 
These findings lead to the idea that the Off-Stream North 
region is genuinely the region with no substructure, i.e., 
part of the smooth halo of M31. 
The loose RGB sequence and the absence of RC 
observed for the Off-Stream North 
also support the idea because stars in the halo are widely distributed 
along the line of sight direction and the features in the CMD, 
such as RGB, RC and BHB, are smoothed out along the magnitude axis.

Conclusions of this section are as follows: 
In addition to the known NW Stream,  
a diffuse substructure is likely to exist  
in the south part of our survey field. 
In the Stream South region, the NW Stream 
and diffuse substructure overlap.  
In contrast, the Off-Stream North region seems to be 
free of substructures and represent a genuine smooth halo of M31. 
In the following analysis, 
we regard the Stream North, the Off-Stream South and the Off-Stream North 
as the representative of the NW Stream, the diffuse substructure 
and the smooth halo, respectively.

\section{Basic Properties of Stellar Populations}\label{sec:basicprop}

\subsection{Tip of Red Giant Branch (TRGB)}\label{sec:trgb}

As shown in Figure~\ref{fig:cmd}, the bright part of RGB of 
M31's halo stars is merged into the higher density 
sequence consisting of numerous MW main sequence stars. 
We can separate the RGB stars from those overlapping 
MW main sequence stars by making use of $NB515$ data. 
TRGB is now visible by 
plotting NBGs in the CMDs of Figure~\ref{fig:4cmd} panels (b). 

Following the recent technique to determine the TRGB magnitude
based on the Bayesian approach 
\citep[e.g.,][]{Conn2011,Tollerud2016,Tanaka2017}, 
we attempt to determine the TRGB magnitude from NBGs,  
which are supposed to be a clean sample of RGB in principle, 
for 4 regions separately. 
The method we adopt is 
a maximum-likelihood estimation based on
the Markov Chain Monte Carlo (MCMC) algorithm, which is
described in detail by \citet{Tanaka2017}. 
In this study, we assume the model luminosity function (LF) as
\begin{eqnarray}
  \phi(m|m_{\scriptsize TRGB},a,b) &=& e^{a(m-m_{\scriptsize TRGB})} + b,\hspace{5mm} (for \hspace{1mm} m \ge m_{\scriptsize TRGB}) \\ 
  			&=& b, \hspace{5mm} (for \hspace{1mm} m < m_{\scriptsize TRGB})
\end{eqnarray}
where $m_{\scriptsize TRGB}$ is the TRGB magnitude. 
We also assume 100\% completeness for our NBG sample. 
The prior distribution assumed for $m_{\scriptsize TRGB}$ is 
a normal distribution with the mean, which is estimated from 
the edge detection algorithm \citep{Sakai1996} applied for the data
in each region, and $\sigma = 0.5$ mag, 
and those for $a$ and $b$ are a uniform distribution between 0 and 2.
The number of chain used in this study is 4. 
We also apply a color-cut,  
$-0.35 (i_{0}-23)+0.7 \le (g-i)_{0} \le -0.25 (i_{0}-23)+1.5$, 
to eliminate those stars deviated from RGBs.  
As shown in panels (b) of Figure~\ref{fig:4cmd}, 
this color-cut, which is represented in green dashed lines, 
clearly excludes outliers and distinguish the TRGB. 

\begin{figure}[t!]
\centering
\includegraphics[width=80mm]{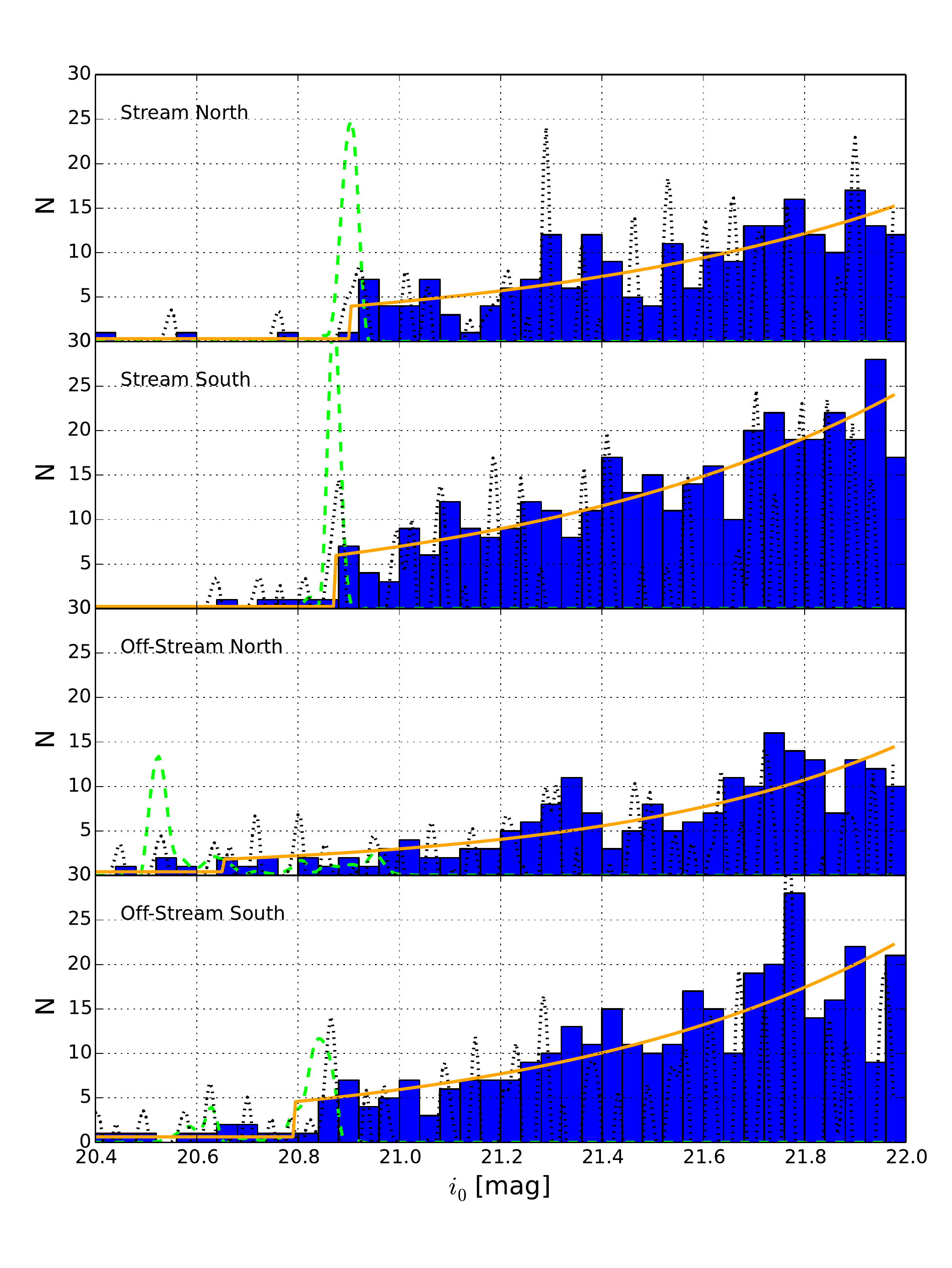} 
\caption{
The blue histogram shows the number distribution of NBGs 
and the dashed 
green
line shows the frequency of $m_{\scriptsize TRGB}$ 
obtained by 20,000 MCMC runs 
after 80,000 unused runs. 
The orange line is the model LF, $\phi(m_{n}|m_{\scriptsize TRGB},a,b)$, 
for which mean values of $m_{\scriptsize TRGB}, a, b$ are used. 
For comparison, we also plot the edge detection filter 
described by \citet{Sakai1996} as black dotted lines. 
}
\label{fig:TRGB}
\end{figure}

Figure~\ref{fig:TRGB} shows the result. 
The blue histogram shows the number distribution of NBGs 
and the dashed
green 
line shows the frequency of $m_{\scriptsize TRGB}$ 
obtained from 20,000 
MCMC
runs after 80,000 unused runs. 
The orange line is the model LF, $\phi(m_{n}|m_{\scriptsize TRGB},a,b)$,  
for which mean values of $m_{\scriptsize TRGB}, a, b$ are used. 
The figure clearly shows that 
the model LF fits the observed distribution of NBGs reasonably well 
and  $m_{\scriptsize TRGB}$ is well determined for Stream 
North and South regions and fairly determined for Off-Stream South. 
On the other hand, the determination of $m_{\scriptsize TRGB}$ is poor for 
Off-Stream North since the number of NBGs found for this region is small
and it is difficult to recognize the sharp rise of LF 
at $m_{\scriptsize TRGB}$. 
For comparison, we also plot the edge detection filter 
described by \citet{Sakai1996} as black dotted lines, 
suggesting that the both estimates are consistent with each other.    

Table~\ref{tab:TRGB} summarizes the distribution of $m_{\scriptsize TRGB}$. 
We take the peak value as the representative TRGB magnitude and 
the 68\% interval of the distribution around the peak as the error. 
We obtain the TRGB magnitudes as 
$i_{0}=$ 20.90$\pm$0.02, 20.87$\pm$0.02 and 20.84$\pm$0.03 mag
for the Stream North, Stream South and Off-Stream South, 
respectively. 

\begin{deluxetable*}{lccccccc}
\tablecaption{Estimated TRGB magnitude for 4 regions. }
\tablewidth{0pt}
\tablehead{ \colhead{Region} & \colhead{Mean} & \colhead{Peak} & \colhead{5\%} & 
	\colhead{25\%} & \colhead{50\%} & \colhead{75\%} & \colhead{95\%}}
\startdata
Stream North     & 20.90 & 20.90 & 20.87 & 20.89 & 20.90 & 20.91 & 20.92 \\
Stream South     & 20.87 & 20.87 & 20.85 & 20.86 & 20.87 & 20.87 & 20.89 \\
Off-Stream South & 20.79 & 20.84 & 20.58 & 20.78 & 20.83 & 20.85 & 20.86 
\enddata
\label{tab:TRGB}
\end{deluxetable*}

The calibration of the absolute magnitude of TRGB is 
intensively studied by \citet{Jang2017} and 
various calibrations are listed in \citet{Jang2017}. 
For the absolute magnitude of the TRGB, 
we adopt the calibration by \citet{Rizzi2007}:
\begin{equation}
M_{I_c} ({\rm TRGB}) = -4.05(\pm0.02) + 0.22(\pm0.01) [ (V - I_c) - 1.6 ]
\label{eq:Rizzi}
\end{equation}

When we use the color conversion formula for the relevant range of
$1.3 < g-i < 1.7$ (see Appendix~\ref{app:colconv}),
\begin{eqnarray}
V - I_c &=& 0.715 (g - i) + 0.317     \label{eq:V-Ic} \\
i - I_c &=& 0.067 (g - i) + 0.426 \ , \label{eq:i-Ic}
\end{eqnarray}
we obtain
\begin{equation}
M_i ({\rm TRGB}) = 0.222 (g - i) - 3.902
\label{eq: TRGB-i}
\end{equation}

For the TRGB of the Stream North, 
we find $i_{0} = 20.90 \pm 0.033$ and $(g-i)_{0} = 1.7$,
where the $(g-i)_{0}$ is the mean color of TRGB,  
thus equation~(\ref{eq: TRGB-i}) yields $(m-M) = 24.42$.
Similarly, we obtain $(m-M) = 24.39$ and $(m-M) = 24.31$ 
for those of the Stream South and Off-Stream South. 

The error of $(m-M)$ is calculated as follows: 
For the random error, 
we take the errors from Rizzi equation (Eq.~\ref{eq:Rizzi}),
the error from the TRGB determination ($\sigma=0.02$), 
and the width of the TRGB color ($\sigma=0.1$) 
and then calculate the root square sum of all the propagated errors. 
For the systematic error, we take the errors from 
the fitting of color conversions 
($\Delta=0.017$ and $\Delta=0.009$ for Eqs.~\ref{eq:V-Ic} and \ref{eq:i-Ic}) 
and the absolute photometric calibration ($\Delta=0.02$) 
and then calculate the sum of all the propagated errors. 
The random and systematic errors are calculated to be 
0.033 and 0.033, respectively, for the Stream North and South 
and those for the Off-Stream South are 0.040 and 0.033. 
The distances derived from TRGBs and their errors 
are summarized in Table~\ref{tab:distance}.

\subsection{Stellar Population of RGB, RC, and RGB bump}\label{sec:isofit}

To determine the absolute luminosities of specific
features in a CMD, such as RC and RGB bump (RGBb), we need
to set the age, $\tau_{\rm age}$, and metallicity, $Z$, 
of stellar population in
both stream and off-stream regions. For this purpose,
we attempt to reproduce both the locus of RGB and the relative position
of RC in a CMD for the ranges of $\tau_{\rm age}$ and $Z$,
based on a Padova isochrone with a fiducial value for 
mass-loss parameter on RGB of $\eta = 0.2-0.5$. 
The color system of the isochrone (SDSS system) is converted to 
the HSC system as described in Appendix~\ref{app:colconv}. 

First, we adopt the distance modulus determined by TRGB method 
described in Section~\ref{sec:trgb} and look for the best-fit isochrone. 
For the Off-Stream South, we find that the isochrone with   
$Z=0.0014$ ([Fe/H] = -1.13), $\log \tau_{\rm age} ({\rm Gyr}) = 10.00$ 
and $\eta = 0.3$ traces the RGB and RC features perfectly   
[see Figure~\ref{fig:4cmd} (4)-(b) and (4)-(c)]. 
On the other hand, it is found to be difficult to fit 
the isochrones to the CMDs of Stream North and Stream South, 
in particular, to the RC features 
[see black thick lines in Figure~\ref{fig:4cmd} (1)-(c) and (3)-(c)]. 
We therefore try to fit the isochrone by changing the distance modulus 
and find that the isochrone with 
$Z=0.0008$ ([Fe/H] = -1.37), $\log \tau_{\rm age} ({\rm Gyr}) = 10.00$ 
and $\eta = 0.3$ traces the RGB and RC features for 
Stream North and Stream South by increasing the distance modulus 
by 0.3 and 0.2 mag, respectively  
[see black dashed lines in Figure~\ref{fig:4cmd} (1)-(c) and (3)-(c)]. 
We note that the `deep dip' in LF is found for 
the Stream North at $i_{0} \simeq 21.14$ mag, which is 
in contrast to the monotonic increase of RGB population toward 
fainter magnitude found for most galaxies.  
This may suggest that the real RGB population in the Stream North 
emerges from $i_{0} \sim 21.14$ mag. The edge detection filter 
suggests the peak at $i_{0} \simeq 21.22$ which is consistent 
with the suggestion from isochrone fitting. 
Note the typical metallicity range estimated from the width of 
RGB is $\sigma_{\rm [Fe/H]} \sim$ 0.2, which is used in what follows.

\subsubsection{Red Clump}\label{sec:rc}

\begin{figure*}[t!]
\centering
\begin{tabular}{cc}
\includegraphics[width=50mm]{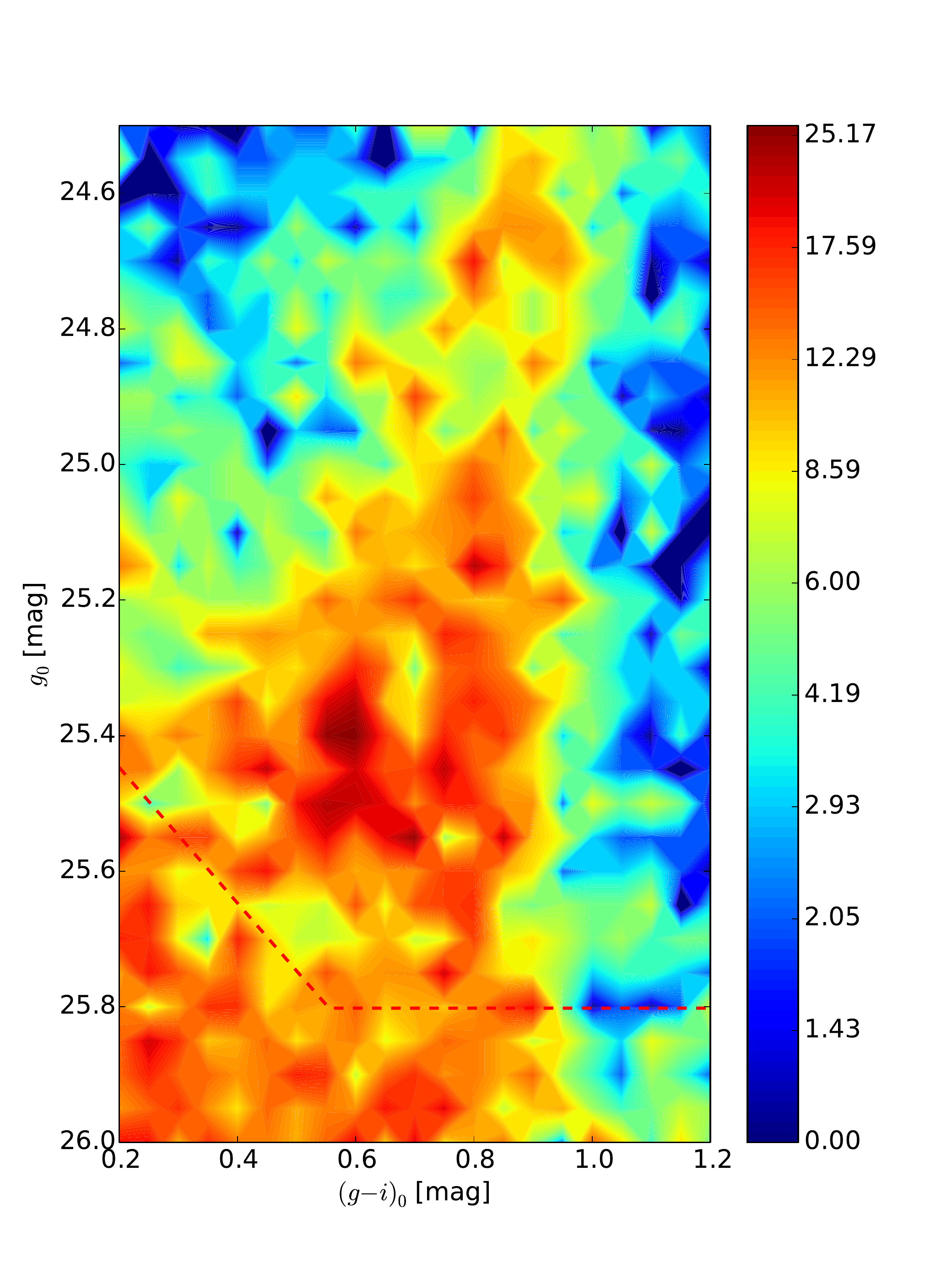} 
& 
\includegraphics[width=50mm]{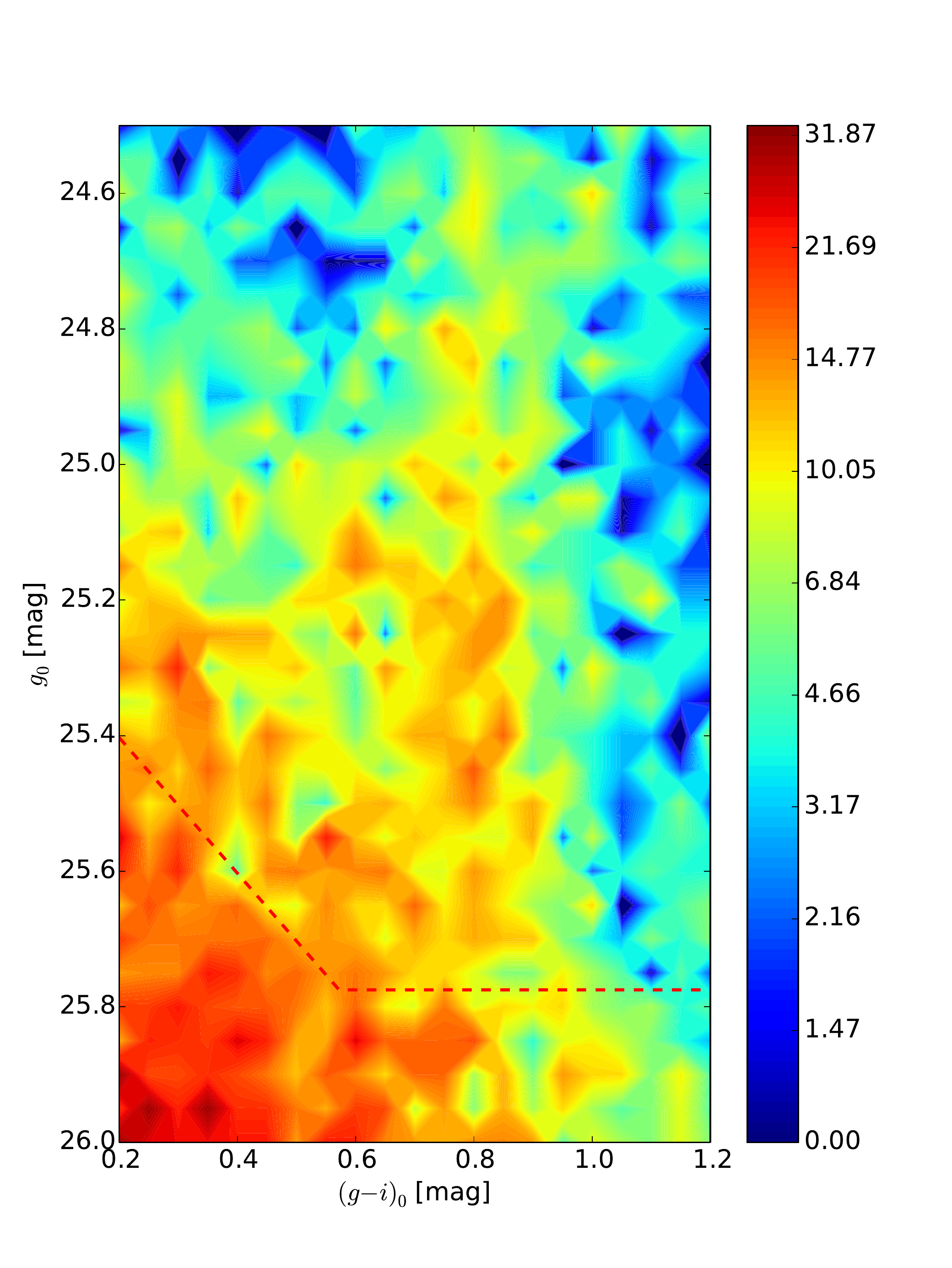} \\
(a) Stream North & (b) Off-Stream North \\
\includegraphics[width=50mm]{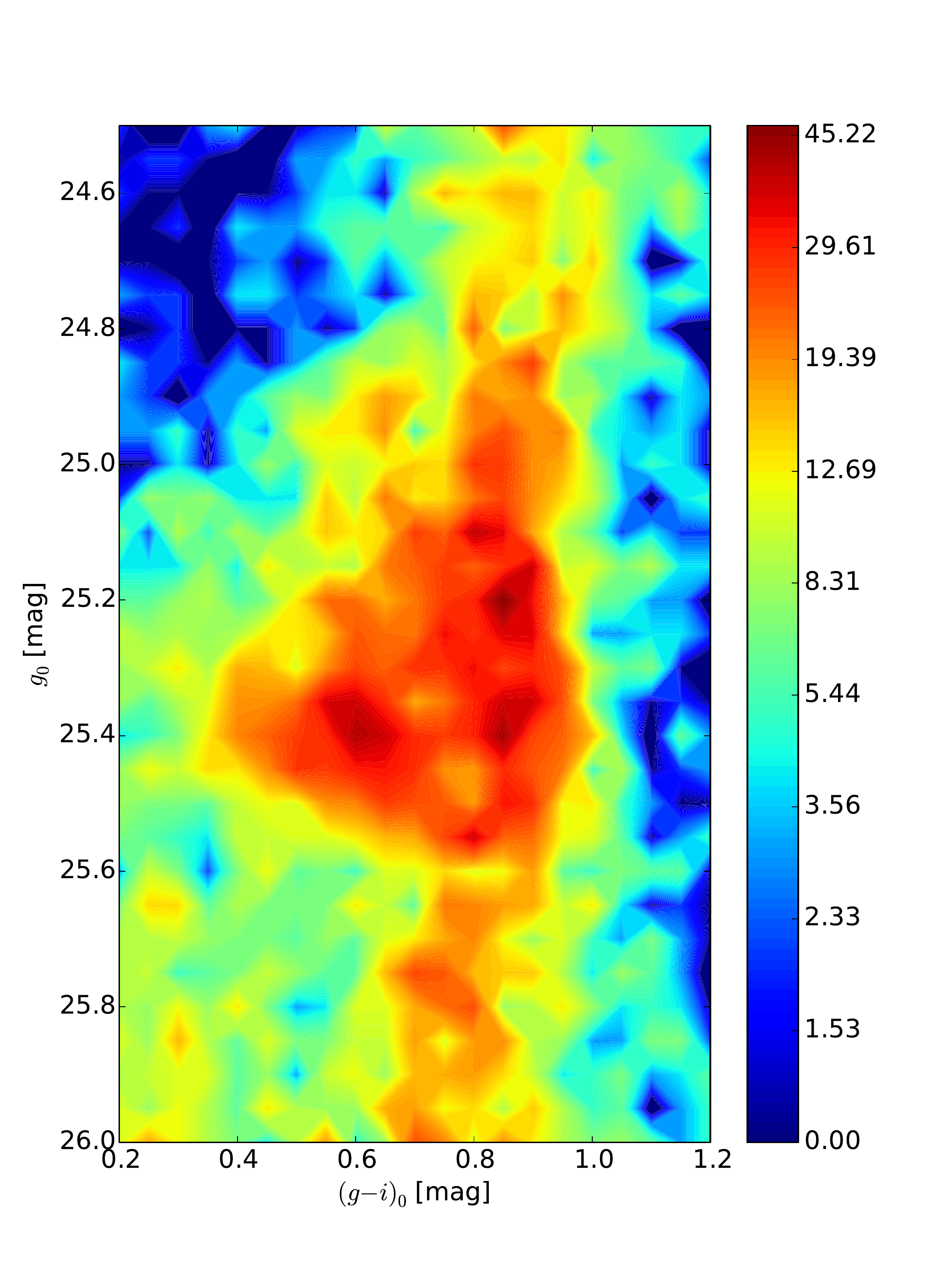} 
&
\includegraphics[width=50mm]{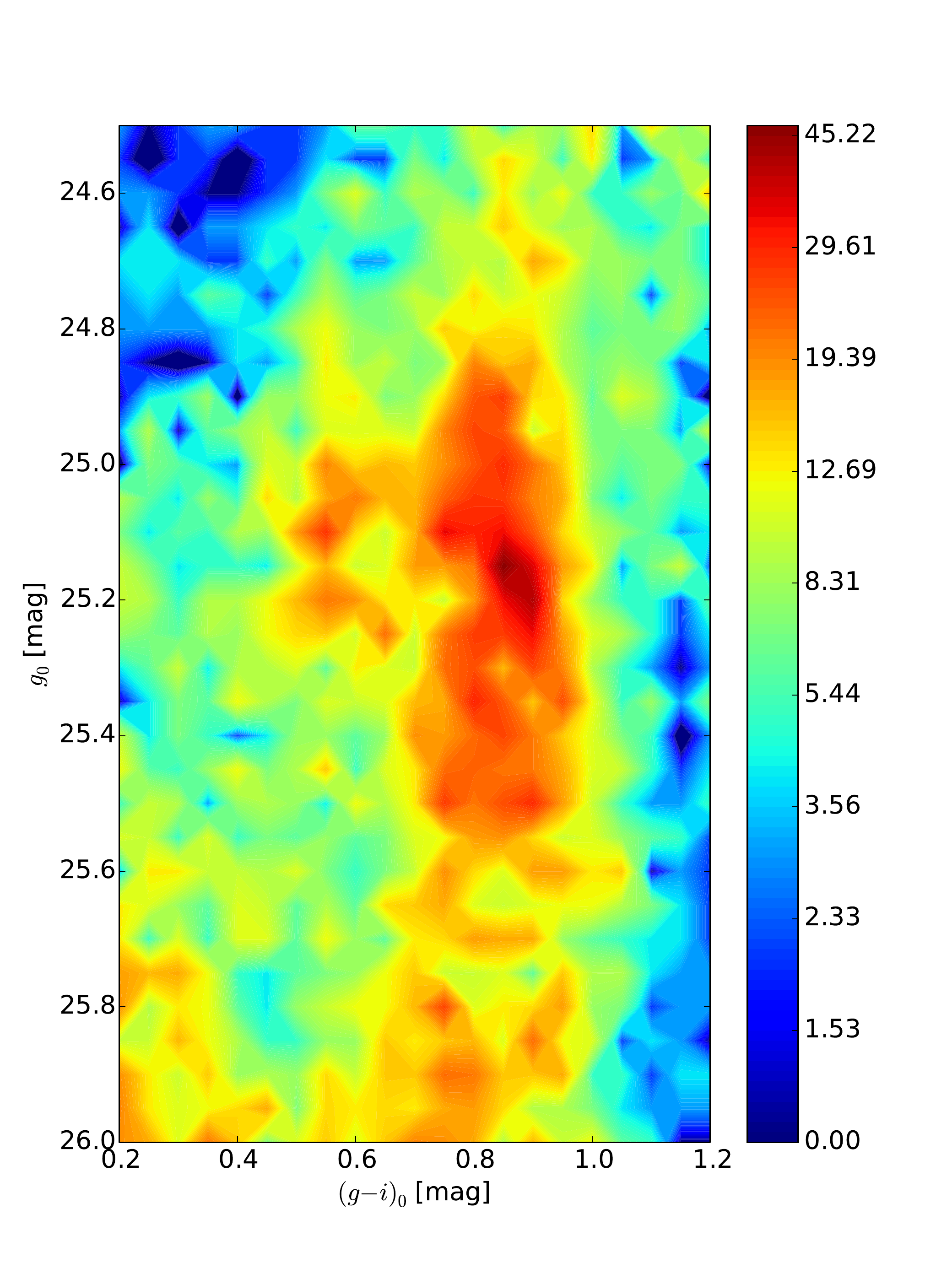} \\
(c) Stream South & (d) Off-Stream South \\
\end{tabular}
\caption{
Zoomed CMDs in the form of density map 
around RC of 4 regions in our survey field. 
From top-left to bottom-right, CMDs of 
(a) Stream North, (b) Off-Stream North, (c) Stream South, 
and (d) Off-Stream South, respectively, are plotted. 
The grid widths used in these figures are 
$\Delta (g-i) = 0.05$ and $\Delta g = 0.05$ mag. 
Note that the vertical axis is $g$-band magnitude. 
}
\label{fig:4rc}
\end{figure*}

Figure~\ref{fig:4rc} shows the zoomed view of the CMDs 
in the form of density map around RC 
where we note that the vertical axis is 
now $g$-band magnitude in this figure.  
Significant peaks of RC at $(g-i)_{0} \simeq 0.6$ 
are found in the CMDs for 
the Stream North, Stream South and the Off-Stream South. 
It is also suggested that the CMD around RC for the Stream South can be 
reproduced by the combination of the CMDs for 
the Stream North and the Off-Stream South. 
No significant peak is found for the Off-Stream North
but this is naturally understood if the Off-Stream North region is 
a representative of a smooth halo (see Section~\ref{sec:4cmd}). 

We measure the color and magnitude of the peak of RC 
for each region. They are measured to be   
$(g-i)_{0} = 0.562 \pm 0.085$ and $g_{0} = 25.456 \pm 0.132$
for the Stream North and 
$(g-i)_{0} = 0.568 \pm 0.080$ and $g_{0} = 25.122 \pm 0.160$
for the Off-Stream South.
Here, we adopt the standard deviation around the peak of RC as errors. 
Since two different structures are suggested to be overlapping 
in the Stream South region and no significant peak is 
found for the Off-Stream North, 
we do not attempt to derive the color and magnitude of RC 
for these two regions.

For the absolute magnitude of RC, we 
adopt the calibration by \citet{Bilir2013}:
\begin{eqnarray}
M_V ({\rm RC}) &=& 0.627 (\pm0.104) (B-V)_0 + 0.046 (\pm0.043) [Fe/H] \nonumber \\
		 & & + 0.262 (\pm0.111).
\label{eq:Bilir}
\end{eqnarray}

When we use the color conversion formula for the relevant range of
$0.4 < g-i < 1.2$ (see Appendix~\ref{app:colconv}),
\begin{eqnarray}
g-V &=& 0.371 (g - i) + 0.068 \label{eq:g-V} \\
B-V &=& 0.709 (g - i) + 0.170 \ , \label{eq:B-V}
\end{eqnarray}
we obtain
\begin{equation}
M_g ({\rm RC}) = 0.8155 (g - i) + 0.046 [Fe/H] + 0.4366
\label{eq: RC-g}
\end{equation}

Using eq.~(\ref{eq: RC-g}) combined with the metallicity of
[Fe$/$H]$=-1.38$ (see Section~\ref{sec:isofit}), 
we obtain $M_g ({\rm RC}) = 0.831$
and thus $(m-M) = 24.63$ for the RC of the Stream North.
For the RC of the Off-Stream South, we obtain $M_g ({\rm RC}) = 0.848$
and thus $(m-M) = 24.27$.

The error of $(m-M)$ is calculated in the similar way 
to the calculation for TRGB.  
For the random error, we take the errors from 
Bilir equation (Eq.~\ref{eq:Bilir}),
the error of the RC magnitude, 
the error of the RC color, 
and the error of the metallicity ($\sigma=0.2$)
and then calculate the root square sum of all the propagated errors. 
For the systematic error, we take the errors from 
the fitting of color conversions 
($\Delta=0.029$ and $\Delta=0.019$ for Eqs.~\ref{eq:g-V} and \ref{eq:B-V}) 
and the absolute photometric calibration ($\Delta=0.02$) 
and then calculate the sum of all the propagated errors. 
The random and systematic errors are calculated to be 
0.191$\sim$0.211 and 0.057, respectively. 
The distances derived from RC and their errors 
are summarized in Table~\ref{tab:distance}.

\subsubsection{RGB Bump}

In Figure~\ref{fig:4rc}, RGBb features are recognized 
at $(g-i)_{0} \simeq 0.8$ for  
the Stream North, Stream South and Off-Stream South. 
This feature is most prominent at the Off-Stream South, 
suggesting that this feature at  
$(g-i)_{0} \simeq 0.8$ and $g_{0} \simeq 25.2$ is 
characteristic for the Off-Stream South, i.e., the diffuse substructure.  
Looking at the CMD of the Stream South, a peak is found at the 
same position as the Stream South, and the other peak is found 
at $(g-i)_{0} \simeq 0.8$ and $g_{0} \simeq 25.4$, which  
is suggested to be the bump feature of the M31 NW Stream and 
should also be found at the Stream North. 
Bearing this in mind, 
we measure the color and magnitude of RGBb feature 
for the Stream North and Off-Stream South regions. 
They are measured to be   
$(g-i)_{0} = 0.830 \pm 0.076$ and $g_{0} = 25.478 \pm 0.097$
for the Stream North and 
$(g-i)_{0} = 0.857 \pm 0.076$ and $g_{0} = 25.112 \pm 0.110$
for the Off-Stream South.

\citet{Alvez+1999} show the table on the $V$-band absolute magnitude of
RGBb, $M_V ({\rm RGBb})$, for stellar population 
with different age and metallicity. 
The interpolation of the table gives $M_V ({\rm RGBb}) = 0.177$ 
and $M_V ({\rm RGBb}) = 0.477$ for the Stream North 
and Off-Stream South, respectively. 
Thus, using eq.~(\ref{eq:g-V}) to estimate $V$-band magnitude from 
the values of $g_{0}$ and $(g-i)_{0}$ of RGBb, 
we obtain the distance modulus $(m-M)$ to each region.
For RGBb of the Stream North, 
our estimate of $(g-i)_{0}=0.830$ and $g_{0}=25.478$ 
yields $V=25.102$, then $(m-M) = 24.93$.
For RGBb of the Off-Stream South, 
our estimate of $(g-i)_{0}=0.857$ and $g_{0}=25.112$ 
yields $V=24.726$, then $(m-M) = 24.25$.
The derived distance moduli are consistent with 
the tendency obtained from RC methods 
that the Stream North (the NW Stream) is located at 
farther than the Off-Stream South (the diffuse substructure). 
Note that \citet{Alvez+1999} relation seems to be very sensitive to 
the metallicity and age of the population and 
the present estimate would have large uncertainty.

\subsection{Summary of Distance Estimates of Stellar Populations}

Table~\ref{tab:distance} summarizes the distance moduli measured by 
different methods presented here.
The Off-Stream South region shows small variation in 
the distance moduli estimated from different methods. 
However, the difference between the distance modulus derived from TRGB 
and that from RC is large for the Stream North region. 
This difference is relaxed if the distance modulus from TRGB 
is increased by 0.2-0.3 mag, which is suggested from the isochrone 
fitting (see Section~\ref{sec:isofit}). 
We therefore prefer to use the distance moduli derived 
from RC method for the following analysis. 

Considering that the distance modulus of M31 is 24.45 mag (776 kpc), 
which is the median of 345 measurements compiled by 
NASA/IPAC Extragalactic Database (NED)\footnote{http://ned.ipac.caltech.edu/}, 
the NW Stream of our survey field is located behind M31 by $\sim$90 kpc 
if we adopt the distance modulus derived from RC for the Stream North. 
In contrast, the diffuse substructure, which is represented by 
the Off-Stream South found by this study 
is located at more than $\sim$30 kpc in front of M31.

\begin{deluxetable*}{llll}
\tablecaption{The summary of distances calculated in this study.}
\tablewidth{0pt}
\tablehead{ \colhead{Field} & \colhead{Stream North} & \colhead{Stream South} & \colhead{Off-Stream South} }
\startdata
(m-M) from TRGB & 24.42 & 24.39 & 24.36 \\
\hspace{10pt}(random error) & \hspace{10pt}$\pm$0.033 & \hspace{10pt}$\pm$0.033 & \hspace{10pt}$\pm$0.040 \\ 
\hspace{10pt}(systematic error) & \hspace{10pt}$\pm$0.033 & \hspace{10pt}$\pm$0.033 & \hspace{10pt}$\pm$0.033 \\ 
\tableline
(m-M) from RC & 24.63 & - & 24.29 \\
\hspace{10pt}(random error) & \hspace{10pt}$\pm$0.191 & - & \hspace{10pt}$\pm$0.211 \\ 
\hspace{10pt}(systematic error) & \hspace{10pt}$\pm$0.057 & - & \hspace{10pt}$\pm$0.057 \\ 
\tableline
(m-M) from RGBb & 24.77 & - & 24.39  
\enddata
\label{tab:distance}
\end{deluxetable*}

\subsection{3D Structure of the M31 NW Stream}\label{sec:3Dstr}

To understand the nature of the M31 NW Stream, 
it is important to obtain the distance information for the M31 NW Stream. 
If the 3D structure of the NW Stream is obtained, 
then this information gives a useful constraint 
on the origin of the stream.

In this section, we investigate the distances to the NW Stream 
by dividing the stream regions (i.e., Stream North and South) into 4 regions. 
The boundaries are set as $\delta = 43.7, 44.8$ and 46.1 and 
we name the regions Stream 1 to 4 from the north to south. 
Due to the issues on the TRGB distances discussed in the previous Section, 
we adopt the RC method as described in 
Section~\ref{sec:rc} to address the distance distribution 
along the NW stream. 

\begin{figure*}[t!]
\centering
\begin{tabular}{cc}
\includegraphics[width=50mm]{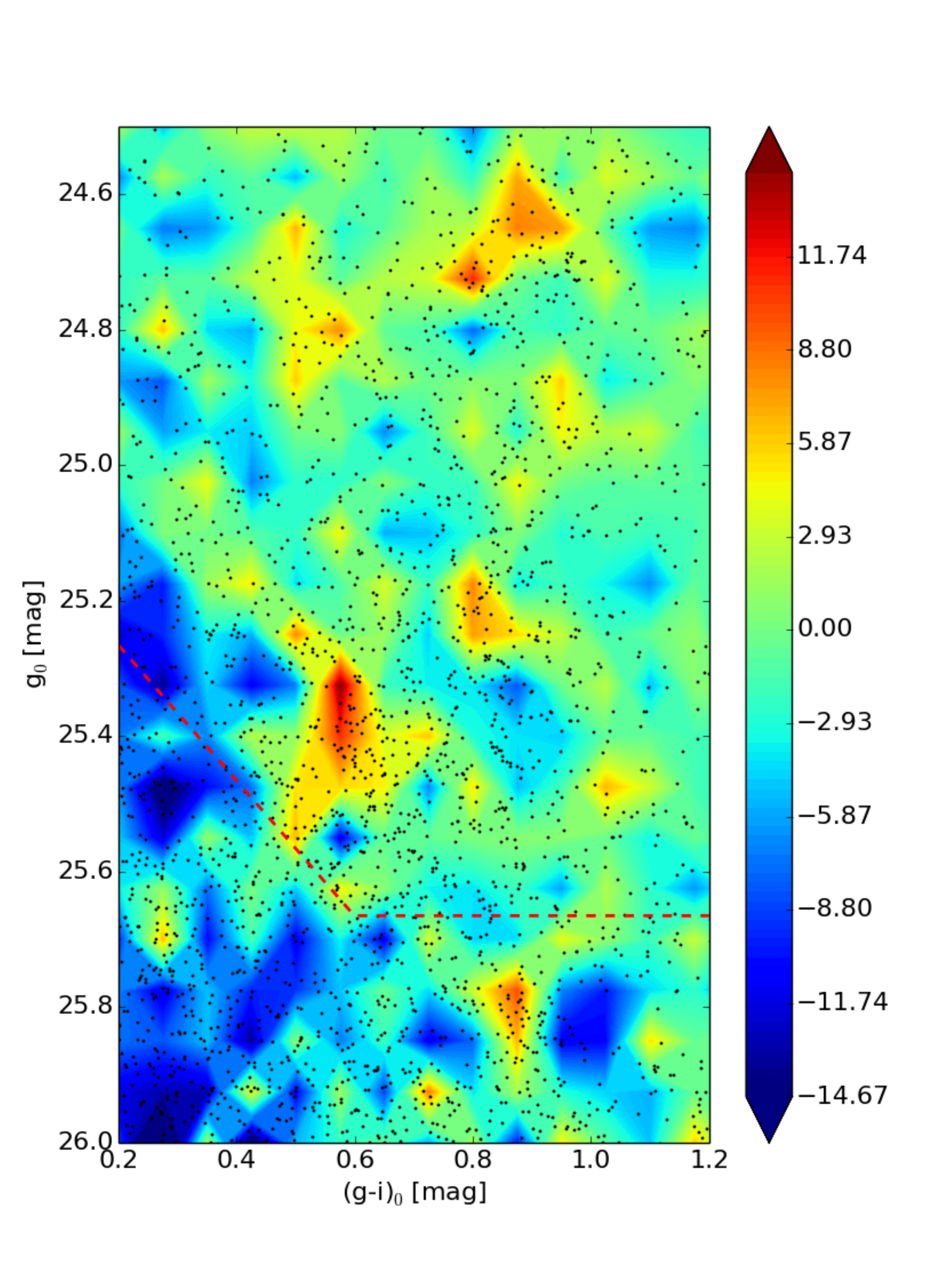} 
& 
\includegraphics[width=50mm]{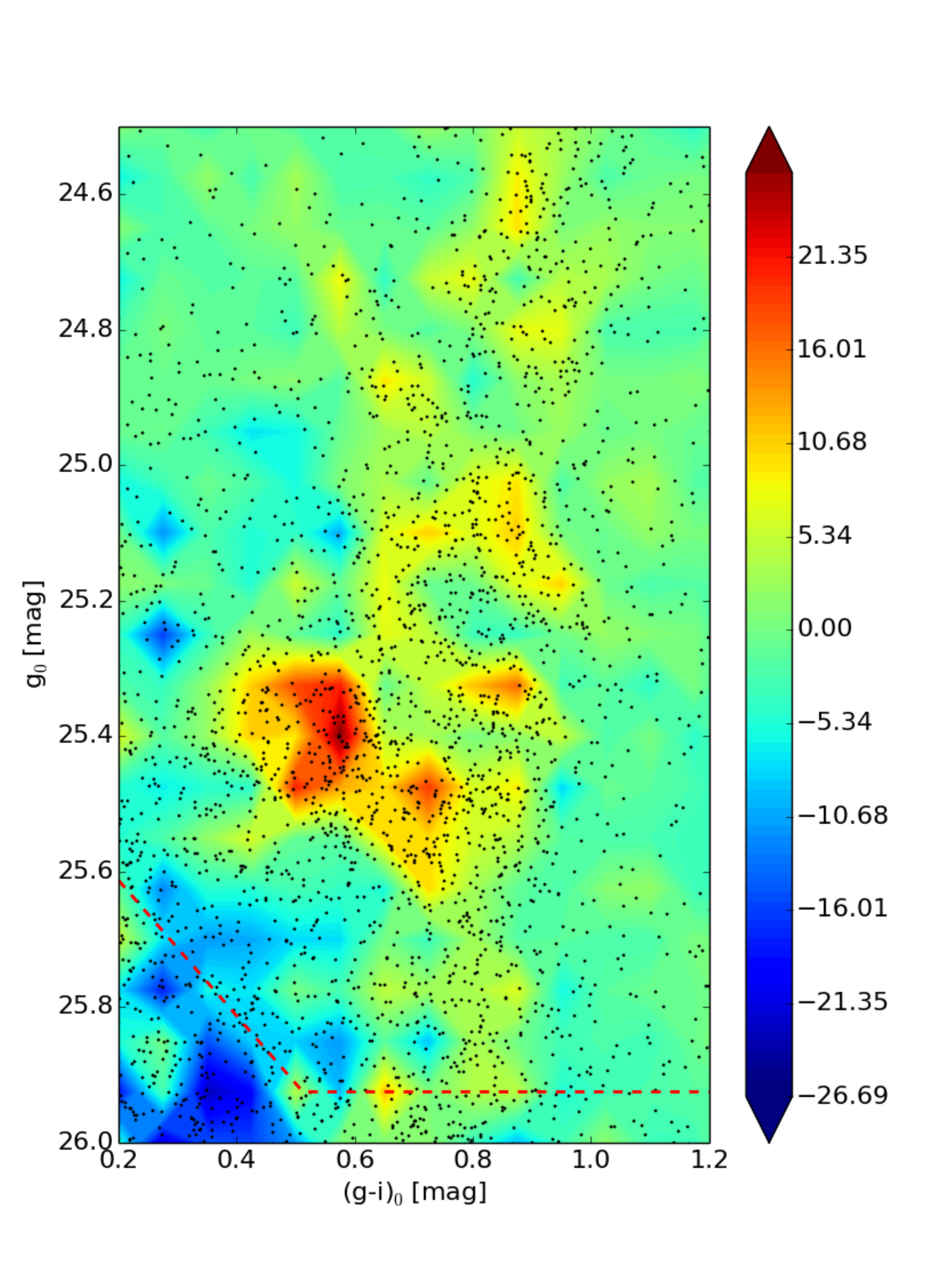} \\
(a) Stream 1 & (b) Stream 2 \\
\includegraphics[width=50mm]{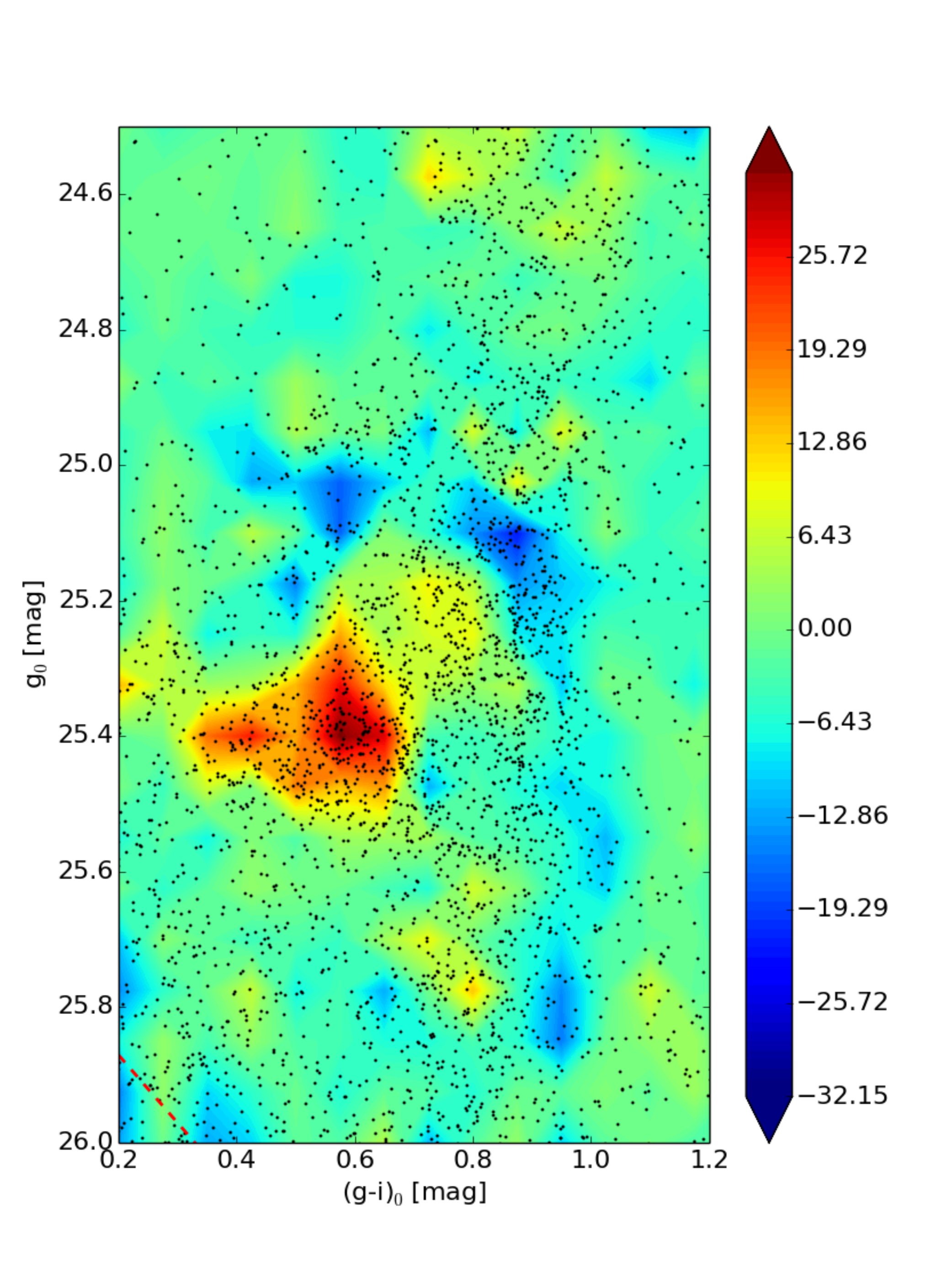}
&
\includegraphics[width=50mm]{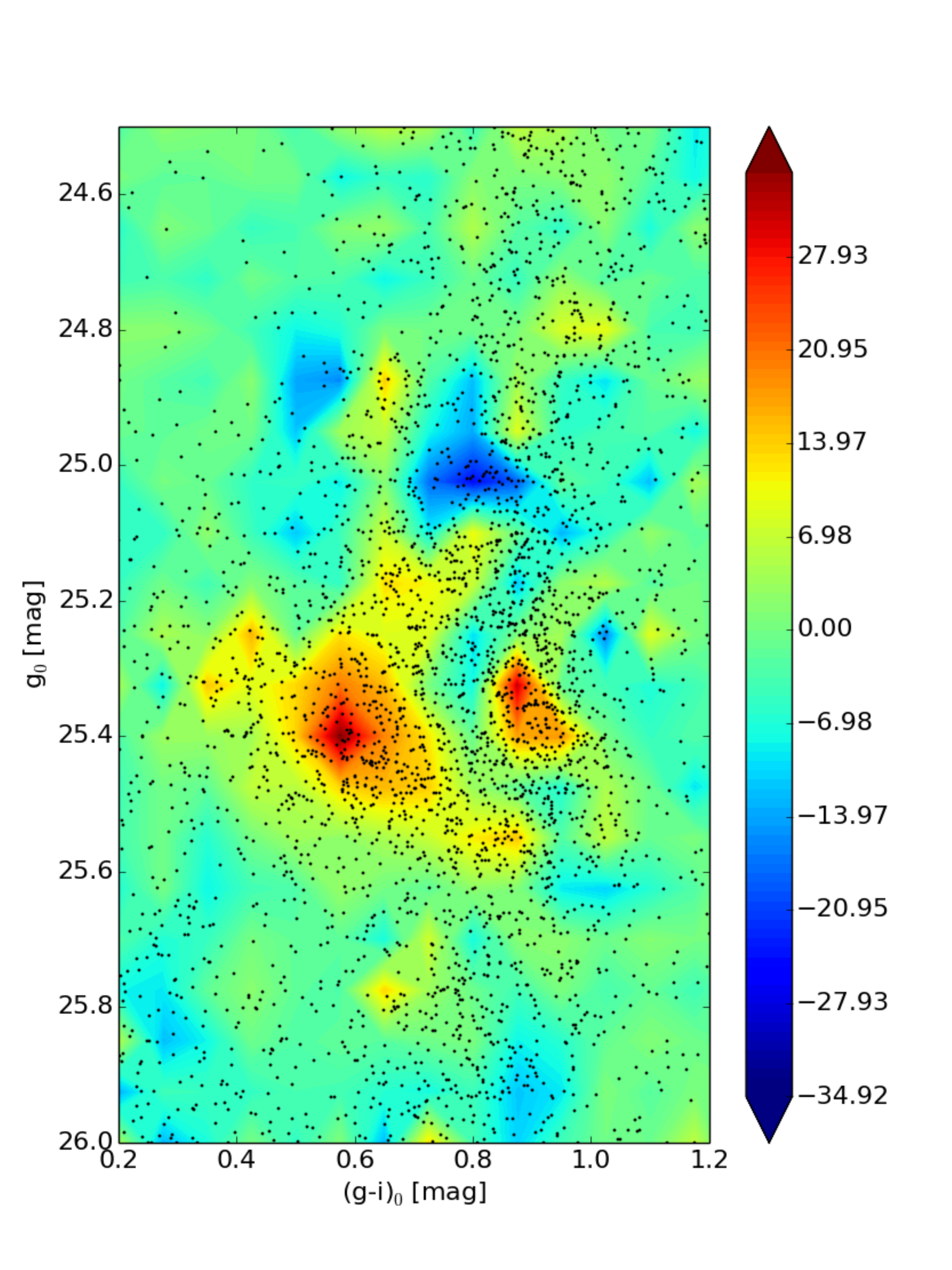} \\
(c) Stream 3 & (d) Stream 4 \\
\end{tabular}
\caption{
Background/foreground-subtracted CMDs (HESS diagram) 
around RC of 4 regions along the NW Stream. 
From top-left to bottom-right, CMDs of 
(a) Stream 1, (b) Stream 2, (c) Stream 3 
and (d) Stream 4, respectively, are plotted. 
In each panel, individual stars in the stream region
are also plotted as black dots. 
The grid widths used in these figures are 
$\Delta (g-i) = 0.075$ and $\Delta g = 0.075$ mag. 
}
\label{fig:4rcstr}
\end{figure*}

Figure~\ref{fig:4rcstr} shows the background/foreground-subtracted 
CMDs (HESS diagrams) around RC. 
The CMD of the background/foreground objects for the Stream 1 region 
is made from those stars which reside outside the stream region 
with the same declination range as the Stream 1 region 
(i.e., $\delta > 46.1$). 
The CMDs of the background/foreground objects 
for the Stream 2-4 regions are made in the same way. 
In each panel, individual stars in the stream regions 
are also plotted as black dots, aiming to show 
the real distribution of stars in each CMD. 
We measure the mean color and magnitude of the RC 
for each region and find  
$(g-i)_{0} = 0.562 \pm 0.085$ and $g_{0} = 25.467 \pm 0.139$,  
$(g-i)_{0} = 0.562 \pm 0.084$ and $g_{0} = 25.447 \pm 0.125$,  
$(g-i)_{0} = 0.569 \pm 0.085$ and $g_{0} = 25.430 \pm 0.119$, and 
$(g-i)_{0} = 0.573 \pm 0.083$ and $g_{0} = 25.420 \pm 0.123$  
for the Stream 1 to 4 regions, respectively.
The distance moduli derived from the RC method are 
24.64, 24.62, 24.59, 24.58 for the Stream 1 to 4, respectively
(see Table~\ref{tab:strdistance}). 
Note that the Stream 1 is shallow in terms of the completeness 
but the peak of the RC seems to be reasonably measured since it is found at 
$\sim 0.2$ mag brighter than the 50\% completeness limit.

As explained in Section~\ref{sec:reduction}, 
the Galactic extinction is corrected for every single star
using the new estimate from \citet{SchlaflyFinkbeiner2011}
which is based on the dust map by \citet{Schlegel1998}. 
The reddening E($B-V$) ranges from $\sim$0 to 0.1318 
within our survey field (see Figure~\ref{fig:extinction}) 
which corresponds to the maximum extinction of $A_{g}=0.4858, A_{i}=0.2440$.   
Therefore, the error in the reddening map would introduce 
additional error for extinction-corrected magnitudes. 
It is noted that the variation in extinction-corrected color
$(g-i)_{0}$ of the RC along the stream is as small as 0.01 mag.
This suggests that our method to correct extinction is carried out properly
given that the stellar population is homogeneous along the stream.

Although the uncertainty in the RC method is large, 
the result indicates that the south part of the NW Stream is 
$\sim$20 kpc closer to us (and M31) relative to the north part 
(see the right panel of Figure~\ref{fig:NWStream3D}).
The angle between the Stream 1 and 4 is $\sim$ 4.2 deg, 
which corresponds to the projected distance of $\sim$ 61 kpc, 
indicating that this configuration seems to be realistic. 
The angle from the Stream 4 to the major axis of M31 
along the stream is roughly 6 deg (see Figure~\ref{fig:pointing}), 
which corresponds to the projected distance of $\sim$ 90 kpc. 
The line of sight distance between Stream 4 and M31 is 
calculated to be $\sim$ 50 kpc. 
If the NW Stream is extended to the south-east direction 
with the same rate as that found in our survey field, 
it is still behind M31 when it crosses the major axis of M31. 
Therefore, we conclude that the NW Stream is 
a part of an orbiting stream around M31 
on the farther side of M31.

\begin{deluxetable*}{lllll}
\tablecaption{The summary of distances of NW Stream.}
\tablewidth{0pt}
\tablehead{ \colhead{Field} & \colhead{Center Coordinates [deg]} & \colhead{(m-M)} & \colhead{(m-M) Error (random/systematic)} & \colhead{Distance [kpc]} }
\startdata
Stream 1 & (2.06, 46.90) & 24.64 & $\pm 0.196 \pm 0.057$ & 847 \\
Stream 2 & (3.00, 45.45) & 24.62 & $\pm 0.186 \pm 0.057$ & 839 \\
Stream 3 & (3.78, 44.25) & 24.59 & $\pm 0.183 \pm 0.057$ & 828 \\
Stream 4 & (4.52, 43.10) & 24.58 & $\pm 0.185 \pm 0.057$ & 824
\enddata
\label{tab:strdistance}
\end{deluxetable*}

\subsection{Shape of the M31 NW Stream}

\begin{figure}[t!]
\centering
\includegraphics[width=80mm]{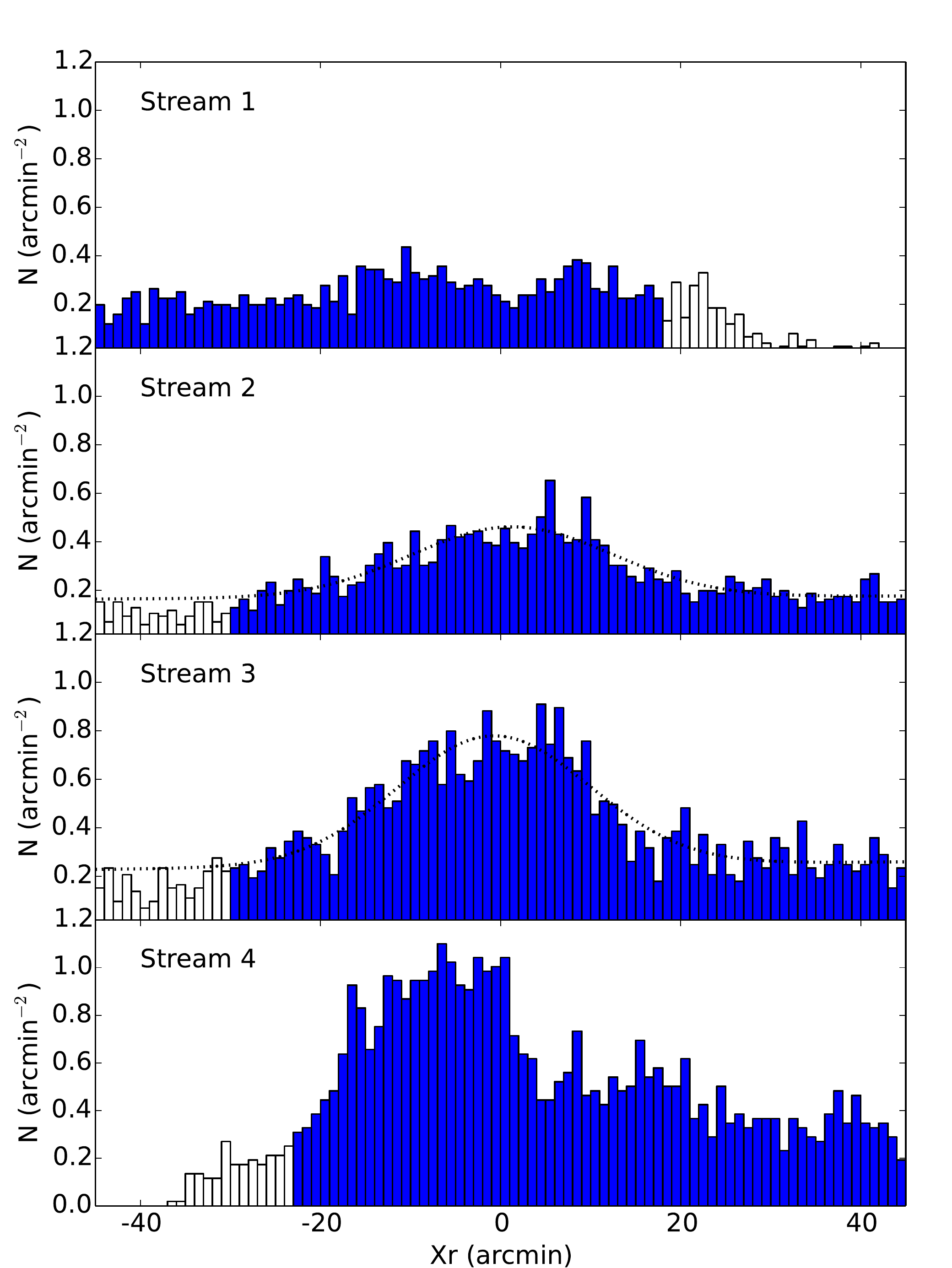} 
\caption{
The number density distributions of RC stars 
along the direction perpendicular to the NW Stream. 
From top to bottom, the histograms for the Stream 1 to 4 are shown. 
The positive Xr corresponds to the south-west side of the NW Stream.
The areas shown in white histograms are 
where our survey coverage is incomplete. 
The best-fit profile, which is a combination of a Gaussian 
and a linear function, is plotted as a dotted line in 
the Stream 2 and 3 fields. 
}
\label{fig:NWStreamCS}
\end{figure}

\begin{figure*}[t!]
\centering
\begin{tabular}{cc}
\includegraphics[width=80mm]{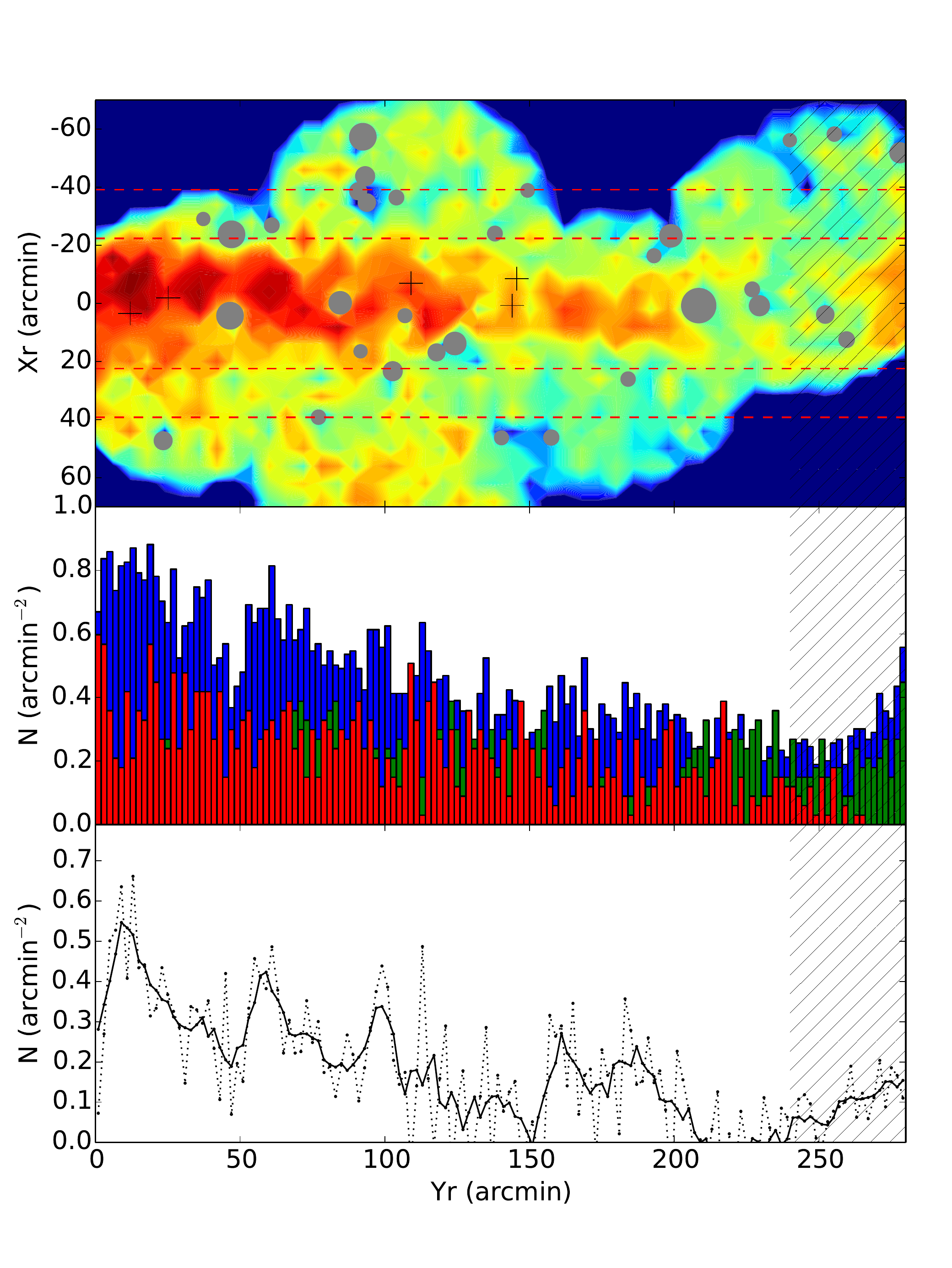}
& 
\includegraphics[width=80mm]{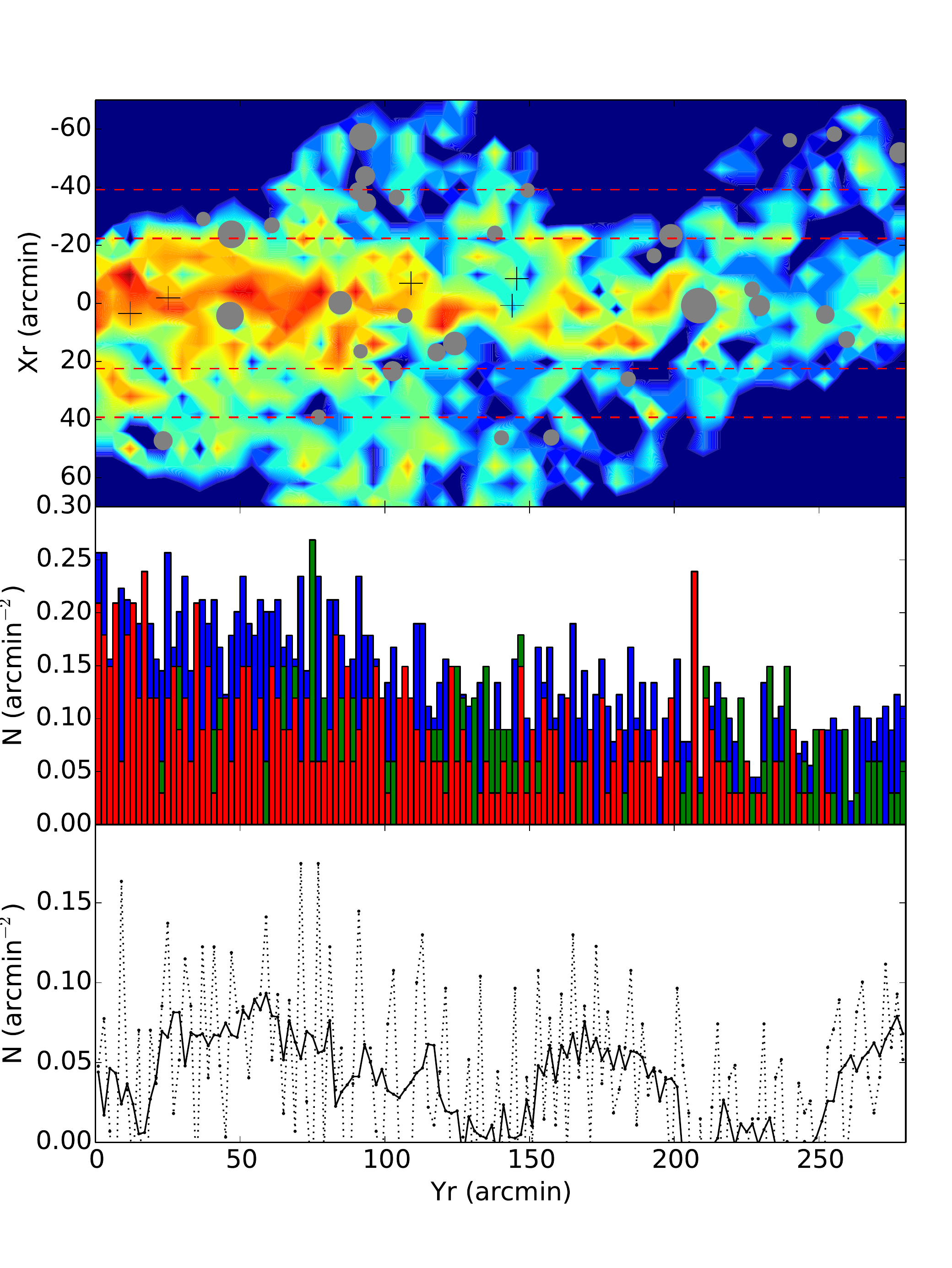} \\
(a) & (b) \\
\end{tabular}
\caption{
(a) The shape of the NW stream represented by RCs:   
(Top)
The spatial distribution of RC stars on the Xr-Yr coordinates system. 
The red dashed lines indicate the division lines 
between blue/green/red histograms, 
i.e., $\pm$2$\sigma$ and $\pm$3.5$\sigma$ of  
the Gaussian fitted for Stream 2 and 3. 
The crosses represent the positions of globular clusters 
found on the NW Stream by \citet{Huxor2014}. 
(Middle)
The number histograms of RC stars 
along the direction of the NW Stream. 
The blue histogram represents the surface number density 
of RC stars within the width of 44.8 arcmin 
(i.e., $\pm$2$\sigma$). 
The green and red histograms represent the surface number density 
of RC stars outside the above region 
to the north and south, respectively.  
The positive Yr corresponds to the north-west side of the NW Stream.
(Bottom)
The best-estimated foreground/background subtracted 
surface number density distribution of RC stars 
along the direction of the NW Stream. 
The foreground/background is estimated 
assuming that the maximum of the green and red histograms 
of middle figure at fixed Yr and is subtracted from the blue histogram. 
The dotted line represents the subtracted number density 
and the thick line represents the moving average over 5 bin (i.e., 10 arcmin). 
Note that the Yr$>$240 region where lower completeness than 50\% 
is suggested from Figure~\ref{fig:space} (b) is hatched in the figure. 
(b) Same figure as (a), but for NBGs. 
}
\label{fig:NWStreamProf}
\end{figure*}

Figure~\ref{fig:NWStreamCS} shows 
the number density distributions of RC stars for the Stream 1 to 4 field 
along the direction perpendicular to the NW Stream, 
where the coordinate Xr increases towards the south-west direction 
as clarified in the top panel of Figure~\ref{fig:NWStreamProf}.  
Although our survey does not fully cover 
the south-west side (i.e., Xr$>$18) of the Stream 1 field and 
the north-east side (i.e., Xr$<$-30$\sim$-24) 
of the Stream 2-4 field, 
figures for Stream 2 and 3 clearly show the number density distribution 
across the NW Stream. 

The overall shape can be fitted by the combination of a Gaussian with 
FWHM of $\sim$ 25 arcmin and a linear function 
for the Stream 2 and 3 fields. 
It is noted that the distribution seems to be slightly skewed: 
the south-west side of the stream 
(i.e., right side of the distribution in the figures) shows 
steep rise ($\sim$ 0.06 / arcmin$^{3}$ for 6 $<$ Xr $<$ 18) 
while the north-east side shows rather shallow rise
($\sim$ 0.03 / arcmin$^{3}$ for -18 $<$ Xr $<$ -6)  
in the Stream 3 field.
Similar trend is also seen for the Stream 2 and 4 fields. 
This may tell something on the formation process 
of the NW Stream. 
The Stream 1 field shows no prominent peak at around Xr = 0, 
but a hint of overdensity is clearly visible from 
Figure~\ref{fig:space}. 
The overdensity seems to be extended to north-west 
beyond our survey fields, thereby further observation to this direction 
is required to investigate the detailed shape for the Stream 1 field.

The middle panel of Figure~\ref{fig:NWStreamProf} 
(a)
shows 
the number density distributions of RC stars along the NW Stream, 
where the coordinate Yr increases towards the north-west direction. 
The blue histogram represents the surface number density 
of RC stars within a width of 44.8 arcmin centered on the stream. 
The width of 44.8 arcmin corresponds roughly to $\pm$2$\sigma$ 
of the Gaussian fitted for Stream 2 and 3. 
The positive Yr corresponds to the north-west side of the NW Stream
as clarified in the top panel of Figure~\ref{fig:NWStreamProf}
(a)
.  
The green and red histograms represent those for outside the stream 
on the north and south side, respectively. 
The histograms are calculated using those RC stars found in 
1.5$\sigma$ width regions, although 
our survey does not always cover these regions. 
Therefore, the red and green histograms should be 
taken as a guide to estimate the foreground/background 
contamination to the surface number density 
of RC stars of the NW Stream. 

The blue histogram outnumbers the green and red histograms 
for the most part of the figure, 
indicating that the stream is found for all through 
our survey field. 
It is also noted that the surface number density  
varies dramatically along the stream 
and several gaps are found in the stream. 
We try to estimate the foreground/background subtracted 
number density distribution of RC stars 
assuming that the maximum of the green and red histograms 
at fixed Yr represents the foreground/background. 
The bottom panel of Figure~\ref{fig:NWStreamProf} 
(a)
shows 
the foreground/background subtracted number density distributions in 
each bin (dotted line) and a moving average over 5 bins (thick line). 
The surface number density is the highest at Yr = 13 arcmin
and is 0.66$\pm$0.10 arcmin$^{-2}$ (Poisson error is assumed). 
The overall surface number density gradually decreases as Yr increases, 
with bumps and dips along the stream. 
Since the typical error ranges from 0.05 to 0.1 arcmin$^{-2}$, 
most of bumps are statistically significant. 
The most significant gap is found at 210$<$Yr$<$240, 
which is also clearly seen in Figure~\ref{fig:space}. 
It is again recovered at Yr$>$240 though the surface number 
density is the lowest among our survey field.  
However, the significance of this rise is uncertain  
since the lower completeness than 50\% for Yr$>$240 is suggested 
from Figure~\ref{fig:space} (b).  
Comparing the top and bottom panels of Figure~\ref{fig:NWStreamProf}
(a)
, 
the peak positions are not coincident with the globular clusters found 
on the NW Stream \citep{Huxor2014}. 

We also made a similar figure for NBGs as shown in
Figure 13 (b). Since the error in surface number density
is calculated to be between 0.02 and 0.05 arcmin$^{-2}$
(Poisson error is assumed),  the significance of the
features seen in the bottom panel of Figure 13 (b) is low
compared to that of Figure 13 (a).
But it  is suggested that the overall shape along
the stream, such as the gap at 210$<$Yr$<$240,
the bump at 150$<$Yr$<$200, and the gradual increase
in the number density distribution toward smaller Yr,
is similar between RCs and NGBs except for Yr$<$30,
where the derived number density seems to be affected
by a clump of NBGs found at (Xr, Yr)=(30, 15).
Unfortunately, the fine structures such as the numbers of
sharp peaks seen in the bottom panel of Figure 13 (a)
are not confirmed by NBGs due to the shortage of
statistical significance.

Recently, many studies on the formation of gaps in the stellar streams  
due to the encounter of $\Lambda$CDM substructures 
have been carried out 
\citep[e.g.,][]{Carlberg2011,Carlberg2012,Erkal2016}. 
Our data provides an excellent observational example to test these simulations. 
It is also noted that the bumps found in 
the bottom panel of Figure~\ref{fig:NWStreamProf} (a)
seems to be periodic, which is reminiscent of the epicyclic bumps 
seen in simulations of star clusters
\citep[e.g.,][]{Kupper2012}. 

\citet{Carlberg2012} investigated the density profile 
of the NW Stream using PAndAS data \citep{Carlberg2011}. 
They showed the density profile with a moving average 
over 5 or 11 bins of 0$^{\circ}$.25. 
This indicates that the resolution of their analysis 
is $>$ 0$^{\circ}$.25, an order lower than ours. 
Therefore, our data are complementary to theirs and 
would be of help to understand the detail of the density profile 
at the small scale.

\section{Discussion}\label{sec:discussion}

\subsection{Properties of Globular Clusters on the NW Stream}\label{sec:gc}

\citet{Huxor2014} reported their discovery of 59 globular clusters (GCs)
and 2 candidates in the halo of M31 based on the PAndAS survey data. 
Follow-up spectroscopic observations carried out by 
\citet{Veljanoski2014} obtained the radial velocities for 78 GCs. 
Now the large sample of GCs are available and the detailed studies 
of GCs can be possible even for M31. 

Among these GCs, 7 GCs show a hint of association with the NW Stream 
and 5 share a clear trend in corrected radial velocity 
as a function of projected radius \citep{Veljanoski2014}, 
suggesting the same origin as the NW Stream. 
Our survey field contains 5 GCs and they are found to be 
located near the center of the stream, suggesting their 
association with the stream (Figure~\ref{fig:space}). 
Out of 5 GCs, 3 GCs located at the south of our survey field (PAndAS-11 to 13) 
clearly trace the high surface number density regions 
as shown in Figure~\ref{fig:space}. 
On the other hand, the rest 2 (PAndAS-9 and 10) are located 
at relatively low surface number density region 
(Yr=140 in Figure~\ref{fig:NWStreamProf}). 

\begin{figure*}[t!]
\centering
\begin{tabular}{cc}
\includegraphics[width=80mm]{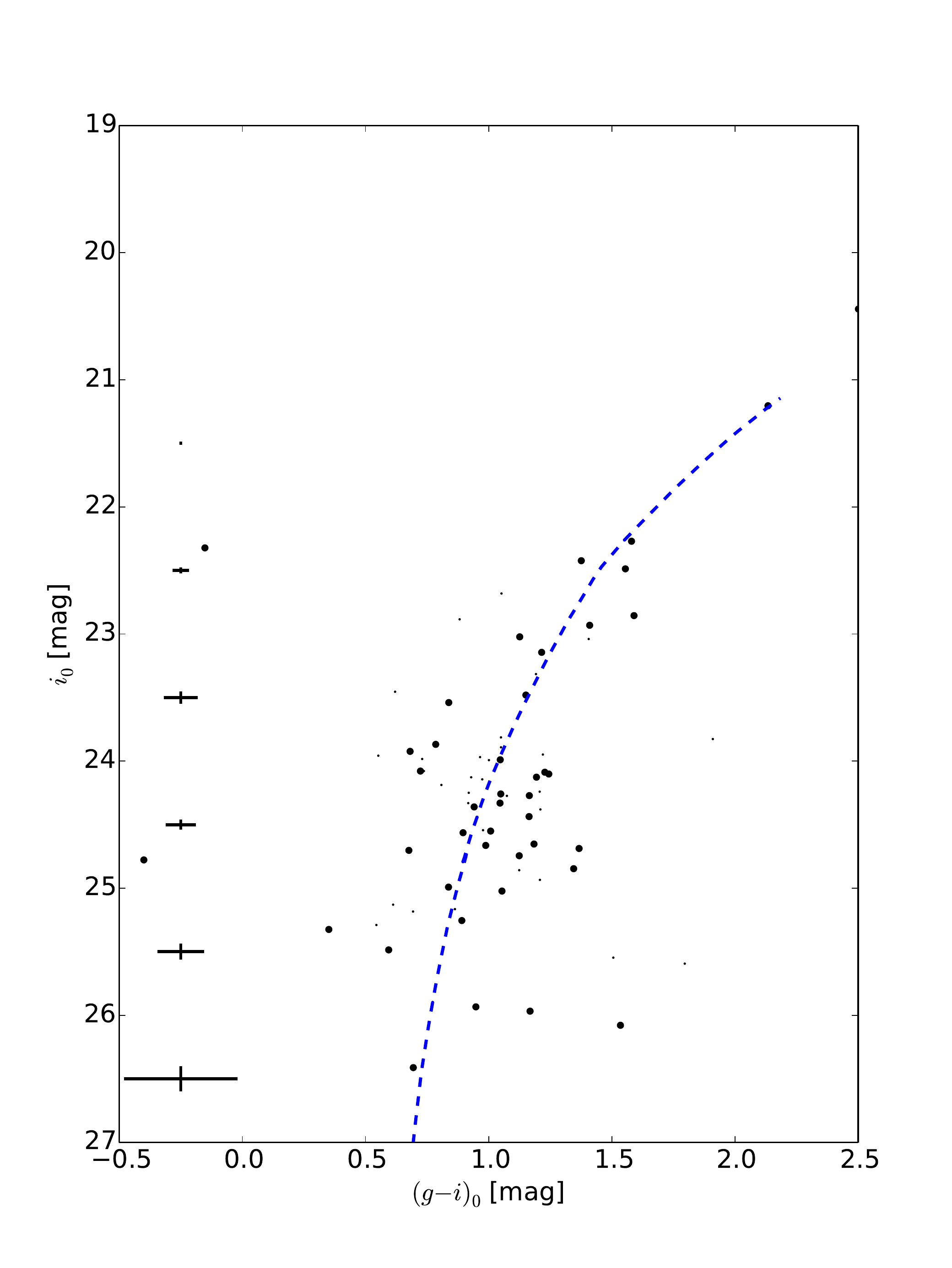} 
& 
\includegraphics[width=80mm]{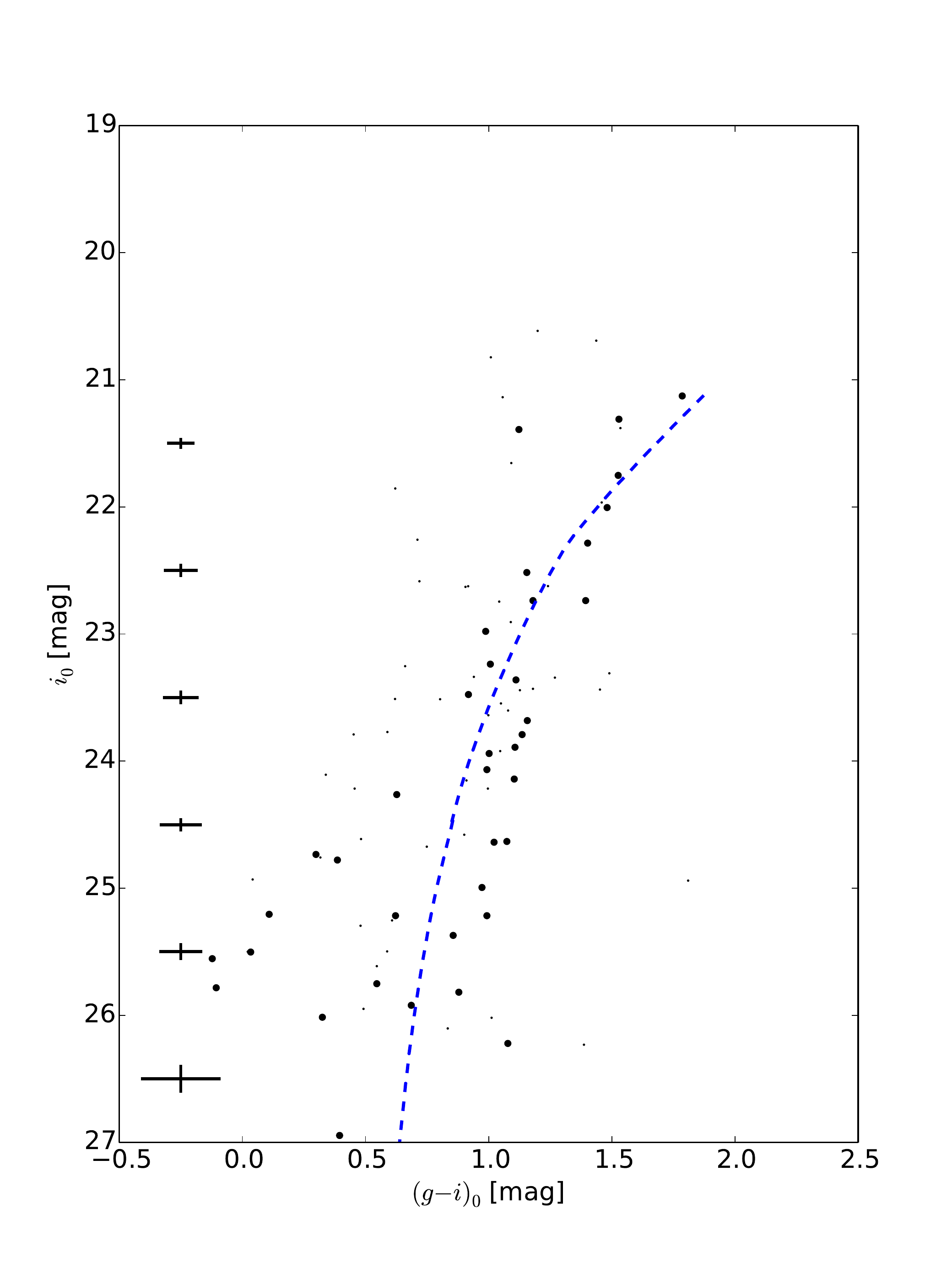} \\
(a) PAndAS-10 & (b) PAndAS-11 \\
\includegraphics[width=80mm]{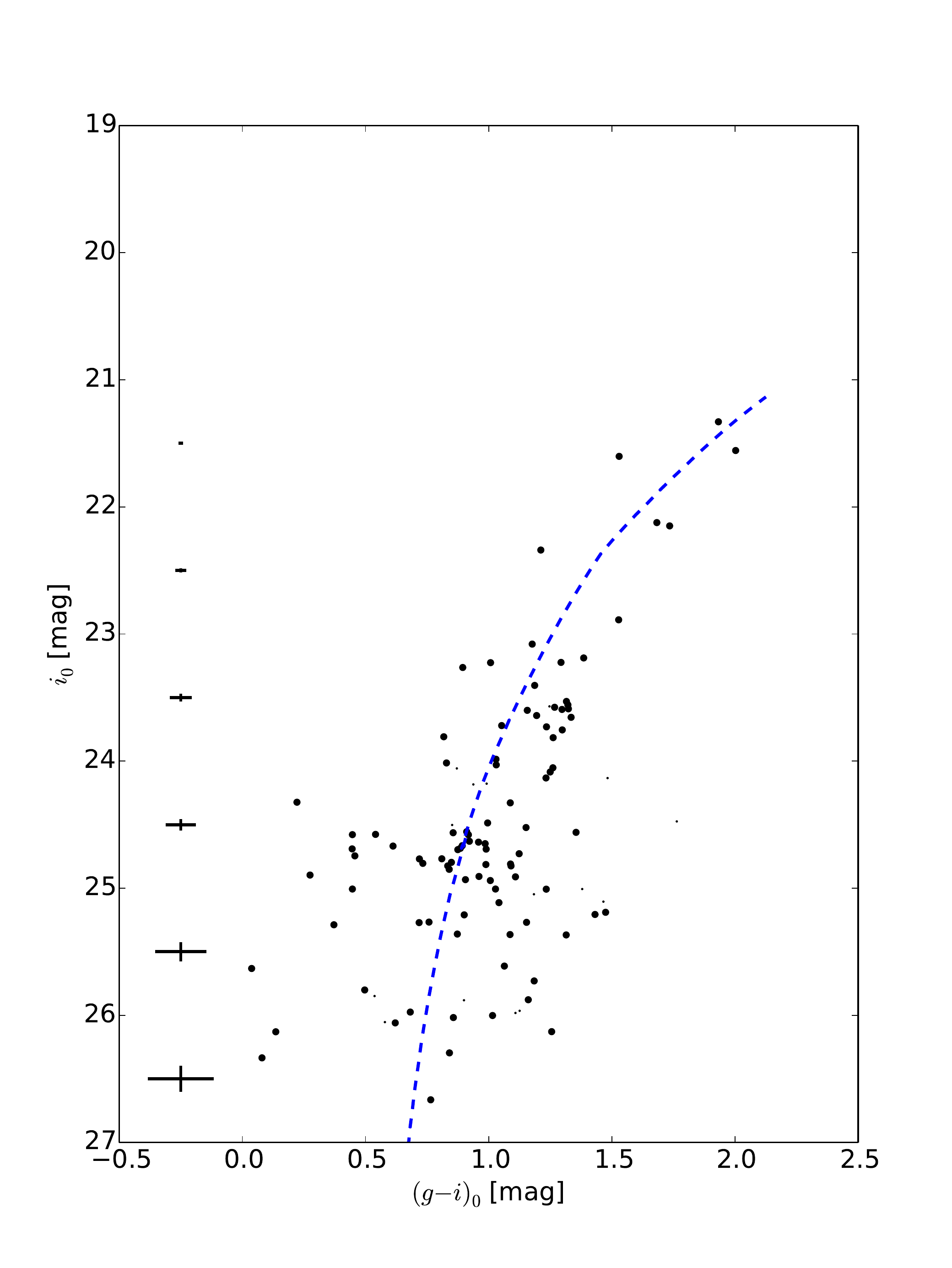}
&
\includegraphics[width=80mm]{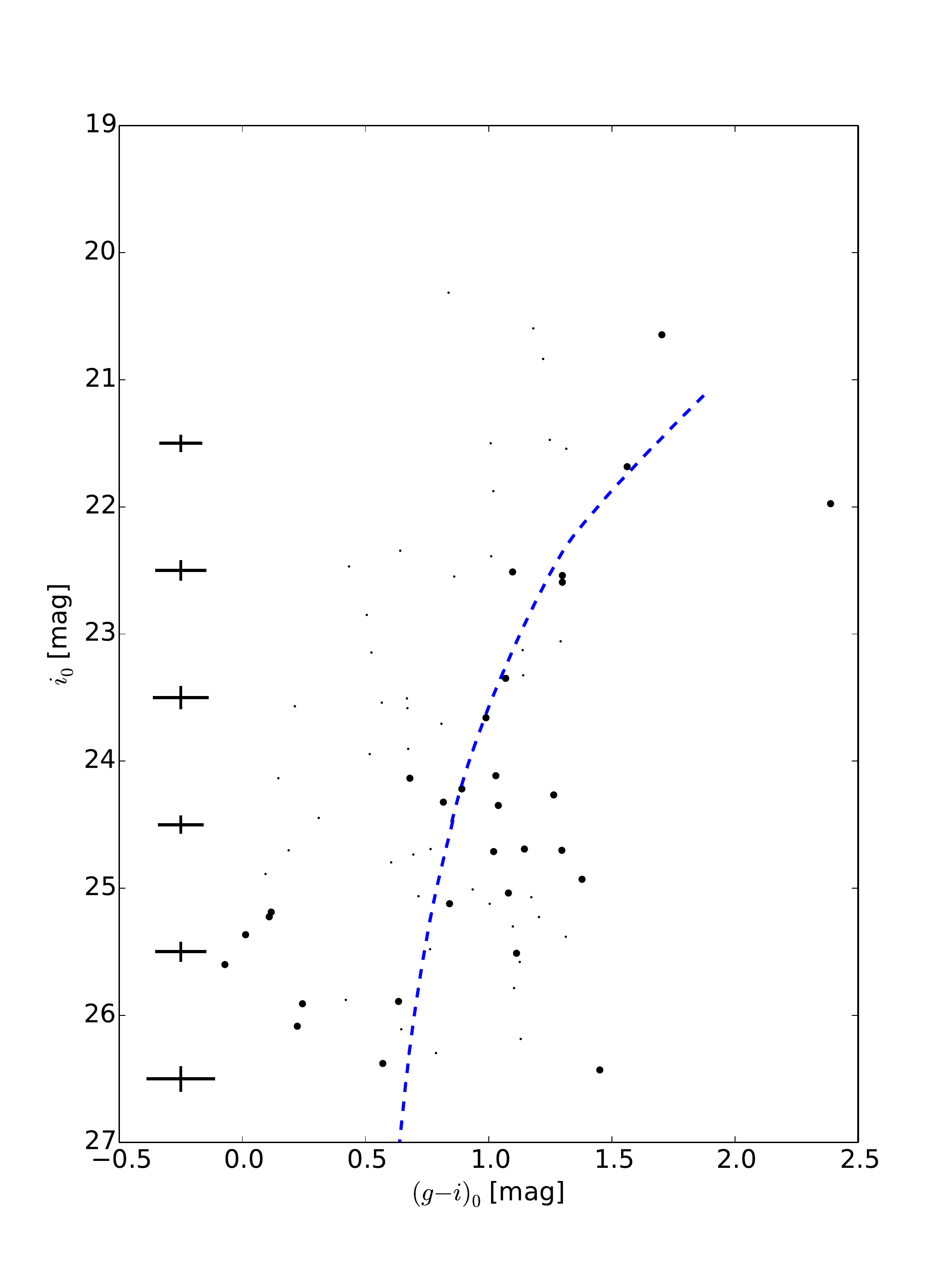} \\
(c) PAndAS-12 & (d) PAndAS-13 \\
\end{tabular}
\caption{
CMDs of GCs found on the NW Stream;   
(a) PAndAS-10, (b) PAndAS-11, (c) PAndAS-12 
and (d) PAndAS-13. 
Large and small dots represent those stars with 
$Sharpness < 1$ and $> 1$, respectively. 
The mean errors are calculated for every 1 mag and plotted 
as black crosses. 
Isochrone of RGB part with ($\log \tau_{\rm age}$, $Z$) = (10.12, 0.002), 
(10.00, 0.001), (10.00, 0.002) and (10.12, 0.001) is overlaid for 
PAndAS-10 to 13, respectively.  
The distance modulus of (m-M)=24.63 is adopted.  
}
\label{fig:GCCMD}
\end{figure*}

Making use of our high image quality data, 
we carry out a detailed study for these GCs 
in combination with the NW Stream 
to investigate the relation between these GCs and the NW Stream. 
Unfortunately, 1 GC (PAndAS-9) is 
buried in the halo of a bright star ($V$=10), 
making it difficult to detect stars belonging to the GC.  
We carried out the detailed analysis for the rest 4 GCs, PAndAS-10 to 13. 
DAOphot PSF photometry software \citep{Stetson1987,Stetson1994} 
is applied for these GCs and CMDs based on DAOphot photometry 
are plotted in Figure~\ref{fig:GCCMD}. 
Those stars with good fitting quality ($Sharpness < 1$) 
are plotted as large dots and others are as small dots. 
The mean photometric errors are calculated for 
every 1 mag range and plotted as crosses at $(g-i)_{0}=-0.25$. 
All the CMDs show clear sequences of RGB stars 
although the scatters are slightly larger than 
that observed for Stream North [Figure~\ref{fig:4cmd} (a)]. 
In each CMD, an isochrone of RGB which best traces the distribution 
of RGB stars [($\log \tau_{\rm age}$, $Z$) = (10.12, 0.002), 
(10.00, 0.001), (10.00, 0.002) and (10.12, 0.001) for 
PAndAS-10 to 13, respectively, with (m-M)=24.63] 
is plotted for guide. 

Comparing these CMDs with that for the Stream North, 
no significant difference in the position of RGB in the CMD
is seen between the Stream North and all the GCs, 
indicating that the stellar populations consisting of GCs   
are similar to that of the stream. 
It is also noted that the brightest RGB star in each GC is 
equal to or fainter than the tip of the isochrone, 
which is consistent with the condition that GCs are located near or farther 
than the distance of (m-M)=24.63, except for PAndAS-13 
in which a star with $i_{0}=20.4$ is found. 
Note that \citet{Veljanoski2014} pointed out that 
PAndAS-13 is displaced from the linear relation in 
radial velocity with the projected radius observed for 
other 5 GCs (PAndAS-4, 9 to 12). 
The results support the idea that both GCs (PAndAS 10-12) and 
the stream are originated from same progenitor system 
and share a same orbit in the halo of M31. 

It is also pointed out that small difference in stellar population 
of GCs is suggested from isochrone fitting.   
This difference is supported by the integrated photometry for these GCs
by \citet{Huxor2014} which shows that 
PAndAS-10 and 12 are redder [$(g-i)_{0}=0.75$ and 0.75, respectively] 
than PAndAS-11 and 13 [$(g-i)_{0}=0.67$ and 0.65, respectively]. 
Therefore, the small difference in stellar population of GCs 
seems to be real. 
Follow-up spectroscopic observations to measure the metallicity 
would give an answer for the stellar population of these GCs.

\subsection{Comparison with Simulation}

\begin{figure}[t!]
\centering
\includegraphics[width=80mm]{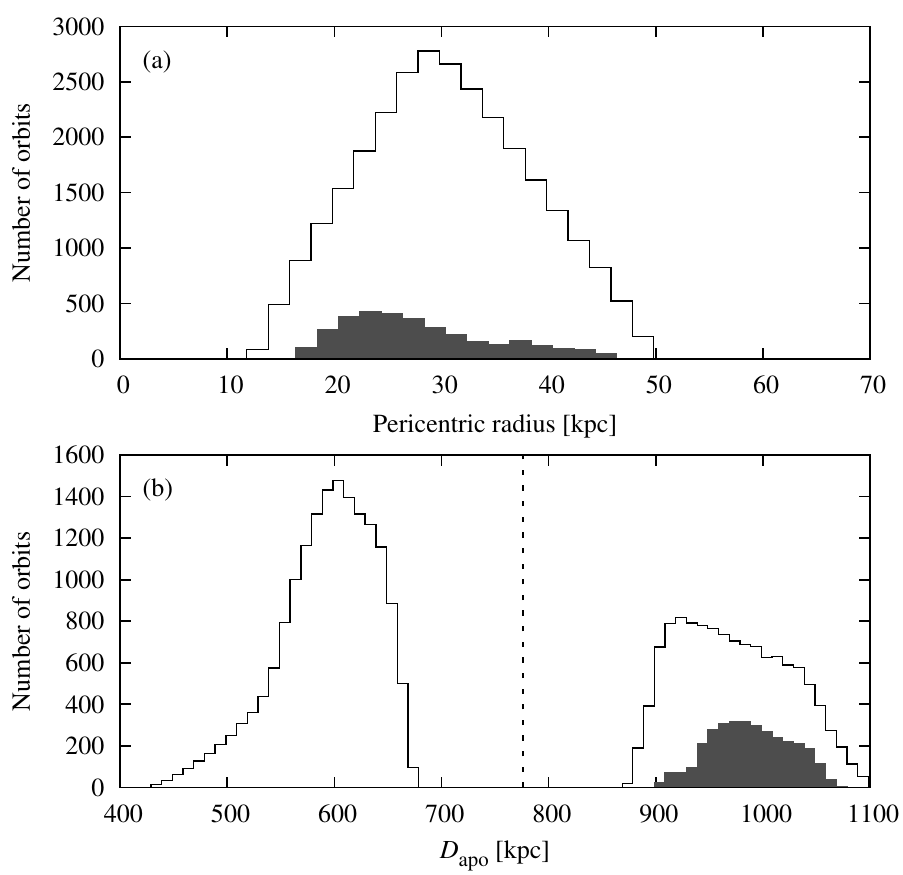}
\caption{
Histogram of (a) the pericentric radii and (b) the apocenter 
$D_{\mathrm{apo}}$ of the potential orbits.
The thin lines indicate the possible orbits basically derived by 
\citet{Kirihara2017b}.
The shaded area corresponds to the new orbits derived by 
this work including the observed distance of the NW Stream.
The vertical dashed line in panel (c) marks the distance to M31 from Earth.
}
\label{fig:apo_distance}
\end{figure}

\citet{Kirihara2017b} recently carried out an extensive numerical simulation 
to constrain the orbits and physical properties of the progenitor of 
the NW Stream, provided that 5 GCs on the NW Stream 
(PAndAS-04, 09, 10, 11 and 12) share the same orbit as the NW Stream, 
thereby making use of the radial velocities of these GCs.
They showed that the allowable orbits of the NW Stream are divided into 
two branches: one that the stream is located in front of M31 with 
line-of-sight distances of apocenter $D_{\rm apo}$ of 430$\sim$680 kpc, 
and the other that the stream is located behind M31 with $D_{\rm apo}$ of 
880$\sim$1100 kpc (see also Figure~\ref{fig:apo_distance} b).

The key to constrain the orbit is the line-of-sight distance to the NW Stream, 
which we derived as 824-847 kpc from RC method 
(see Table~\ref{tab:strdistance}). 
Combined with our distance measurement, 
we carried out test-particle simulations.
A test-particle is launched from the position of the GC PAndAS-12 
in the gravitational potential of M31.
The potential consists of a Hernquist bulge \citep{Hernquist1990}, 
an exponential disk and an NFW dark matter halo \citep{Navarro1996}
following \citet{Kirihara2017a}.
The M31 bulge is set to have the scale radius of $0.61$~kpc and 
the total mass of $3.24 \times 10^{10}M_{\odot}$.
The scale height, radial scale length, total mass and central surface density 
of the M31 disk are $0.6$~kpc, $5.4$~kpc, $3.66 \times 10^{10}M_{\odot}$ 
and $2.0\times 10^8 M_{\odot}\;{\rm kpc}^{-2}$, respectively.
The inclination and position angle of M31's disk are $77^{\circ}$ and 
$37^{\circ}$, respectively \citep{Geehan2006}.
The scale radius and scale density of the NFW halo are $7.63$~kpc and 
$6.17 \times 10^7 M_{\odot} \rm{kpc}^{-3}$, respectively.
The radial velocity of the test-particle is set to 472~km~s$^{-1}$, 
which is the observed value of PAndAS-12 \citep{Veljanoski2014}.
Its initial distance and proper motion velocities are systematically 
varied following \citet{Kirihara2017b}.
To find the acceptable ranges of orbital parameters, we conduct 
a $\chi^2_{\nu}$ analysis for the observed position of the NW Stream 
\citep[see table~1 of][]{Kirihara2017b} and radial velocities of the 5 GCs.
Successful orbits satisfy the 1$\sigma$ confidence level for the both 
criteria and have experienced two or more apocentric passages within 12~Gyr.
We also use the NW Stream distance obtained from RC for Stream 1-4.
Additional constraints in the test-particle simulations are 
(1) the test-particle passes within 0.5$^{\circ}$ from the center of each field 
and (2) the distance matches the observation within the observed uncertainty.
Although the distance data derived by RC for Stream 1-4 have rather large 
errors, they give a strong constraint on potential orbits for 
the progenitor of the NW Stream as shown in Figure~16, 
and the number of allowable orbits is 3,290 in 5,068,617 orbit models.

The pericentric distance of the progenitor's orbit is one of 
the great keys to limit the physical properties of the progenitor.
Figure~16a shows the number distribution of the successful orbits 
as a function of pericentric radius.
The histogram has a peak between a pericentric distance of 20~kpc and 30~kpc.
It is clear that no orbit can approach M31's center within 17~kpc.
Even if we adopt the constraint for 3$\sigma$ confidence level, 
any orbit cannot reach 17~kpc from the M31's center.
The result provides an update criteria for the physical properties of 
the progenitor model.
Following \citet{Kirihara2017b}, we estimate the stripped mass of 
a progenitor using Hill radius at the pericenter that defines 
the tidal radius of the satellite in a gravitational potential of 
the host system.
The half-light radius of the progenitor should be greater than $200$~pc.

We demonstrate the formation process of the NW Stream using 
an $N$-body simulation.
We construct the progenitor dwarf galaxy as a Plummer model 
using the 
{\scriptsize {\MakeUppercase{MAGI}}} (Miki \& Umemura in prep.).
The total mass $M_\mathrm{tot}$ and scale radius $r_{\rm s}$ of 
the Plummer distribution are set to $M_\mathrm{tot}=5\times10^7M_{\odot}$ 
and $r_{\rm s}=1$~kpc, respectively.
The total number of particles is 65,536, and we use the gravitational 
octree code 
{\scriptsize {\MakeUppercase{GOTHIC}}} \citep{Miki2017} to run the simulation.
We adopt the Plummer softening parameter of 16~pc and the accuracy control 
parameter of $2^{-7}$.
The initial phase-space coordinates of the progenitor for 
the successful orbit are 
$(\xi,\eta,d,V_{\xi},V_{\eta},V_{\rm los})=(-22^{\circ}.18,-0^{\circ}.60, 
860.07{\rm kpc}, -9.88, -23.37 , -302.15)$.
The unit of velocity is km~s$^{-1}$.
In this model, the progenitor travels from north to south 
along the NW Stream with a perigalactic distance of 25.65~kpc.
This demonstration is an updated version of their Case~A simulation 
\citep[see Figure~4 of][]{Kirihara2017b}.

\begin{figure*}[t!]
\centering
\includegraphics[width=160mm]{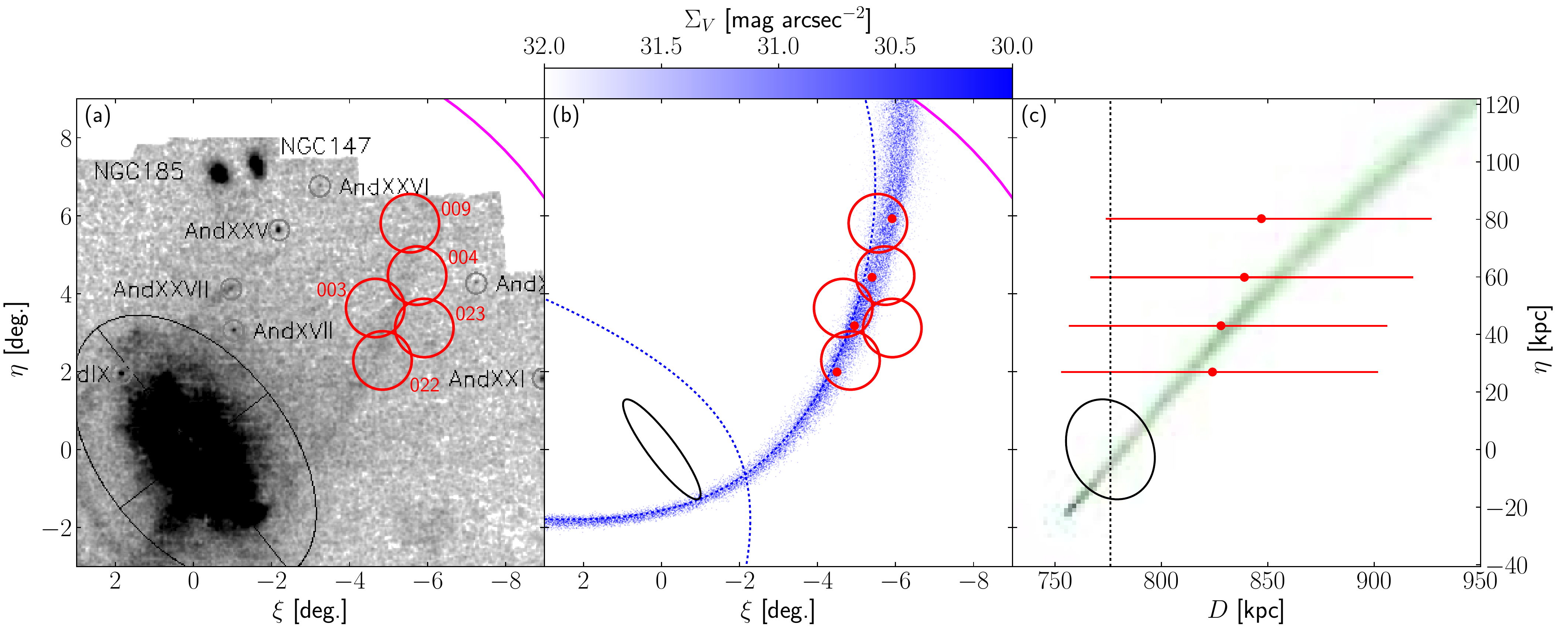} 
\caption{
The 3D schematic picture of the NW Stream.
The optical image is shown in panel (a), 
which is the same as Fig.~\ref{fig:pointing}. 
Particle distribution in panels (b) and (c) is a result of 
the $N$-body simulation.
M31 is represented as an ellipse whose major diameter is 190 arcmin 
assuming inclination and position angle of 77$^{\circ}$ and 37$^{\circ}$ 
\citep{Geehan2006}, respectively.
The HSC pointings are shown by red circles in panel (b).
The magenta curve presents 150~kpc from the M31 center.
The vertical black dashed line in panel (c) represents the distance to M31.
The positions and distances of the Stream 1 to 4 are shown 
in red squares together with the error bars.
}
\label{fig:NWStream3D}
\end{figure*}

Figure~\ref{fig:NWStream3D} shows an example of 
the $N$-body simulation of the NW Stream 
with our distance measurement, demonstrating that the simulation 
well represents the observation.
To draw the distance distribution of the simulated NW Stream, 
we use the distribution of $\xi<0$ on the sky.
Together with the results for GCs, our observation strongly supports 
their simulation with the NW Stream as a background of M31.
On the other hand, the latest PAndAS view \citep{Richardson2011} suggests 
that the extension of the NW Stream is not clearly seen in their Figure~1 
and may diminish or disappear when it passes the pericenter.
This view is consistent with the case B simulation of \citet{Kirihara2017b}.
Therefore, the search for the southern extension of the NW Stream is important.
Also, additional observation in the further northern field is important 
because the simulated stream extends outside of the PAndAS field.
Future fine-tuning of the simulation is planned to derive more accurate 
properties of the NW Stream (Kirihara et al. in preparation).
The key parameters from our survey are,
\begin{itemize}
\item The luminosity distribution along the stream: 
According to \citet{Kirihara2017b}, the $N$-body simulation can reproduce 
the observed features (i.e., the position of the NW Stream and 
the radial velocities for 5 GCs) for several Myr to 1~Gyr and 
it is difficult to determine the `current time' during the orbital motion 
of the NW Stream in the halo of M31.
The promising information to solve this degeneracy is the luminosity 
distribution along the stream.
The reason that the simulated NW Stream has a higher surface brightness 
near the GC PAndAS-12 than other portions. This trend is attributed to 
the selection of the `current time'.
Moreover, the position of the surviving central core of the progenitor 
gives strong constraint because it depends on the properties and 
orbit of the progenitor \citep{Kirihara2017b}.
As shown in Figures~\ref{fig:space} and \ref{fig:NWStreamProf}, 
our data provide a sufficient information 
and would be helpful to reveal the nature of the NW Stream.
\item The width of the stream: As shown in Figure~\ref{fig:apo_distance}, 
the number of potential orbits of the NW Stream is still over 
several thousands although it is significantly reduced by making use of 
the line-of-site distance to the stream.
Therefore, other information to further constrain the orbits is required.
By incorporating the width of the stream, the test-particle simulations 
will further constrain the allowable orbits.
Figure~\ref{fig:NWStreamCS} will be used for this purpose. 
\end{itemize}

In relation to above, it is emphasized that the further observation 
along the NW Stream to both North and South directions be important.
If the faint end of the stream is observed at the North as suggested 
from the PAndAS map, it will be a very strong constraint.
The observation in short exposure mode with NB515 filter is effective 
to answer the question.

\subsection{Diffuse Substructure}

The surface brightness of RC stars found at the diffuse substructure 
ranges from 34.8 to 36 mag arcsec$^{-2}$ in $g$ band 
(the median and brightest part is 
35.5 and 34.8 mag arcsec$^{-2}$, respectively). 
The total surface brightness of the diffuse substructure 
can be estimated from that of RC stars 
assuming the isochrone and initial mass function (IMF). 
The offset between the total surface brightness and 
the surface brightness of RC stars 
is calculated to 1.69 mag in $g$ band 
for $Z=0.0014$ and $\log \tau_{\rm age} ({\rm Gyr}) = 10.00$ 
population with the Salpeter IMF. 
Therefore, the diffuse substructure would be observed as 
33.1 -- 34.3 mag arcsec$^{-2}$. 

The surface brightness of 33 mag arcsec$^{-2}$ 
is below the detection limit of the previous surveys 
based on the resolved stellar photometry 
such as PAndAS and \citet{Tanaka2010}. 
Our result clearly shows that RC stars can be a powerful probe 
of the diffuse substructures down to $\sim$33 mag arcsec$^{-2}$. 
Since the number of substructures is predicted to increase 
as the surface brightness goes faint \citep[e.g.,][]{Bullock2005}, 
the actual appearance of the halo structure 
would be revealed in detail by tracing the RC stars. 
Only recently, the existence of an ultra-faint shell, 
having the common origin with the progenitor dwarf galaxy of 
the Giant Southern Stream, was predicted in the north-western 
area of M31 \citep{Kirihara2017a}.
The faint shell has similar properties with those of 
the diffuse substructure (surface brightness, distance and metallicity) 
while there is a spatial offset of around 2~deg.
It suggests that the diffuse substructures may be remnants of 
past galactic mergers.
We therefore emphasize that photometric surveys down to the RC magnitude 
are very important to understand the assembly process 
of many subsystems occurred in the outer part of the galaxy and 
to test the galaxy formation scenarios.

\section{Conclusions}\label{sec:conclusion}

We have carried out a wide and deep imaging survey of the northwest part of 
the halo in M31 using HSC on the Subaru Telescope.
The survey covers 9.2 deg$^{2}$ field consisting of 
5 HSC pointings in $g$ and $i$ bands 
as well as a narrow band filter $NB515$.
The reduced images in $g$, $i$ and $NB515$ bands are deep enough 
with the mean 50\% completeness limit of 26.31, 25.69 and 24.71 mag, 
respectively. 
The extinction corrected CMD of our survey field 
exhibits characteristic features; a band of dwarf stars of 
the MW disk, a narrow sequence of a diffuse stellar 
stream in the MW halo, the broad sequence consisting of 
the RGB stars in the M31 halo, 
a diffuse but distinct concentration of BHB stars, 
and a significant RC feature.  

The spatial distribution of RC stars shows a prominent stream feature, 
which is the known NW Stream. 
This substructure is also confirmed in the spatial distribution of RGB stars 
in the M31 halo, which are selected using the $NB515$ photometry. 
The method using the tip of RGB is applied to the $NB515$-selected RGB stars. 
We obtain the distance modulus to the NW Stream based on the TRGB method  
to be 24.42$\pm$0.033(random)$\pm$0.033(systematic) mag, 
but the isochrone fitting to the CMD of the Stream North 
suggests a larger distance modulus by 0.2-0.3 mag. 
The distance estimates by RC method show the distance moduli  
to be 24.63$\pm$0.191(random)$\pm$0.057(systematic) and 
24.29$\pm$0.211(random)$\pm$0.057(systematic) mag
for the Stream North and Off-Stream South, respectively, 
indicating that the NW Stream 
is located behind the main body of M31, 
whereas the diffuse substructure is located in front of M31. 
We also estimate the line-of-sight distances along the NW Stream 
and find that the south part of the NW Stream is $\sim$20 kpc 
closer to us relative to the north part. 
The number density distribution across the NW Stream represented by RC stars 
is found to be fitted by a Gaussian with FWHM of $\sim$25 arcmin, 
but slightly skewed to the south-west side. 
The number density distribution of RC stars along the NW Stream shows 
the complicated structure including the number of bumps and dips 
and a significant gap. 

The stellar populations of globular clusters found on the NW Stream, 
PAndAS 10-13, are investigated based on the CMDs and 
they are suggested to be similar to that consisting of the NW Stream. 
The 3D structure of the NW Stream revealed in this study is 
compared with the recent simulation carried out by \citet{Kirihara2017a},  
and gives a definite constraint on the allowable orbit of the NW Stream. 
The surface brightness of the diffuse substructure found 
in the south part of the survey field is estimated to be 
$\sim$33 mag arcsec$^{-2}$ in $g$ band, which is below the 
detection limit of the previous surveys of the M31 halo. 
We have found that 
the RC stars are a powerful probe to assess the diffuse substructures 
which are expected to be numerous and reveal the real appearance 
of the M31 halo.

\acknowledgments

This work is supported in part by JSPS Grant--in--Aid for Scientific 
Research (No. JP25287062, JP15K05037, JP25800098, JP25400222 and JP17H01101),  
MEXT Grant--in--Aid for Scientific Research on 
Innovative Areas (No. JP15H05892, JP16H01090 for K.H., 
for JP16H01086 for M.C.) and 
Grant--in--Aid for JSPS Fellows (T.K. 26.348). 
PG was supported by NSF grant AST-1412648. 
MGL and ISJ were supported by the National Research Foundation (NRF) 
grant funded by the Korea Government (NRF-2017R1A2B4004632).

We would like to thank all the staff of Subaru Telescope, in particular 
Drs. Fumiaki Nakata, Tsuyoshi Terai, Shintaro Koshida, 
Francois Finet, Akito Tajitsu, 
for their excellent support during the observation, 
and all the staff of HSC software developing group, 
in particular Dr. Hisanori Furusawa and Dr. Naoki Yasuda,  
for their advice during the processing of our M31 data. 
We would like to express our appreciation to Dr. Satoshi Kawanomoto 
for his effort in developing NB515 filter. 
We are grateful to Dr. Paul Price for his kind support to 
calibrate our data to the PS1 system. 
We appreciate many valuable comments and suggestions from the referee
which have improved this paper significantly. 

The numerical computations were carried out on the HA-PACS System 
in the Center for Computational Sciences, University of Tsukuba, Japan.

The Hyper Suprime-Cam (HSC) collaboration includes the astronomical
communities of Japan and Taiwan, and Princeton University.  The HSC
instrumentation and software were developed by the National
Astronomical Observatory of Japan (NAOJ), the Kavli Institute for the
Physics and Mathematics of the Universe (Kavli IPMU), the University
of Tokyo, the High Energy Accelerator Research Organization (KEK), the
Academia Sinica Institute for Astronomy and Astrophysics in Taiwan
(ASIAA), and Princeton University.  Funding was contributed by the FIRST 
program from Japanese Cabinet Office, the Ministry of Education, Culture, 
Sports, Science and Technology (MEXT), the Japan Society for the 
Promotion of Science (JSPS),  Japan Science and Technology Agency 
(JST),  the Toray Science  Foundation, NAOJ, Kavli IPMU, KEK, ASIAA,  
and Princeton University.
This paper makes use of software developed for 
the Large Synoptic Survey Telescope. We thank the
LSST Project for making their code freely available. 
The Pan-STARRS1 (PS1) Surveys have been made
possible through contributions of the Institute for Astronomy, 
the University of Hawaii, the Pan-STARRS
Project Office, the Max-Planck Society and its participating institutes, 
the Max Planck Institute for
Astronomy and the Max Planck Institute for Extraterrestrial Physics, 
The Johns Hopkins University,
Durham University, the University of Edinburgh, 
Queen's University Belfast, the Harvard-Smithsonian
Center for Astrophysics, the Las Cumbres Observatory Global 
Telescope Network Incorporated, the
National Central University of Taiwan, 
the Space Telescope Science Institute, the National Aeronautics
and Space Administration under Grant No. NNX08AR22G 
issued through the Planetary Science Division
of the NASA Science Mission Directorate, 
the National Science Foundation under Grant
No.AST-1238877, the University of Maryland, 
and Eotvos Lorand University (ELTE).

\appendix

\section{TRANSFORMATION BETWEEN THE HSC AND OTHER STANDARD PHOTOMETRIC SYSTEMS}\label{app:colconv}

Photometric transformations are required to compare our results 
with previous studies which were carried out in 
different filter systems.  
In this section, the color conversion formulae which are 
relevant to our study are summarized. 
The method to obtain the formulae is similar to those used in  
previous studies such as \citet{Yagi2010} and \citet{Fukugita1995}. 
We calculate the colors of stars    
using the Bruzual-Persson-Gunn-Stryker (BPGS) Atlas 
by convolving the transmission curve of a filter system
[Johnson-Cousins system for \citet{Bessell1990} and 
 SDSS system for \citet{Doi2010}] to the stellar spectra. 
The transmission curves for HSC can be obtained from the Subaru Telescope 
HSC website.\footnote{http://www.naoj.org/Observing/Instruments/HSC/index.html}

\begin{figure*}[t!]
\centering
\begin{tabular}{cc}
\includegraphics[width=80mm]{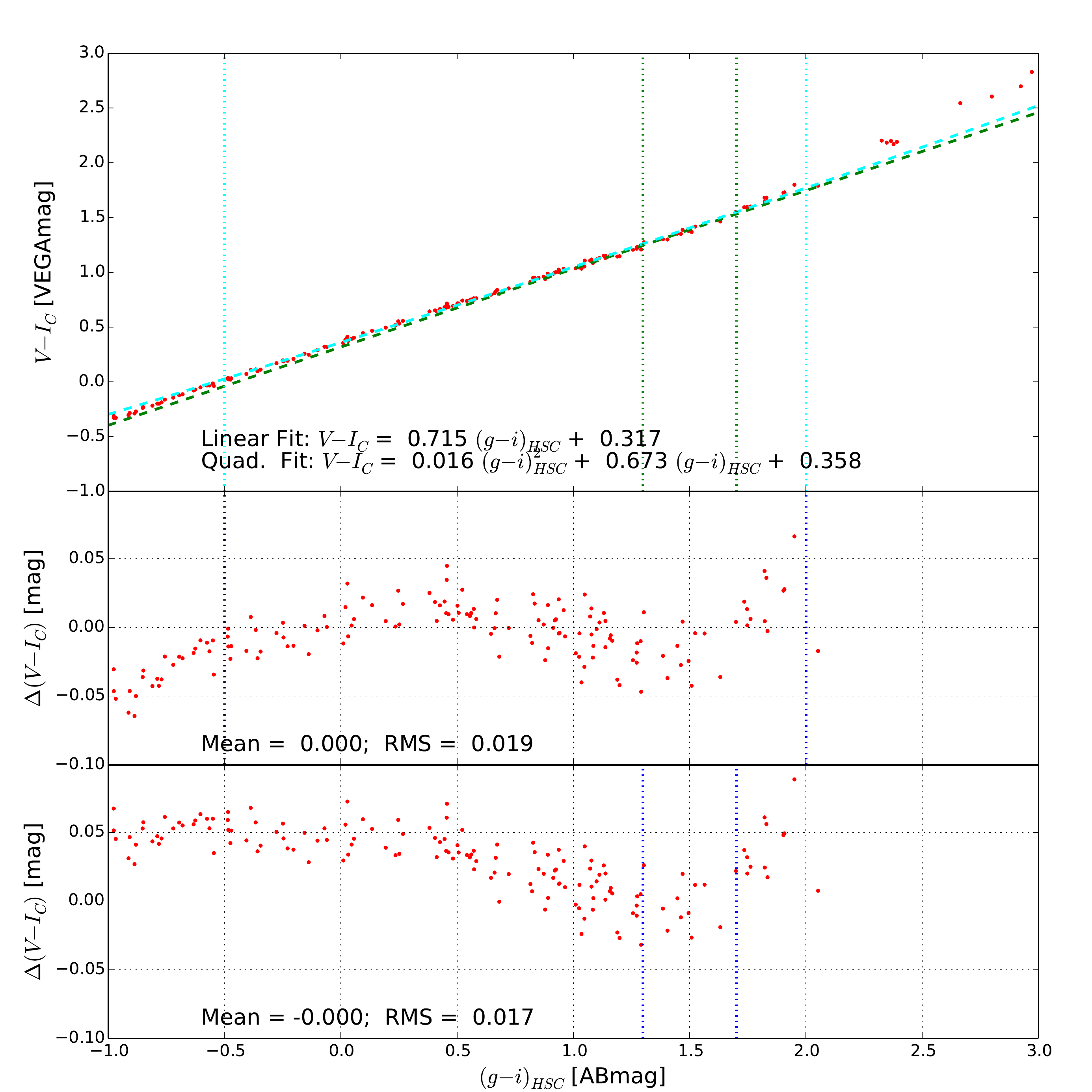}
& 
\includegraphics[width=80mm]{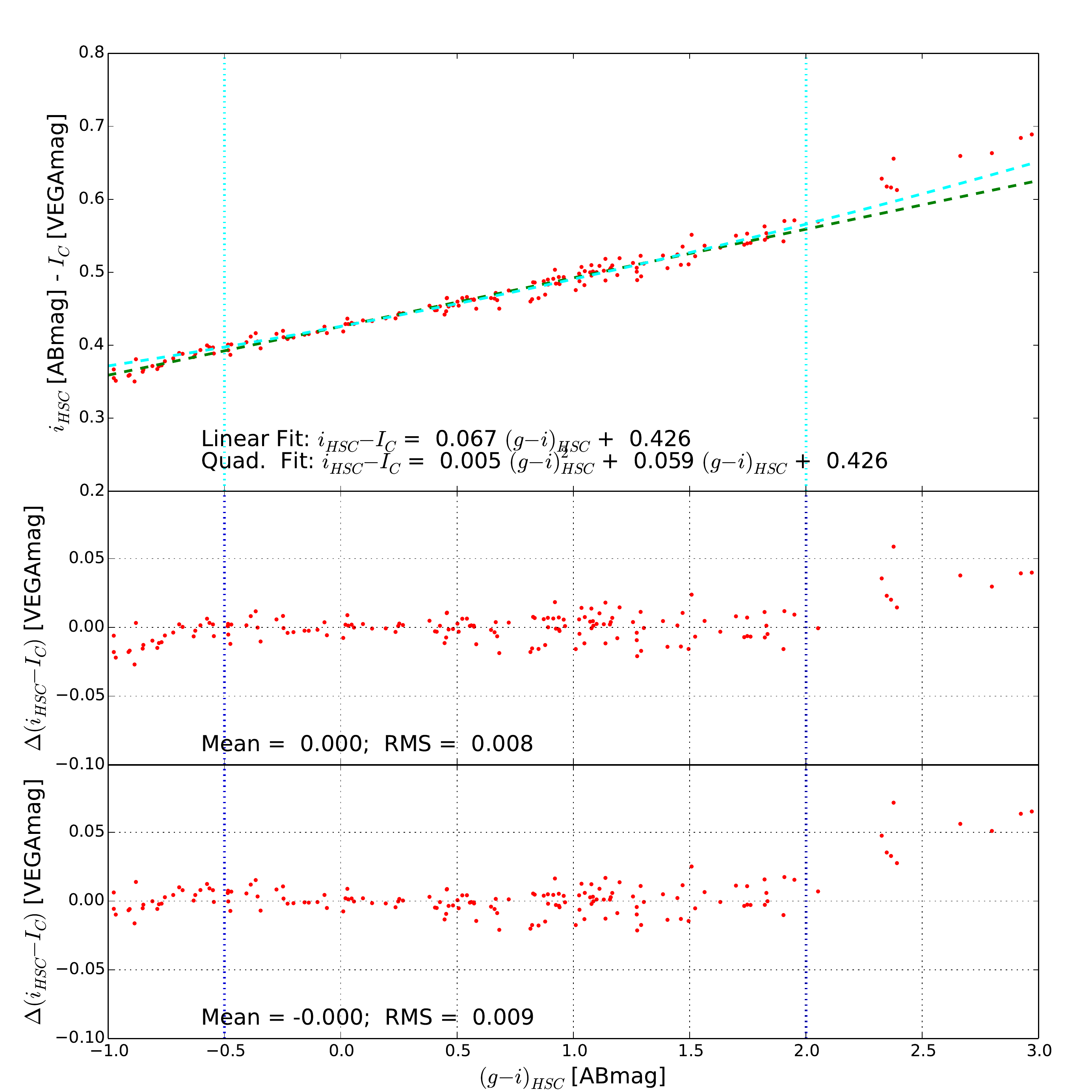} \\
(a) $(g-i)_{HSC}$ versus $V-I_{C}$ & (b) $(g-i)_{HSC}$ versus $i_{HSC}-I_{C}$ \\
\includegraphics[width=80mm]{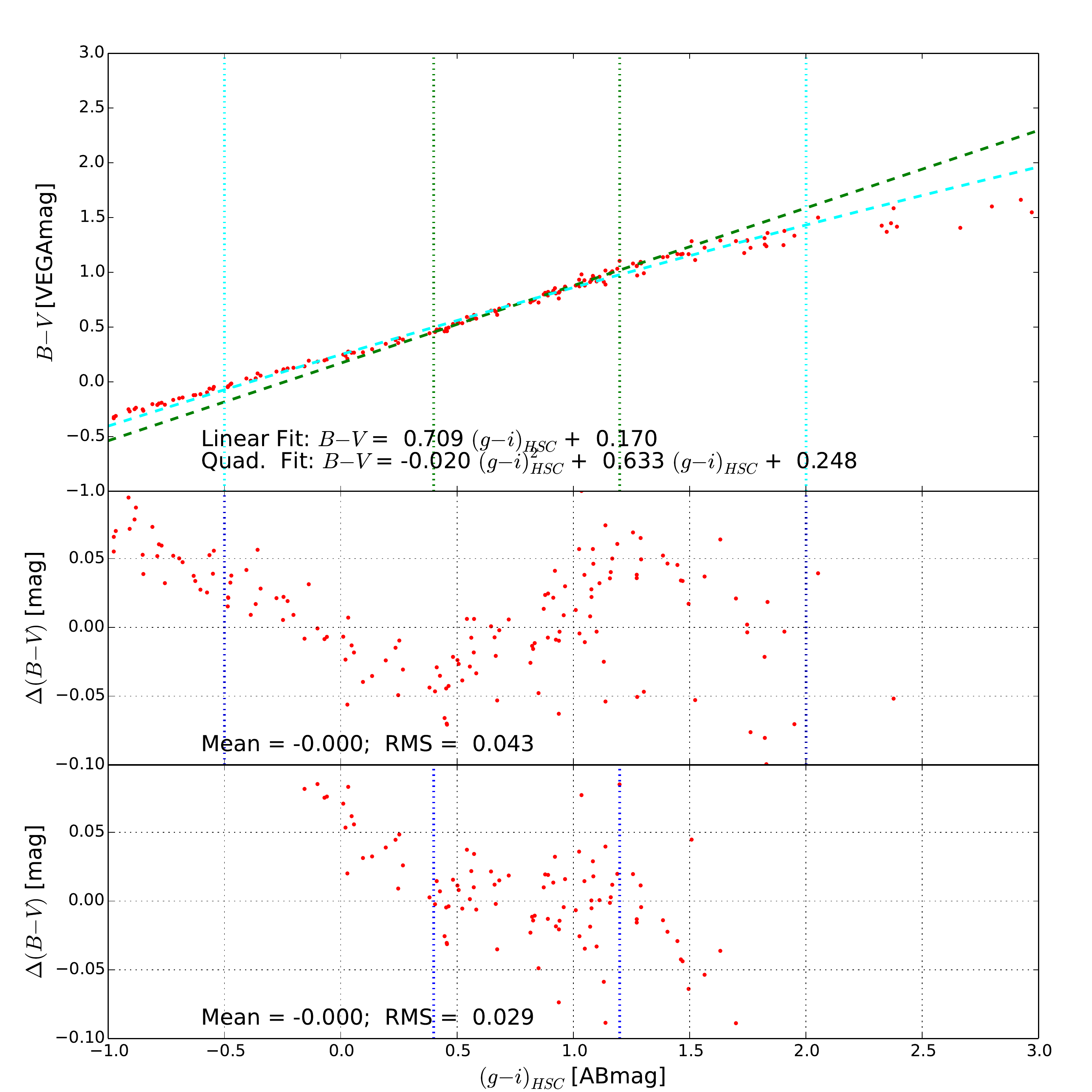} 
&
\includegraphics[width=80mm]{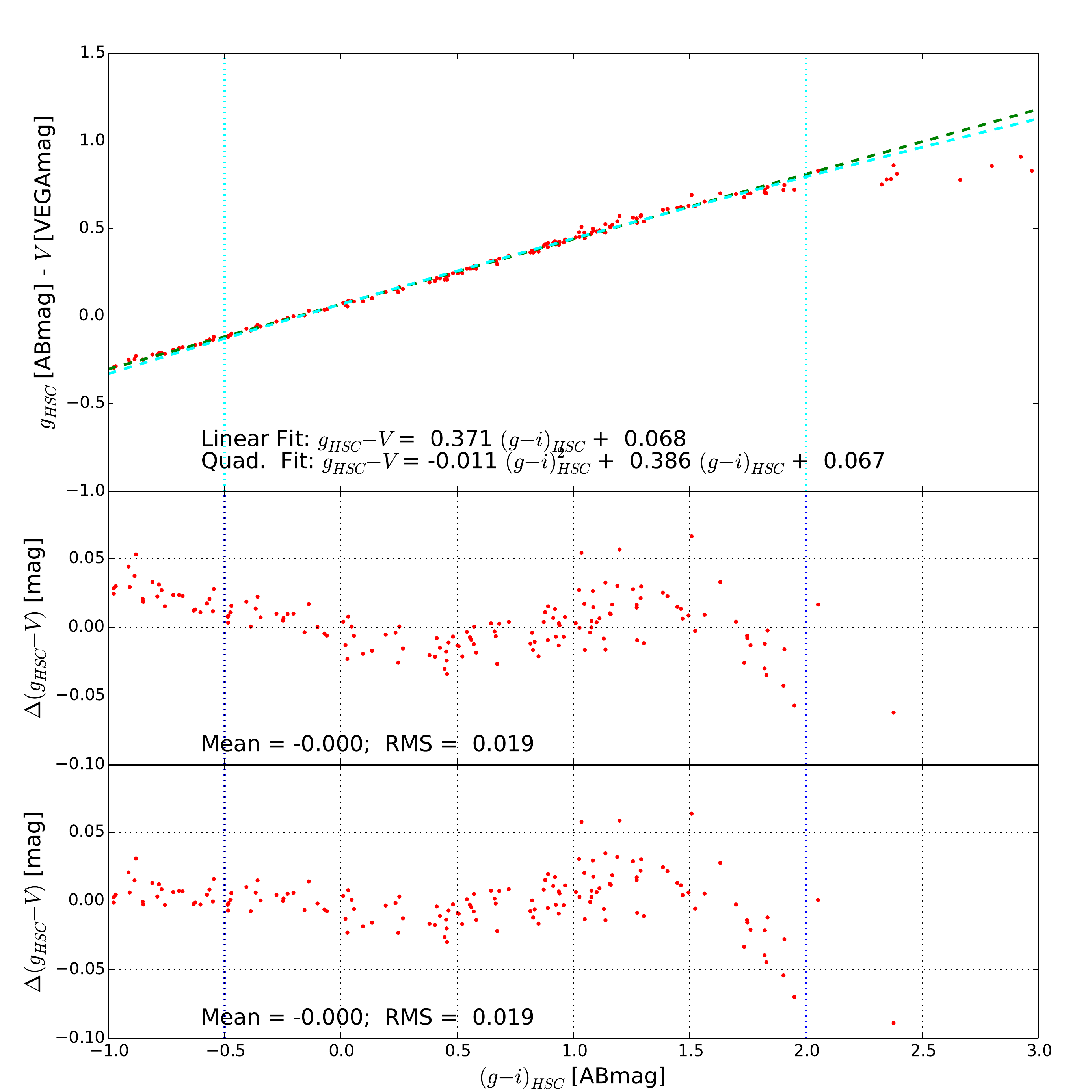} \\
(c) $(g-i)_{HSC}$ versus $B-V$ & (d) $(g-i)_{HSC}$ versus $g_{HSC}-V$ \\
\end{tabular}
\caption{
Color conversion diagrams between 
(a) $(g-i)_{HSC}$ and $V-I_{C}$, (b) $(g-i)_{HSC}$ and $i_{HSC}-I_{C}$, 
(c) $(g-i)_{HSC}$ and $B-V$, and (d) $(g-i)_{HSC}$ and $g_{HSC}-V$. 
For each of the parts (a)--(d): 
(Top) The calculated colors for BPGS stars are plotted against $(g-i)_{HSC}$.
The best-fit functions, linear expression and quadratic expression, 
are plotted as green and cyan dashed lines and the color range  
used for the fitting is plotted as vertical dotted lines with same color, 
respectively. 
(Middle) The residuals of calculated color from the quadratic fit. 
The blue vertical dotted lines indicate the color range 
used for the fitting. 
(Bottom) Same as for the middle panel, but for the linear fit. 
}
\label{fig:colconv}
\end{figure*}

Figure~\ref{fig:colconv} shows our calculation. 
The top panel of each of the figure sections (a)--(d) 
shows the calculated colors 
($V-I_{C}$, $i_{HSC}-I_{C}$, $B-V$, $g_{HSC}-V$) for BPGS stars
plotted against $(g-i)_{HSC}$. 
The best-fit functions, linear expression and quadratic expression, 
are plotted as green and cyan dashed lines, respectively. 
The fitting to the quadratic expression is performed 
for $-0.5<(g-i)_{HSC}<2.0$, but that to the linear expression is 
for a more limited color range ($1.3<(g-i)_{HSC}<1.7$) 
which is relevant to this study. 
In the middle and bottom panels, the residuals from the quadratic 
and linear fits, respectively, are plotted against $(g-i)_{HSC}$.

The fitting for $i_{HSC}-I_{C}$ is performed very well 
with root mean square (RMS) residuals of 
0.008--0.009 and those for $g_{HSC}-V$ and $V-I_{C}$ 
are fair with an RMS of $\sim0.019$. 
That for $B-V$ is the worst but the RMS is reasonable 
if the color range is limited to $0.4<(g-i)_{HSC}<1.2$ 
which is sufficient for this study.  
In this study, we use
\begin{eqnarray}
  V-I_{C}       &=& 0.715 (g-i)_{HSC} + 0.317 \text{\quad for $1.3<(g-i)_{HSC}<1.7$}, \\
  i_{HSC}-I_{C} &=& 0.067 (g-i)_{HSC} + 0.426, \\
  B-V           &=& 0.709 (g-i)_{HSC} + 0.170 \text{\quad for $0.4<(g-i)_{HSC}<1.2$}, \\
  g_{HSC}-V     &=& 0.371 (g-i)_{HSC} + 0.068, 
\end{eqnarray}
for the analysis. 

\begin{figure*}[t!]
\centering
\begin{tabular}{cc}
\includegraphics[width=80mm]{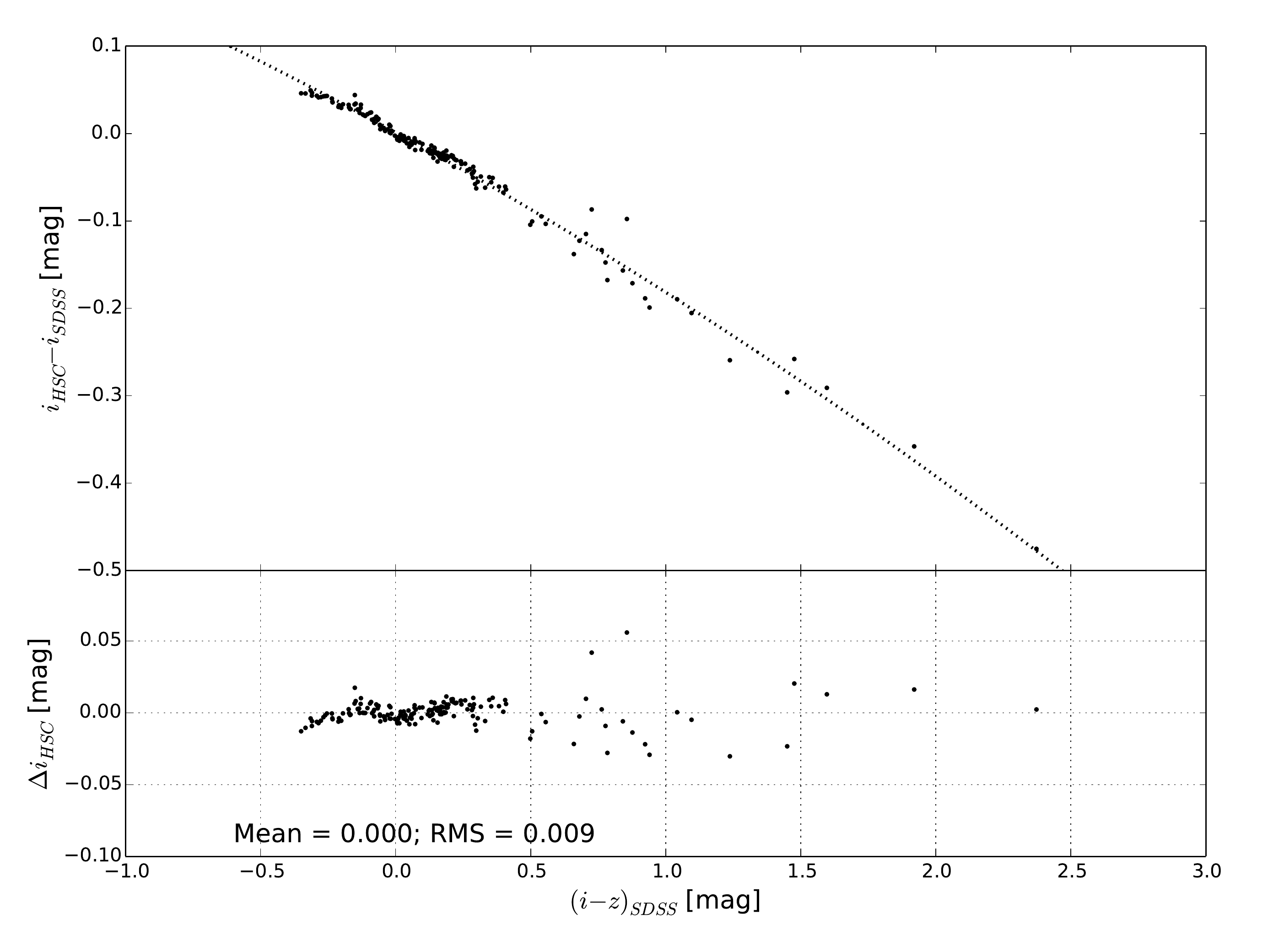}
& 
\includegraphics[width=80mm]{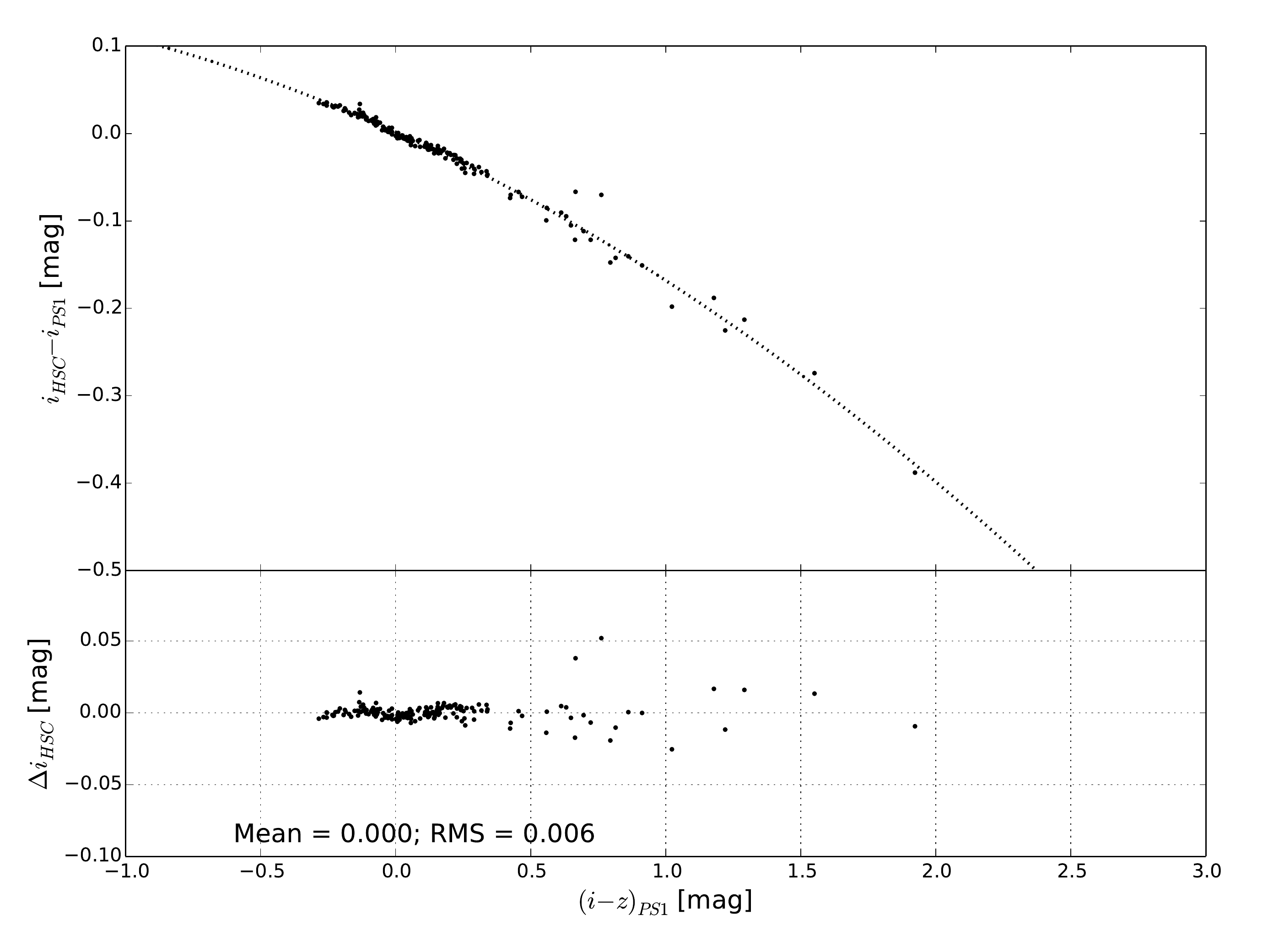} \\
(a) $(i-z)_{SDSS}$ versus $i_{HSC}$ & (b) $(i-z)_{PS1}$ versus $i_{HSC}$ \\
\includegraphics[width=80mm]{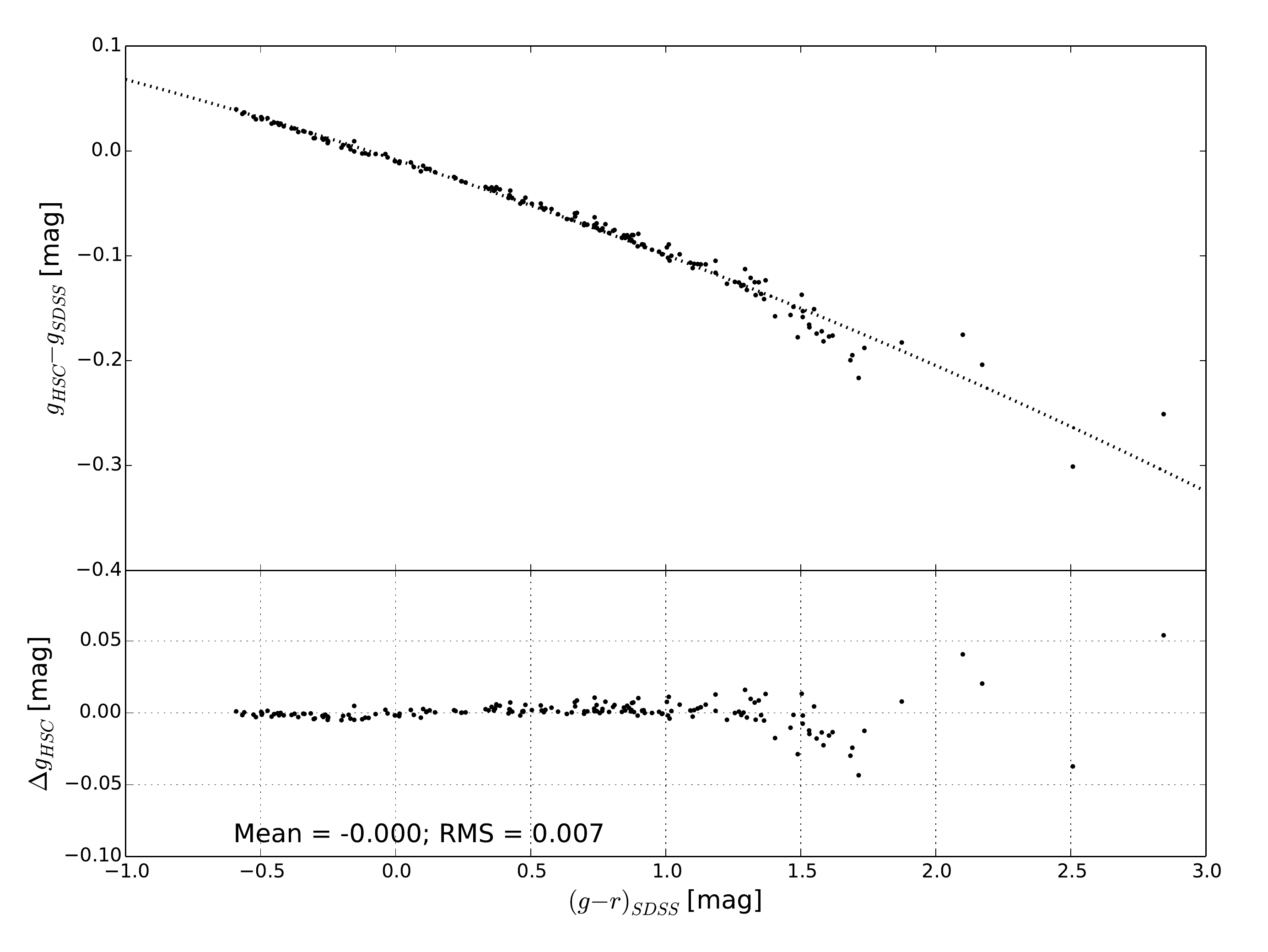} 
&
\includegraphics[width=80mm]{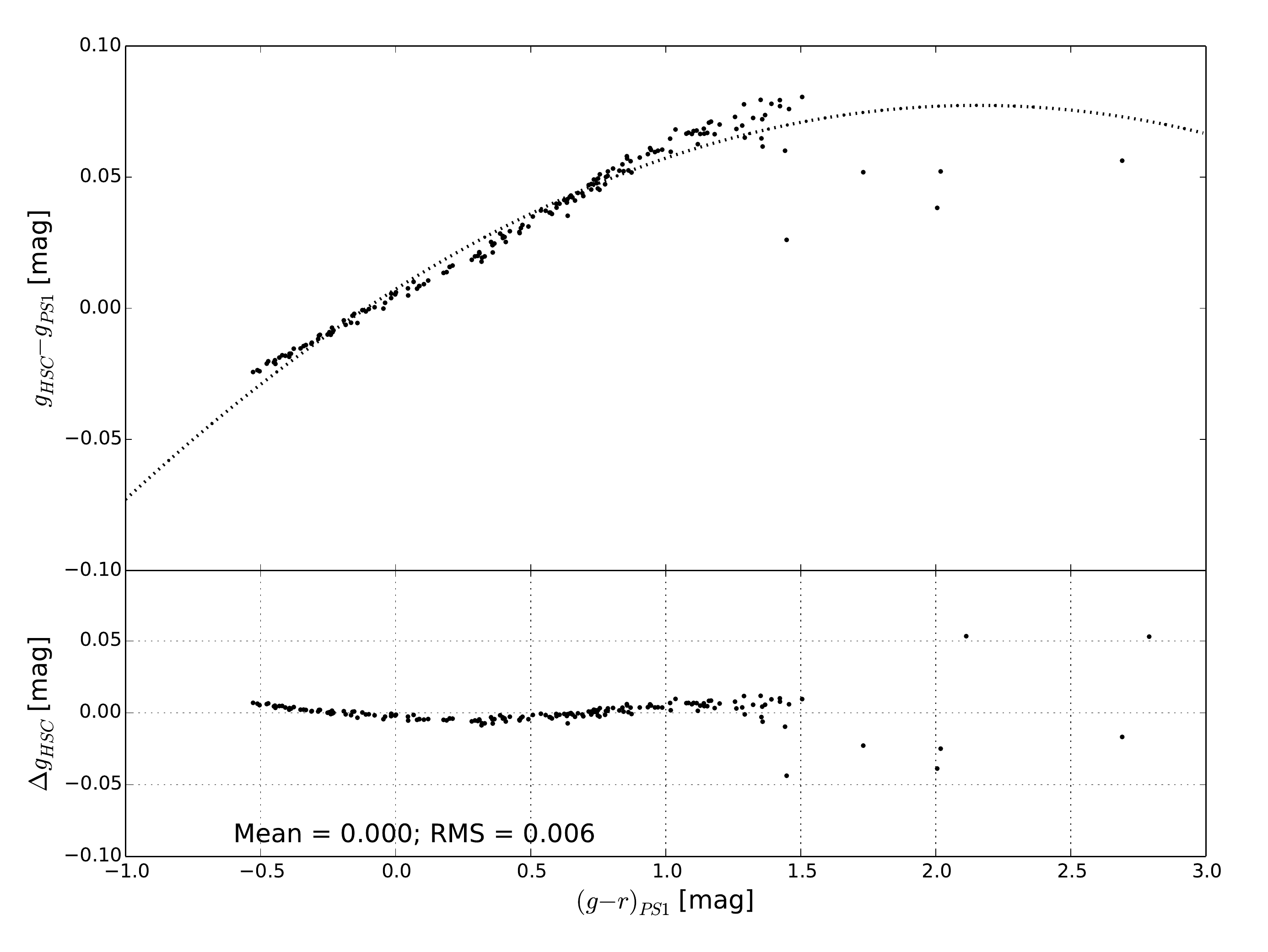} \\
(c) $(g-r)_{SDSS}$ versus $g_{HSC}$ & (d) $(g-r)_{PS1}$ versus $g_{HSC}$ \\
\end{tabular}
\caption{
Color conversion diagrams between 
(a) $(i-z)_{SDSS}$ and $i_{HSC}$, (b) $(i-z)_{PS1}$ and $i_{HSC}$, 
(c) $(g-r)_{SDSS}$ and $g_{HSC}$, and (d) $(g-r)_{PS1}$ and $g_{HSC}$. 
For each of the parts (a)--(d): 
(Top) The magnitude difference between 
HSC and either SDSS or PS1 systems of $g$ and $i$ band (i.e., color terms)
calculated for BPGS stars as a function of color.  
The dotted line shows the quadratic expression of the data, 
which is one of equations \ref{eq:i_SDSS}-\ref{eq:g_PS1}. 
(Bottom) The magnitude difference between HSC magnitude 
and that calculated with one of equations \ref{eq:i_SDSS}-\ref{eq:g_PS1}. 
The RMS value is calculated for stars with moderate color 
[i.e., $(g-r)_{SDSS,PS1} < 2.0$ or $(i-z)_{SDSS,PS1} < 1.0$] 
and listed in the figure.
}
\label{fig:colconv2}
\end{figure*}

We also use the color conversion formulae from the SDSS system 
to the HSC system, 
\begin{eqnarray}
  i_{HSC} &=& i_{SDSS} +0.00130204 -0.16922042 (i_{SDSS}-z_{SDSS}) -0.01374245 (i_{SDSS}-z_{SDSS})^{2}, \label{eq:i_SDSS} \\
  g_{HSC} &=& g_{SDSS} -0.00816446 -0.08366937 (g_{SDSS}-r_{SDSS}) -0.00726883 (g_{SDSS}-r_{SDSS})^{2}, \label{eq:g_SDSS} 
\end{eqnarray}
which are adopted in hscPipe, when converting the isochrone described 
in the SDSS system. 
During the image reduction, the HSC images are calibrated against 
PS1 using the following color conversion formulae, 
\begin{eqnarray}
  i_{HSC} &=& i_{PS1} +0.00166891 -0.13944659 (i_{PS1}-z_{PS1}) -0.03034094 (i_{PS1}-z_{PS1})^{2}, \label{eq:i_PS1} \\
  g_{HSC} &=& g_{PS1} +0.00730066 +0.06508481 (g_{PS1}-r_{PS1}) -0.0151057 (g_{PS1}-r_{PS1})^{2}, \label{eq:g_PS1} 
\end{eqnarray}
The top panels of the four sections (a)--(d) of 
Figure~\ref{fig:colconv2} show the magnitude difference between 
HSC and either SDSS or PS1 systems of $g$ and $i$ band (i.e., color terms)
calculated for BPGS stars as a function of color.  
These panels indicate that the magnitude difference can be 
represented as quadratic expressions of color for wide color ranges.  
The bottom panels of the four sections (a)--(d) of 
Figure~\ref{fig:colconv2} show the difference 
between HSC magnitude and that calculated using equations 
\ref{eq:i_SDSS}--\ref{eq:g_PS1}. 
The RMS is calculated for stars with moderate color 
[i.e., $(g-r)_{SDSS,PS1} < 2.0$ or $(i-z)_{SDSS,PS1} < 1.0$] 
and found to be less than 0.01 mag.

\end{document}